\documentclass[%
reprint,
superscriptaddress,
 amsmath,amssymb,
 aps,
]{revtex4-1}

\usepackage{graphicx}
\usepackage{float}
\usepackage{braket}
\usepackage{subcaption}
\usepackage{color, ulem}
\usepackage{gensymb}
\usepackage{xcolor}

\newcommand{\VHr}{\ensuremath{V_{\mathrm{H}}(\mathbf{r})}}

\begin{document}

\title{Atomistic Hartree theory of twisted double bilayer graphene near the magic angle}

\author{Christopher T. S. Cheung}
\affiliation{Departments of Physics and Materials and the Thomas Young Centre for Theory and Simulation of Materials, Imperial College London, South Kensington Campus, London SW7 2AZ, UK\\}
\author{Zachary A. H. Goodwin}
\affiliation{Departments of Physics and Materials and the Thomas Young Centre for Theory and Simulation of Materials, Imperial College London, South Kensington Campus, London SW7 2AZ, UK\\}
\author{Valerio Vitale}
\affiliation{Departments of Physics and Materials and the Thomas Young Centre for Theory and Simulation of Materials, Imperial College London, South Kensington Campus, London SW7 2AZ, UK\\}
\author{Johannes Lischner}
\affiliation{Departments of Physics and Materials and the Thomas Young Centre for Theory and Simulation of Materials, Imperial College London, South Kensington Campus, London SW7 2AZ, UK\\}
\author{Arash A. Mostofi}
\affiliation{Departments of Physics and Materials and the Thomas Young Centre for Theory and Simulation of Materials, Imperial College London, South Kensington Campus, London SW7 2AZ, UK\\}

\date{\today}

\begin{abstract}
Twisted double bilayer graphene (tDBLG) is a moir\'e material that has recently generated significant interest because of the observation of correlated phases near the magic angle. We carry out atomistic Hartree theory calculations to study the role of electron-electron interactions in the normal state of tDBLG. In contrast to twisted bilayer graphene (tBLG), we find that such interactions do not result in significant doping-dependent deformations of the electronic band structure of tDBLG. However, interactions play an important role for the electronic structure in the presence of a perpendicular electric field as they screen the external field. Finally, we analyze the contribution of the Hartree potential to the crystal field, i.e. the on-site energy difference between the inner and outer layers. We find that the on-site energy obtained from Hartree theory has the same sign, but a smaller magnitude compared to previous studies in which the on-site energy was determined by fitting tight-binding results to ab initio density-functional theory (DFT) band structures. To understand this quantitative difference, we analyze the ab initio Kohn-Sham potential obtained from DFT and find that a subtle interplay of electron-electron and electron-ion interactions determines the magnitude of the on-site potential. 
\end{abstract}

\maketitle

\section{Introduction}
\label{sec:intro}

Twistronics~\cite{Carr_2017} is concerned with the effects that occur when stacking low-dimensional van der Waals materials and introducing a relative twist angle between them~\cite{moiresim,CarrRev2020}. The prototypical example of such a system is twisted bilayer graphene (tBLG), where a twist angle between two graphene sheets creates an emergent honeycomb pattern on a much larger length scale than the graphene honeycomb lattice~\cite{GBWT,Bistritzer12233,LDE,NSCS,PhysRevB.82.121407,Tritsaris_2020}. The physics of tBLG is rich, and exhibits superconductivity in proximity to correlated insulators~\cite{NAT_I,NAT_S,TSTBLG,SOM}, highly tunable van Hove singularities~\cite{NAT_CO,NAT_SS,NAT_MEI}, Dirac revivals~\cite{Wong2020,Zondiner2020}, strange metallic behaviour~\cite{SMTBLG,Polshyn2019strange}, and nematic order~\cite{NAT_CO,NAT_MEI,Caonematicity2020}. This has motivated the investigation of other graphene-based moir\'e materials, such as twisted double bilayer graphene (tDBLG)~\cite{KoshinoMikito2019Bsat,BIBI,haddadi2019moir,PhysRevLett.123.197702,STSCTDB,Samajdar2020}. 

In transport experiments on magic-angle tDBLG, correlated insulators have been observed at a doping of two electrons per moir\'e unit cell in applied electric fields~\cite{BIBI,cao2019electric,TCT,PhysRevLett.123.197702,He2021}. For the undoped system, a substantial band gap emerges in the presence of a perpendicular electric field; this band gap is not caused by electron interactions~\cite{KoshinoMikito2019Bsat}. Signatures of superconductivity have also been reported, but robust superconductivity has not yet been confirmed in this system~\cite{BIBI,cao2019electric,TCT,PhysRevLett.123.197702,He2021,Choi2021Dichotomy}. Scanning tunneling microscopy (STM) experiments have also observed correlated insulating states~\cite{Liu2021,Crommie2021} and nematic ordering~\cite{Carmen2021} near the magic angle. 

The electronic structure of tDBLG has been studied using a variety of methods including continuum theories~\cite{KoshinoMikito2019Bsat} and atomistic tight-binding models~\cite{haddadi2019moir}. In Ref.~\citenum{haddadi2019moir}, it was found that the band gap of the undoped system obtained from tight binding differs from the result of first-principles density functional theory (DFT). To obtain better agreement, a phenomenological on-site energy of approximately $-$30~meV was added to the inner layers in the tight-binding calculations. This on-site potential has been referred to as the ``intrinsic symmetric polarisation'' (ISP) or crystal field. Similar results have been found in Refs.~\citenum{RickhausPeter2019GOiT} and \citenum{CulchacF.J.2020Fbag}. These works suggested that the difference in chemical environments between the inner and outer layers of tDBLG can result in charge transfer between layers which in turn gives rise to significant electron-electron interaction effects in the normal state.

Long-ranged electron-electron interactions have been found to play a important role in the normal state of tBLG~\cite{EE,Cea2019,Rademaker2019,PHD_4,Bascones2020,PHD_6}. Specifically, Hartree theory calculations revealed that such interactions induce pronounced deformation in the electronic band structure which explain the experimentally observed pinning of the Fermi level as function of doping~\cite{EE,Cea2019,Rademaker2019,PHD_4,Bascones2020,PHD_6}. 

In contrast to tBLG, no Fermi level pinning has been observed in tDBLG, in agreement with continuum model calculations of the electronic structure~\cite{Liu2021,Crommie2021,Carmen2021}. However, such continuum model calculations only capture the long-range interaction between electrons, but fail to accurately describe the interaction between nearby electrons on different layers. To address this issue, it would be highly desirable to carry out atomistic Hartree theory calculations of tDBLG.

In this paper, we investigate the role of electron-electron interactions in tDBLG within atomistic Hartree theory. We find that the electronic band structure of tDBLG is not sensitive to electron or hole doping -- in stark contrast to tBLG. Next, we consider the effect of a perpendicular electric field and find that electron-electron interactions play an important role in screening the externally applied field. Finally, we evaluate the contribution of the Hartree potential to the crystal field and find that it has the correct sign, but is too small compared to previous findings. To understand this discrepancy, we analyze the Kohn-Sham potential of an ab initio DFT calculation of tDLBG which produces a crystal field in good agreement with previous results. This suggests that ab initio calculations are required for a quantitatively accurate description of the crystal field in tDBLG.

\section{Methods}

We study commensurate moir\'e unit cells of tDBLG consisting of twisted AB stacked bilayers. The bilayers are initially stacked directly on top of each other, similar to the structure of graphite, and the top bilayer is rotated anticlockwise about an axis normal to the bilayers that passes through a carbon atom in each bilayer. The moir\'e lattice vectors are $\textbf{R}_{1} = n\textbf{a}_{1} + m \textbf{a}_{2}$ and $\textbf{R}_{2} = -m\textbf{a}_{1} + (n + m) \textbf{a}_{2}$~\cite{LDE}, where $n$ and $m$ are integers that specify the moir\'e unit cell in terms of the graphene lattice vectors $\textbf{a}_{1}$ and $\textbf{a}_{2}$.

We relaxed these tDBLG structures using classical force fields as implemented in the LAMMPS software package~\cite{LAMMPS}. The AIREBO-Morse potential~\cite{AIREBO} was used for intralayer interactions; while the Kolmogorov-Crespi potential~\cite{KC} was used for interlayer interactions. More details can be found in Ref.~\citenum{Xia2020}. 

The electronic structure of tDBLG was investigated with an atomistic Hartree theory. The model outlined here is very closely related to that discussed in Ref.~\citenum{PHD_4}. For completeness, we provide all details here. The atomistic Hamiltonian that we solve is given by
\begin{equation}
\mathcal{\hat{H}} = \sum_{i}\varepsilon_{i}\hat{c}^{\dagger}_{i}\hat{c}_{i} + \sum_{ij}[t(\boldsymbol{\tau}_{i} - \boldsymbol{\tau}_{j})\hat{c}^{\dagger}_{j}\hat{c}_{i} + \text{H.c.}],
\label{eq:H}
\end{equation}
where $\hat{c}^{\dagger}_{i}$ and $\hat{c}_{i}$ are, respectively, the electron creation and annihilation operators associated with the p$_{z}$-orbital on atom $i$, and $\varepsilon_{i}$ is its on-site energy. 

The hopping parameters $t(\boldsymbol{\tau}_{i} - \boldsymbol{\tau}_{j})$ between atoms $i$ and $j$ (located at $\boldsymbol{\tau}_{i/j}$) are determined using the Slater-Koster rules~\cite{SK,EPG}
\begin{equation}
t(\textbf{r}) = \gamma_1e^{q_{\sigma}(1 - |\textbf{r}|/d)}\cos^{2}\varphi - \gamma_0e^{q_{\pi}(1 - |\textbf{r}|/a)}\sin^{2}\varphi,
\end{equation}
where $\gamma_{1} = 0.48$~eV and $\gamma_{0} = 2.81$~eV~\cite{haddadi2019moir} correspond, respectively, to $\sigma$- and $\pi$-hopping between p$_{z}$-orbitals, with associated decay parameters $q_{\sigma} = 7.43$ and $q_{\pi} = 3.14$~\cite{LDE,NSCS}. Also, $a = 1.397~\textrm{\AA}$ is the pristine carbon-carbon bond length, $d = 3.35~\textrm{\AA}$ is the pristine interlayer separation parameter, and $\varphi$ is the angle between the $z$-axis and the vector connecting atoms $i$ and $j$, and captures the angle-dependence of hoppings. Hoppings between carbon atoms that are separated by more than $10~\textrm{\AA}$ are neglected~\cite{EDS}. The Slater-Koster tight-binding parameters are based on a best fit to DFT low-energy bandstructures of graphene and bilayer graphene~\cite{LDE,NSCS,EPG}, with a slightly larger $\pi$-hopping parameter as introduced in Ref.~\citenum{haddadi2019moir} to improve agreement with the low-energy DFT bandstructure of tDBLG at large twist angles.

We decompose the on-site energy $\varepsilon_{i}$ in Eq.~\eqref{eq:H} into two contributions, $\varepsilon_{i} = \varepsilon_{\alpha_i} + \varepsilon_{i}^{\textrm{el}}$. The first term, $\varepsilon_{\alpha_i}$, is constant within each layer (with $\alpha_i$ denoting the layer in which atom $i$ resides) and represents the effect of an applied electric field. The second term, $\varepsilon_{i}^{\textrm{el}}$, is the contribution to the on-site energy from electron interactions and is determined self-consistently according to

\begin{equation}
\varepsilon_i^{\textrm{el}} = \int d\textbf{r} \phi_z^2(\textbf{r}-\boldsymbol{\tau}_i)\VHr,
\end{equation}
where $\phi_z(\textbf{r})$ is the p$_z$ orbital of the carbon atoms, and the Hartree potential \VHr\ is determined from the electron density $n(\mathbf{r})$ and the screened electron-electron interaction $W(\mathbf{r})$ via
\begin{equation}
\VHr = \int d\textbf{r}' W(\textbf{r}-\textbf{r}') [n(\textbf{r}') - n_0(\textbf{r}')].
\label{eq:VH}
\end{equation} 
Here, $n_0(\mathbf{r})$ is a reference electron density that ensures overall charge neutrality and whose subtraction in Eq.~\eqref{eq:VH} avoids double counting of Hartree interactions in the uniform system~\cite{NSCS}. The electron density is determined through
\begin{equation}
n(\textbf{r}) = \sum_{n\textbf{k}} f_{n\textbf{k}} |\psi_{n\textbf{k}}(\textbf{r})|^2,
\label{eq:el_den}
\end{equation}
where 
\begin{equation}
\psi_{n\mathbf{k}}(\mathbf{r}) = \dfrac{1}{\sqrt{N_{\textrm{k}}}}\sum_{\textbf{R}j}c_{n\textbf{k}j}e^{i\textbf{k}\cdot\textbf{R}}\phi_{z}(\textbf{r}-\boldsymbol{\tau}_j-\textbf{R})
\end{equation}
denotes the Bloch eigenstates of Eq.~\eqref{eq:H}, with subscripts $n$ and $\mathbf{k}$ denoting the band index and the crystal momentum, respectively. Also, $N_{\textrm{k}}$ is the number of k-points in the summation of the electron density, and $f_{n\textbf{k}}=2\Theta(\varepsilon_\mathrm{F}-\varepsilon_{n\mathbf{k}})$ is the spin-degenerate occupancy of state $\psi_{n\mathbf{k}}$ with eigenvalue $\varepsilon_{n\mathbf{k}}$ (where $\varepsilon_\mathrm{F}$ is the Fermi energy). Inserting the Bloch states into Eq.~\eqref{eq:el_den} gives
\begin{equation}
n(\textbf{r}) =\sum_{j}n_j\chi_j(\textbf{r}),
\end{equation}
where $\chi_j(\textbf{r}) = \sum_\mathbf{R} \phi_{z}^2(\textbf{r}-\boldsymbol{\tau}_j-\textbf{R})$ (with $\mathbf{R}$ denoting the moir\'e lattice vectors) and the total number of electrons on the $j$-th p$_z$-orbital in the unit cell being determined by $n_j = \sum_{n\mathbf{k}} f_{n\textbf{k}}|c_{n\mathbf{k}j}|^2/N_{\textrm{k}}$. 

The reference density is taken to be that of a uniform system, $n_0(\textbf{r}) = \bar{n} \sum_j \chi_j(\textbf{r})$, where $\bar{n}$ is the average of $n_j$ over all atoms in the unit cell, which is related to the filling per moir\'e unit cell $\nu$ through $\bar{n}=1+\nu/N$, where $N$ is the total number of atoms in a moir\'e unit cell~\cite{Rademaker2019}.


\begin{figure*}[ht]
\centering
\begin{subfigure}[b]{0.405\textwidth}
\centering
\includegraphics[width=\textwidth]{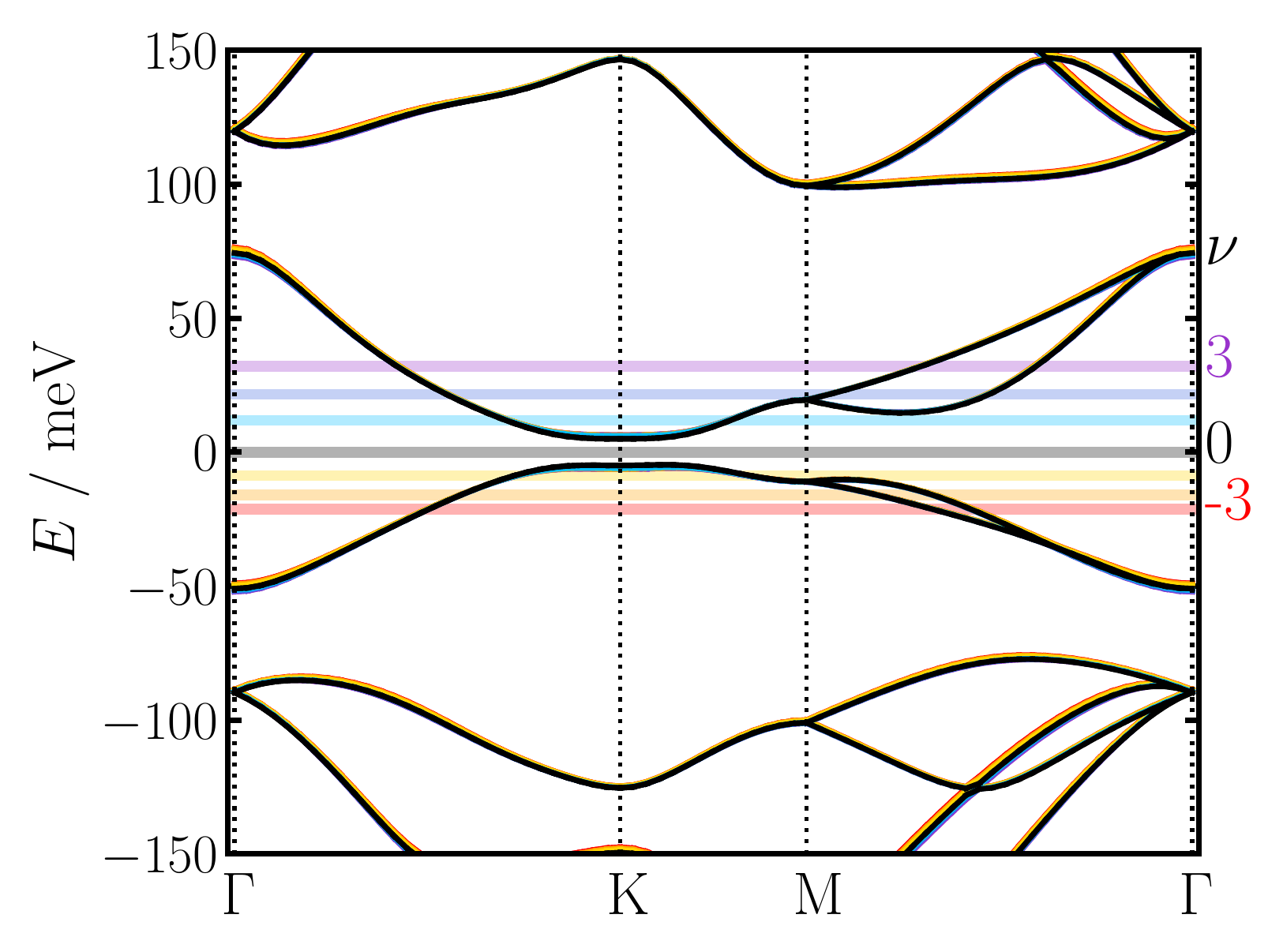}
\end{subfigure}
\centering
\begin{subfigure}[b]{0.4\textwidth}
\centering
\includegraphics[width=\textwidth]{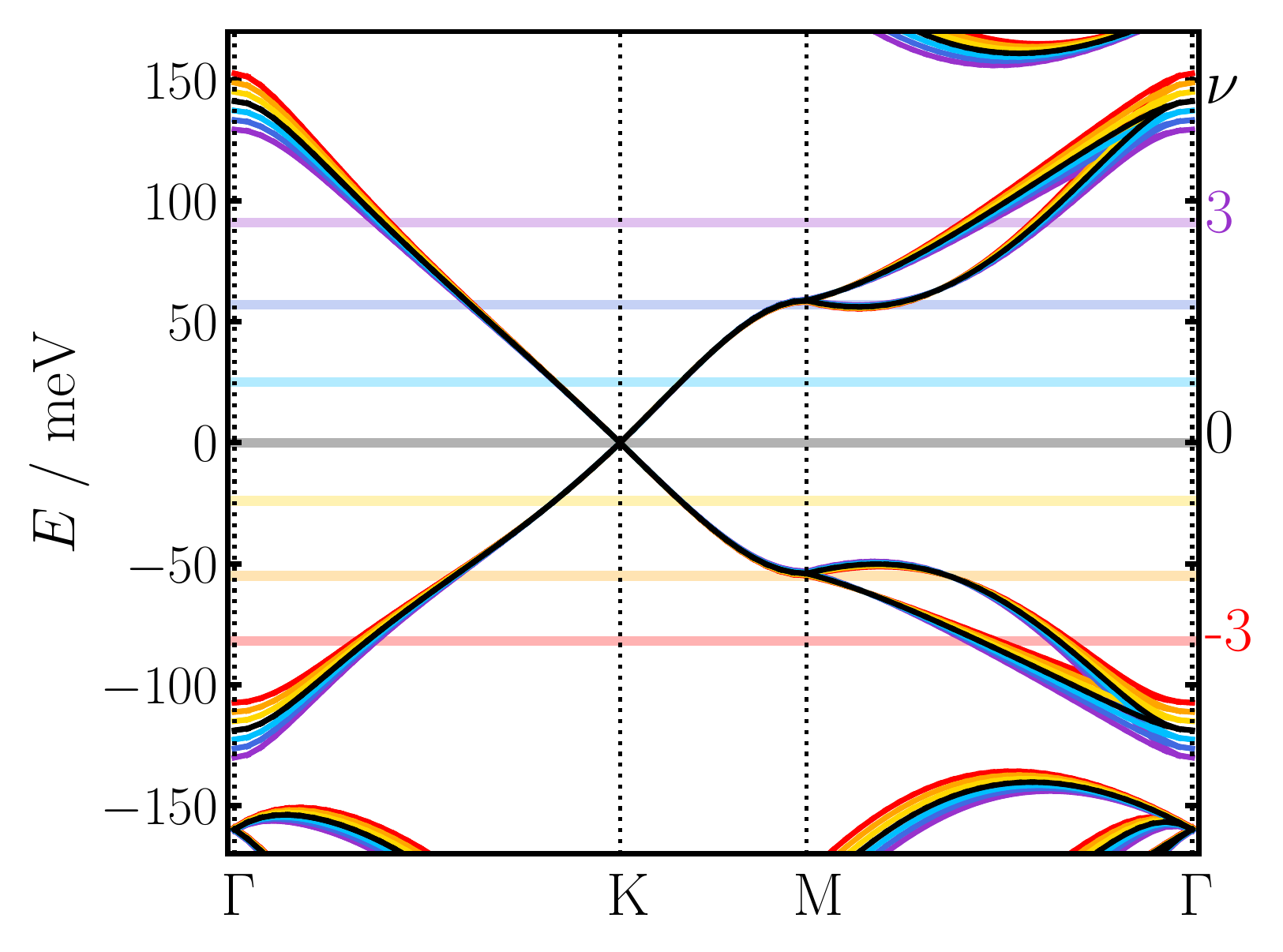}
\end{subfigure}
\centering
\begin{subfigure}[b]{0.405\textwidth}
\centering
\includegraphics[width=\textwidth]{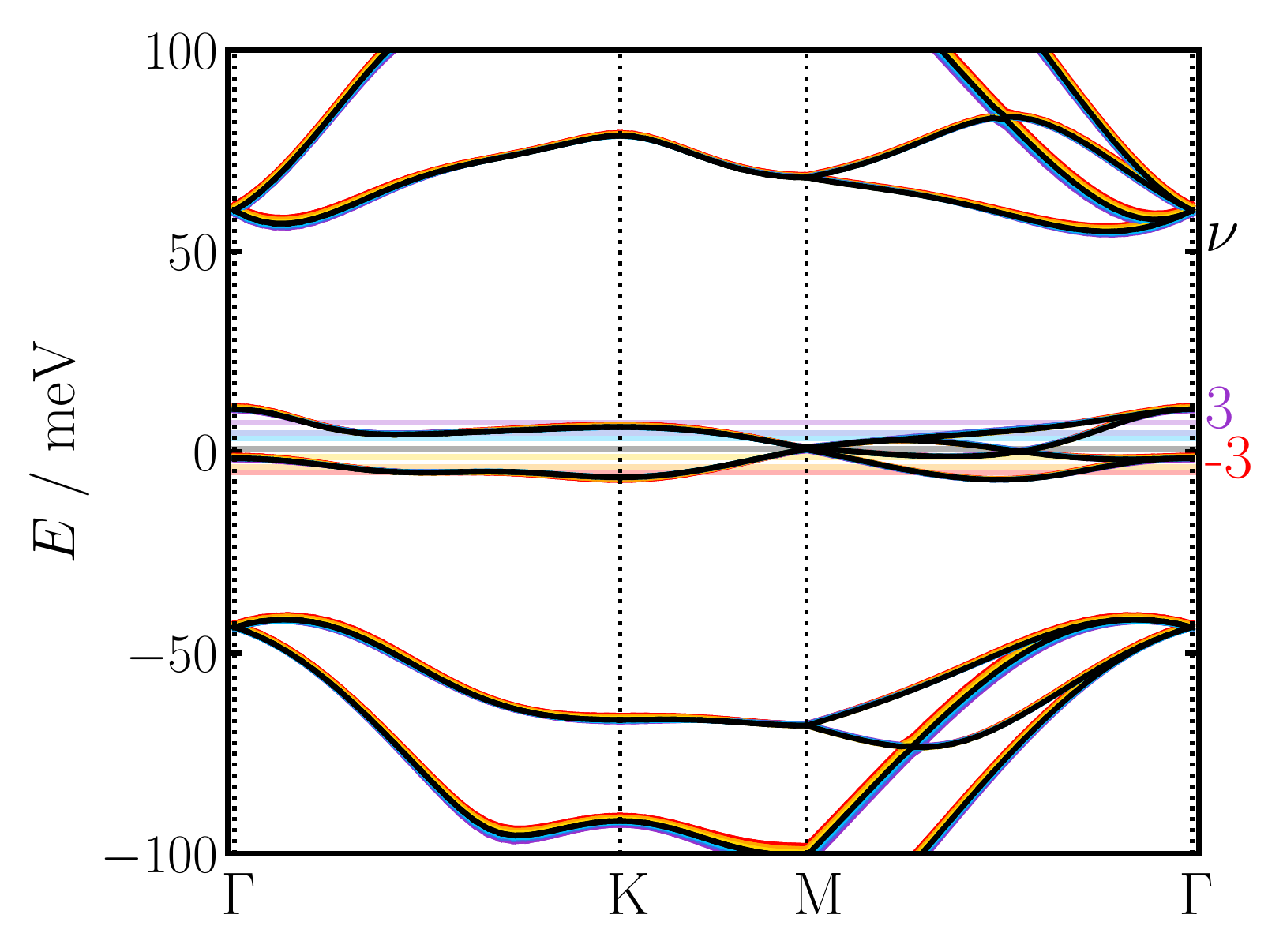}
\end{subfigure}
\centering
\begin{subfigure}[b]{0.4\textwidth}
\centering
\includegraphics[width=\textwidth]{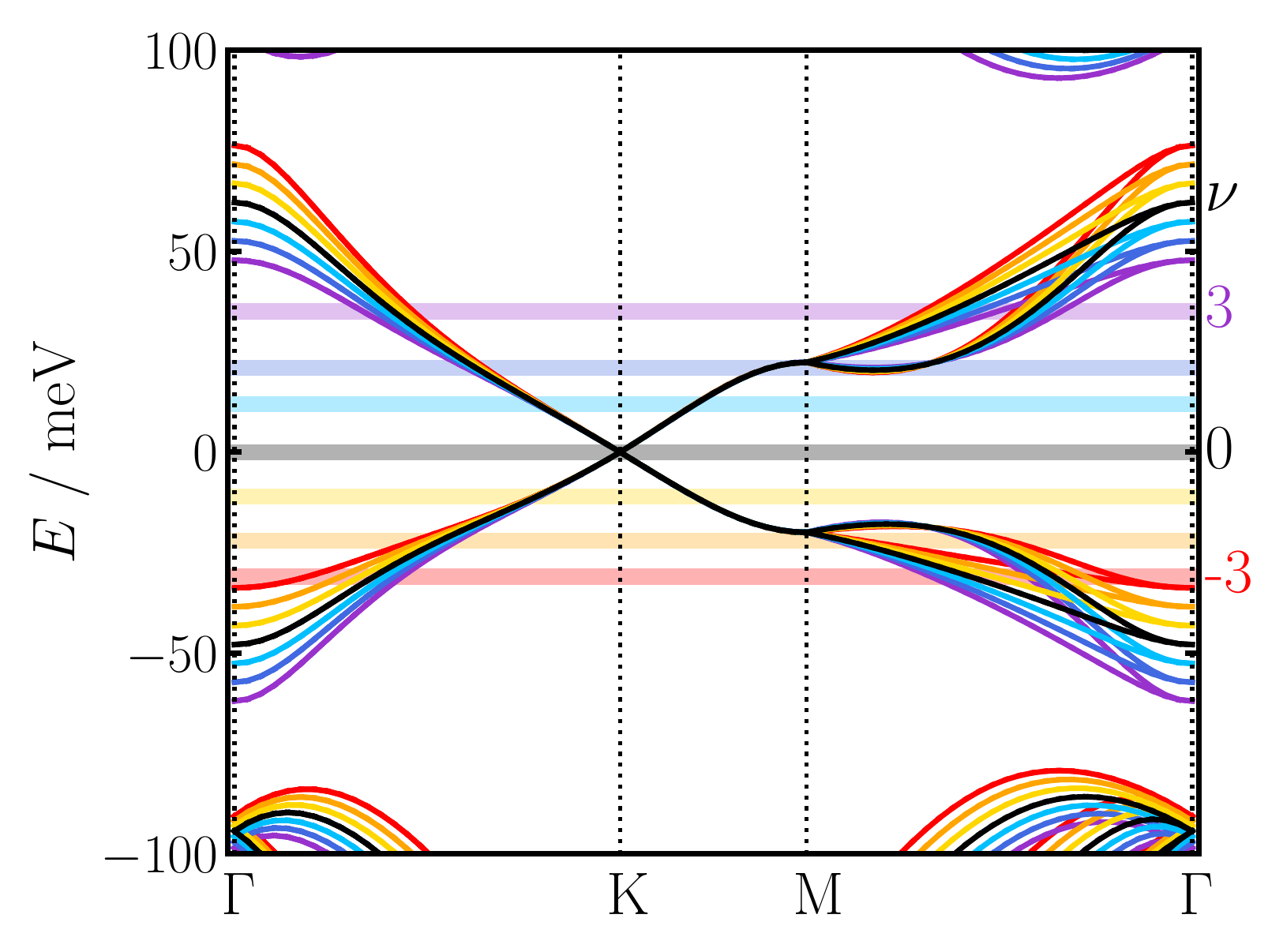}
\end{subfigure}
\caption{Left panels: Hartree theory band structure of doped ($-3\le \nu \le 3$) tDBLG at a twist angle of $\theta = $1.89$\degree$ (top) and $\theta = $1.41$\degree$ (bottom) with $\epsilon_\mathrm{bg}=4$. The horizontal lines denote the Fermi energy at each doping level. The band structure at charge neutrality is shown in black and the band structures at all doping levels are practically identical to it. Note all bands are aligned such that the zero of energy occurs in the middle of the band gap at the K-point. Right panels: results for tBLG, at the same twist angles, generated using the method outlined in Ref.~\citenum{PHD_6}.}
\label{fig:BS_D}
\end{figure*}

In transport experiments, there is often a metallic gate above and below the tDBLG, with a hexagonal boron nitride (hBN) substrate separating the gates from tDBLG. These metallic gates add or remove electrons from tDBLG and can also create electric fields across the system. These gates also screen the electron interactions in tDBLG, and taking this effect into account has been shown to be important in tBLG~\cite{PHD_3,Stepanov2020untying,Saito2020independent}. Therefore, we use a double metallic gate screened interaction
\begin{equation}
W(\textbf{r}) = \dfrac{e^2}{4\pi\epsilon_0\epsilon_\mathrm{bg}}\sum_{m=-\infty}^{\infty} \dfrac{(-1)^{m}}{\sqrt{|\textbf{r}|^2 + (2m\xi)^2}},
\label{WMG}
\end{equation} 
where $\xi$ is the thickness of the hBN dielectric substrate, which provides a background dielectric constant $\epsilon_\mathrm{bg}$ and separates tDBLG from the metallic gate on each side~\cite{MGS,PHD_1,PHD_3}. We set $\xi = 10$~nm in all calculations, and $\epsilon_\mathrm{bg} = 4$, which corresponds to experiments in which tDBLG is encapsulated in hBN~\cite{Laturia2018}, unless otherwise stated. Note that we do not consider the case of the hBN being closely aligned with the graphene layers which induces significant changes to the electronic structure of the graphene layers~\cite{Cea2020hBN}.

In our atomistic model, we neglect contributions to the electron density from overlapping p$_{z}$-orbitals that do not belong to the same carbon atom, which is equivalent to treating $\phi^2_z(\textbf{r})$ as a delta-function. Therefore, we calculate the Hartree on-site energies using
\begin{equation}
\varepsilon_i^{\textrm{el}} = \sum_{j\textbf{R}}(n_j - \bar{n}) W_{\textbf{R}ij},
\label{eq:H_pot}
\end{equation}
with $W_{\textbf{R}ij}=W(\textbf{R}+\boldsymbol{\tau}_j-\boldsymbol{\tau}_i)$. If $\textbf{R}=0$ and $i=j$, we set $W_{0,ii}=U/\epsilon_\mathrm{bg}$ with $U=17$~eV~\cite{SECI}. 

To obtain a self-consistent solution of the Hartree theory, we use a $6\times 6$  k-point grid to sample the first Brillouin zone to converge the density in Eq.~\eqref{eq:el_den} and we sum over a $21\times 21$ supercell of moir\'e unit cells to converge the on-site energies of Eq.~\eqref{eq:H_pot}. Linear mixing of the electron density is performed with a mixing parameter of 0.1 or less (i.e., 10 percent of the new potential is added to 90 percent of the potential from the previous iteration). Typically, the Hartree potential converges to an accuracy of better than 0.1~meV per atom within 100 iterations. For doping levels where tDBLG is metallic, smaller mixing values and a larger number of iterations are sometimes needed to reach this convergence threshold.

\begin{figure*}[ht]
\centering
\begin{subfigure}[b]{0.4\textwidth}
\centering
\includegraphics[width=\textwidth]{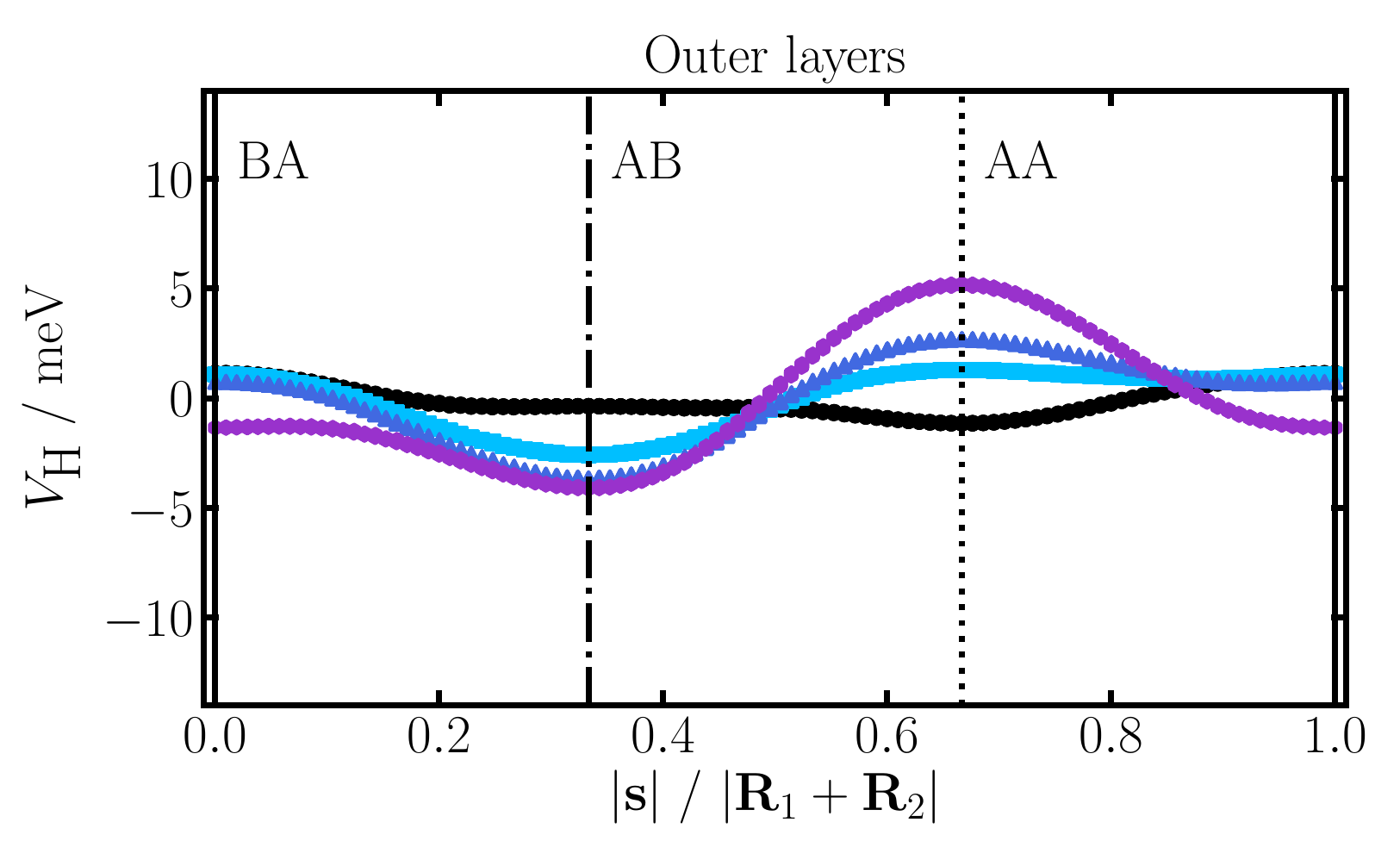}
\end{subfigure}
\begin{subfigure}[b]{0.4\textwidth}  
\centering 
\includegraphics[width=\textwidth]{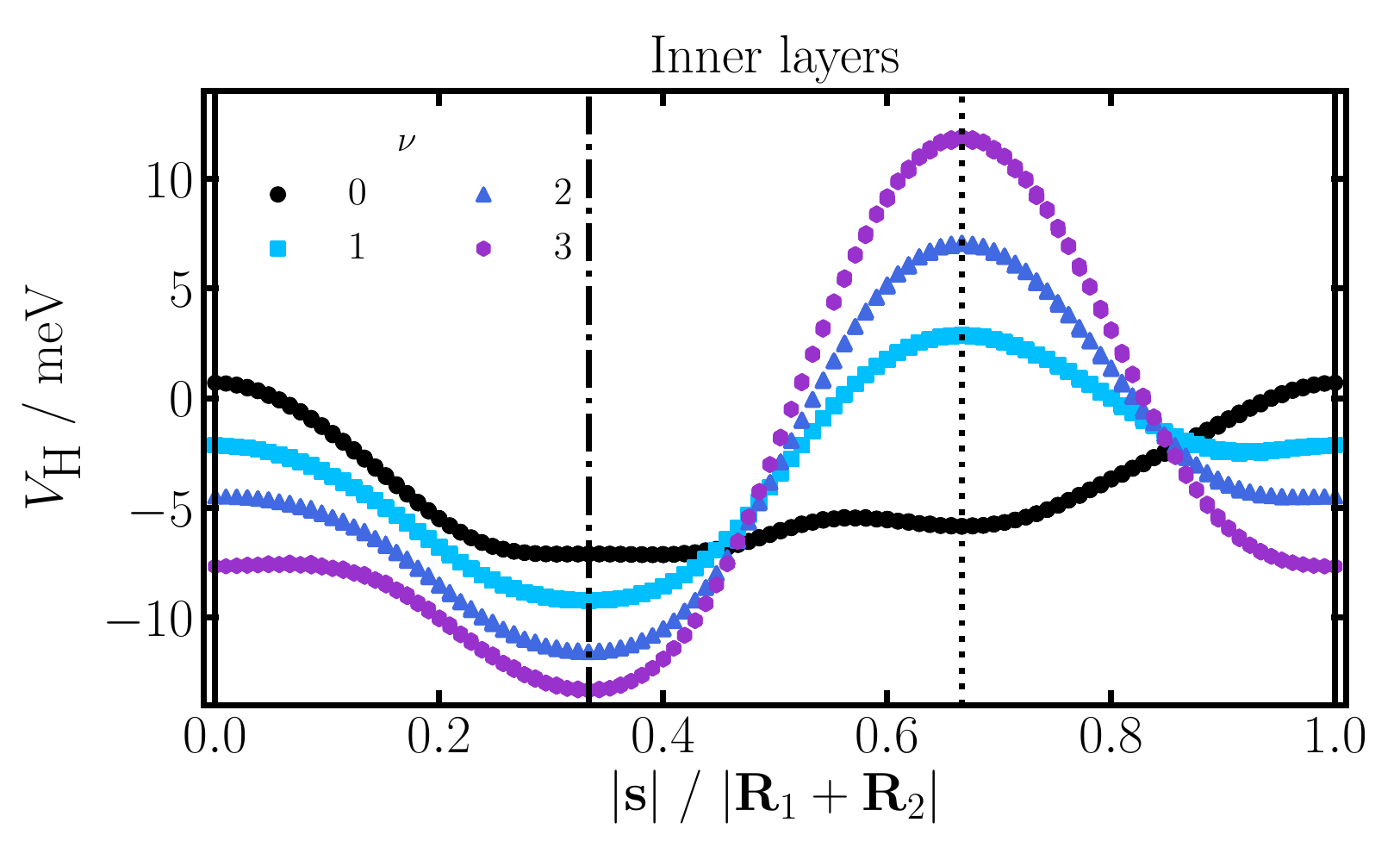}
\end{subfigure}
\centering
\begin{subfigure}[b]{0.4\textwidth}
\centering
\includegraphics[width=\textwidth]{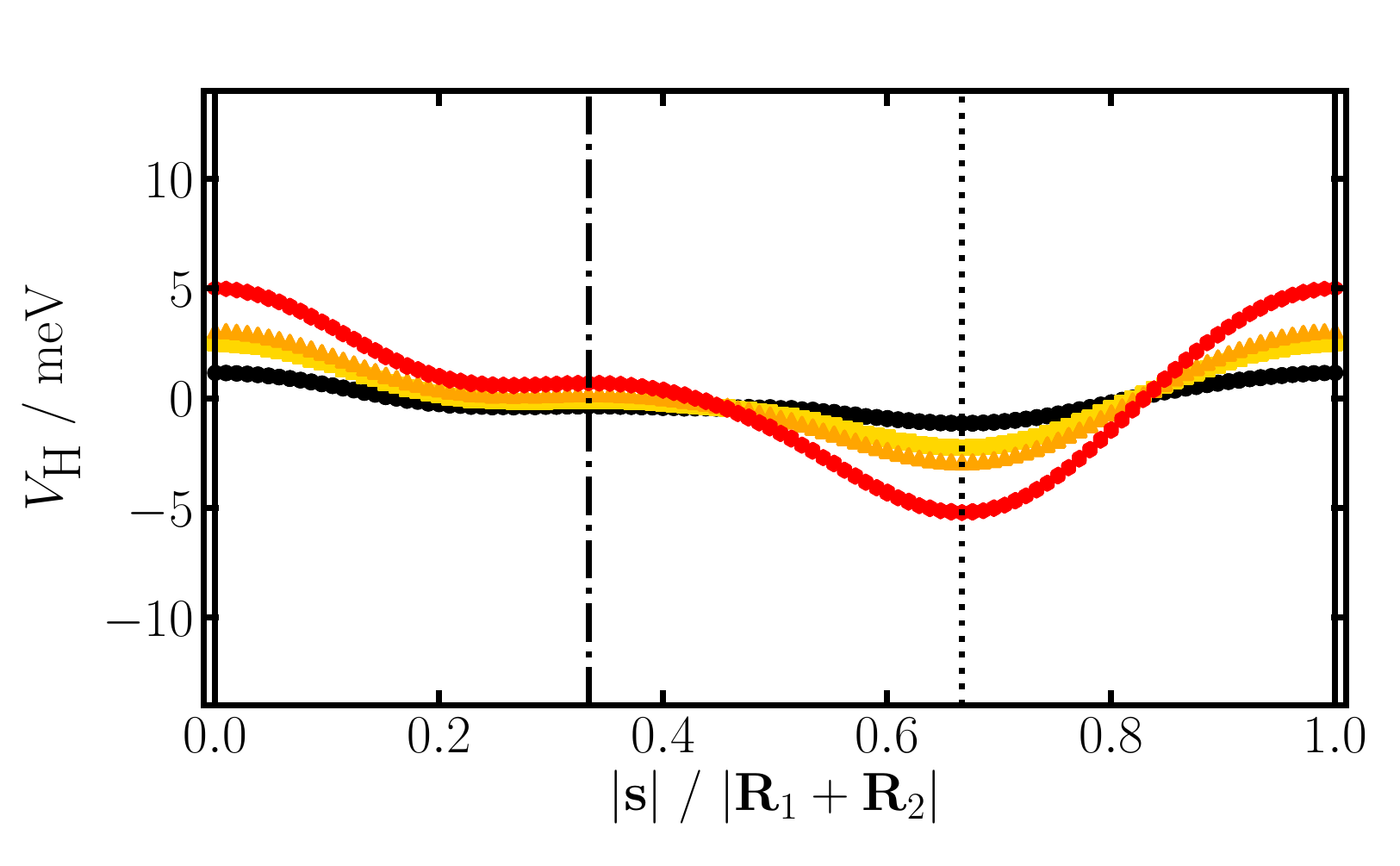}
\end{subfigure}
\begin{subfigure}[b]{0.4\textwidth}  
\centering 
\includegraphics[width=\textwidth]{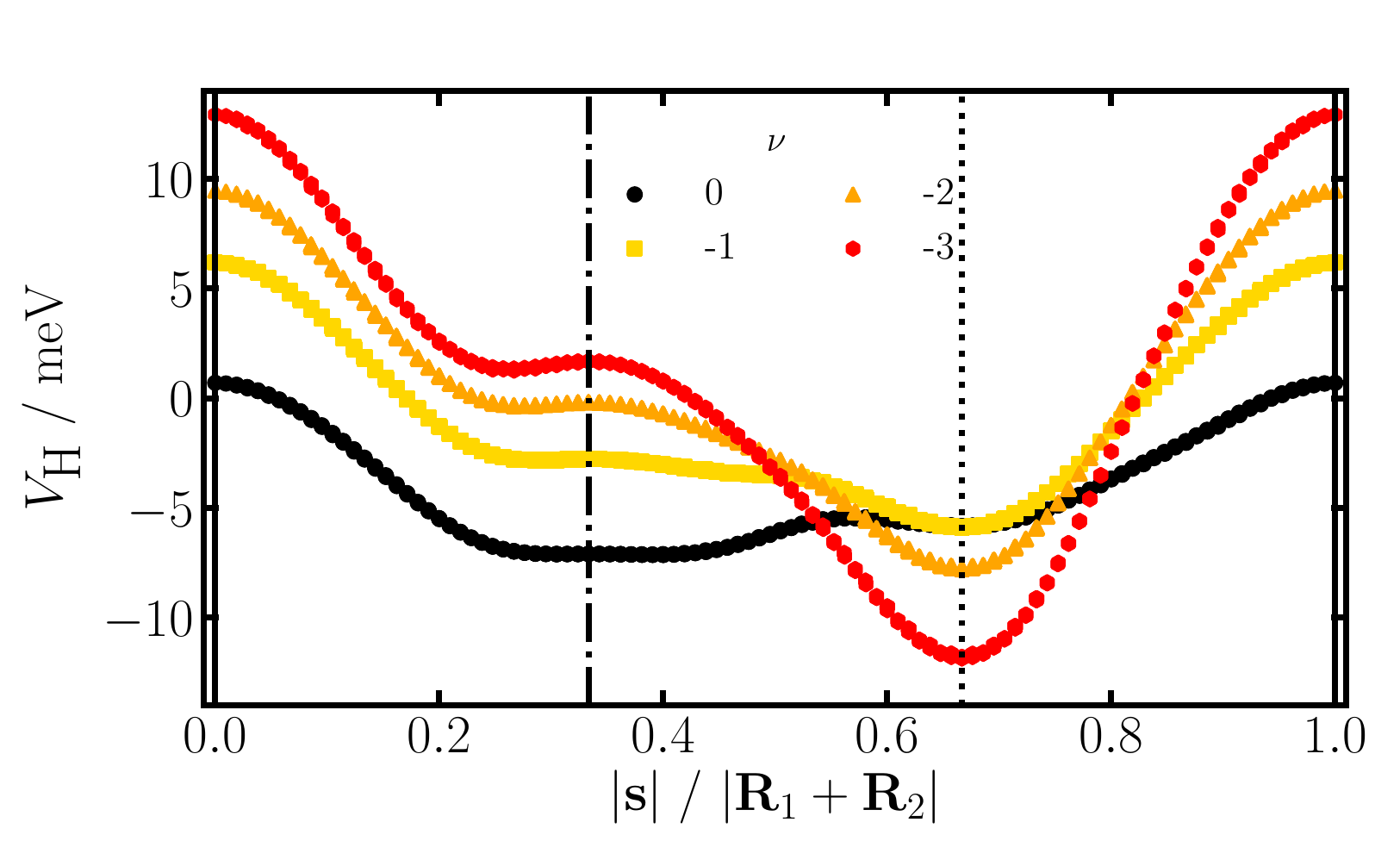}
\end{subfigure}
\caption{Locally-averaged Hartree potential along the diagonal of the moir\'e unit cell for electron and hole doped tDBLG with $\theta = $1.89$\degree$ and $\epsilon_\mathrm{bg}=4$. The variable $|\textbf{s}|$ measures the distance from the BA site of the moir\'e unit cell to a point $\mathbf{s}$ along its long diagonal. The vertical solid lines correspond to BA stacking of the inner layers, dotted-dashed lines to the AB stacking, and dotted lines to AA stacking. The left panels show results for the outer layers and the right panels are for the inner layers. Results for electron doped systems ($\nu>0$) are shown in the upper panels and results for hole doped systems ($\nu<0$) are shown in the lower panels. The Hartree potential in each layer has been locally averaged: the value at each atomic position was obtained by averaging the Hartree potential at this site with the Hartree potentials of the three nearest neighbours. This removes the atomic-scale oscillations of the Hartree potential.}
\label{fig:pot_strc}
\end{figure*}

\begin{figure*}[ht]
\centering
\begin{subfigure}[b]{0.3\textwidth}
\centering
\includegraphics[width=\textwidth]{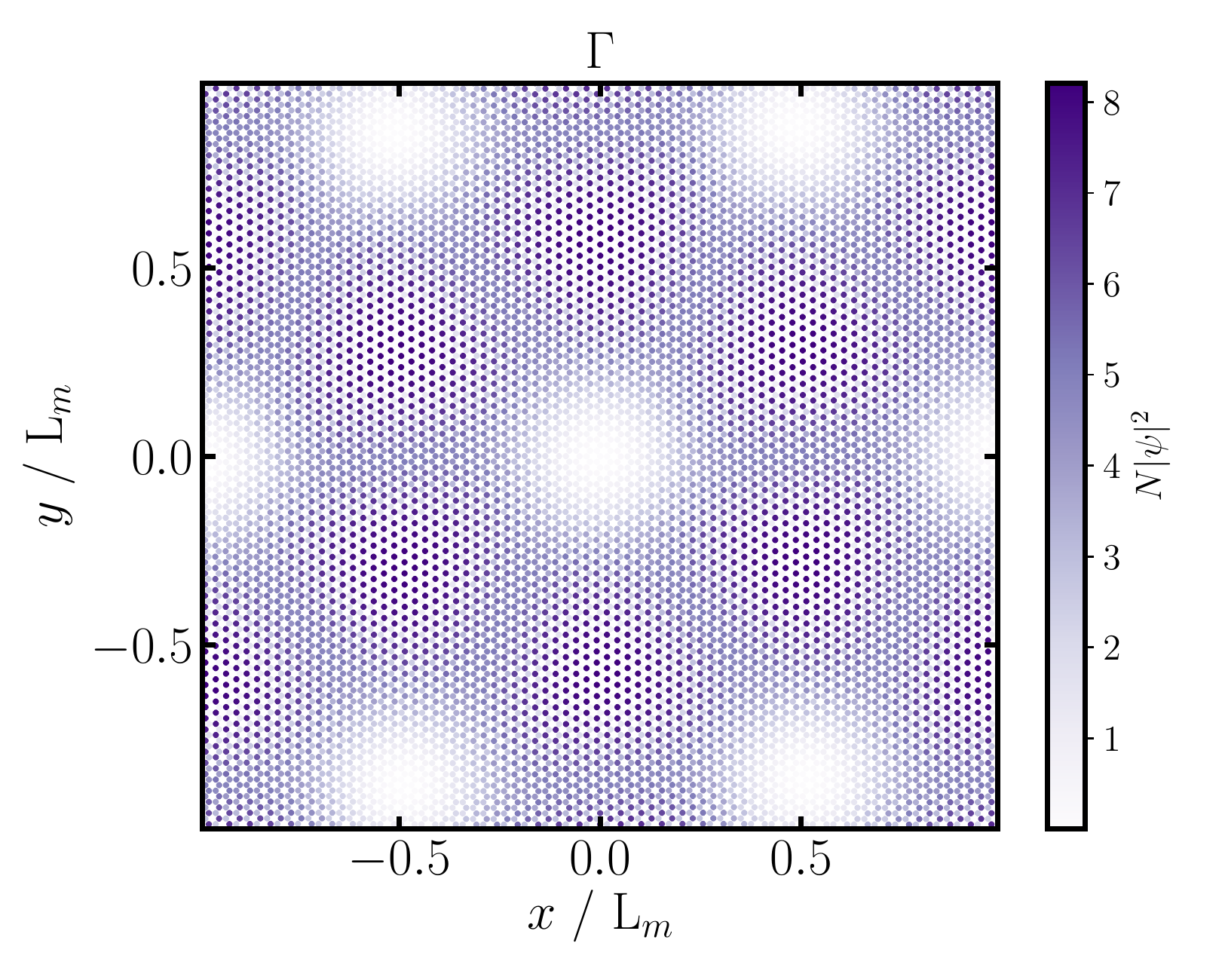}
\end{subfigure}
\centering
\begin{subfigure}[b]{0.3\textwidth}
\centering
\includegraphics[width=\textwidth]{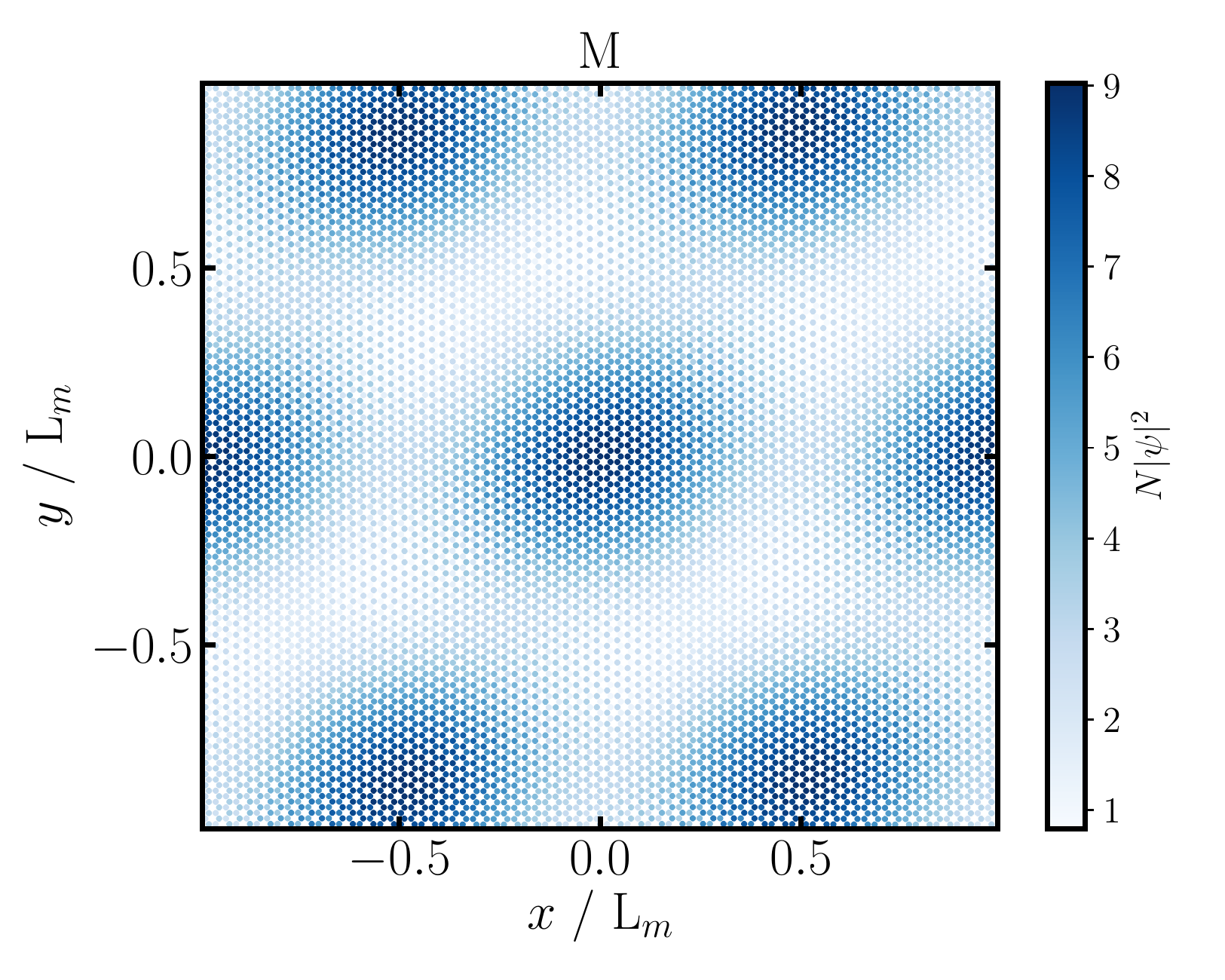}
\end{subfigure}
\centering
\begin{subfigure}[b]{0.3\textwidth}
\centering
\includegraphics[width=\textwidth]{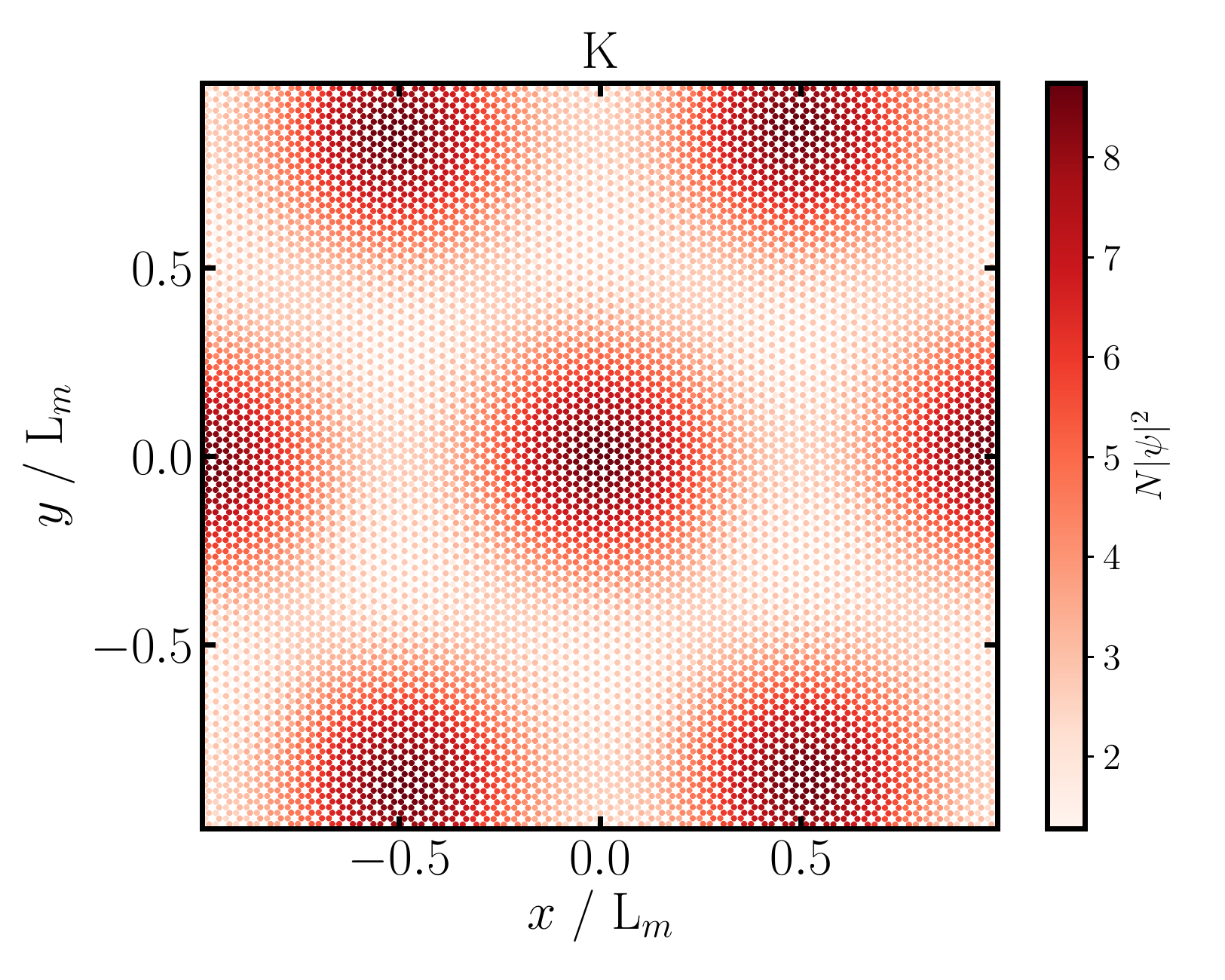}
\end{subfigure}
\caption{Square modulus of the flat-band wavefunctions (multiplied by the number of atoms in the moir\'e unit cell $N$) of tBLG at a twist angle of 1.89$^\circ$ at different crystal momenta. The $\Gamma$, M and K points are shown in the left, middle and right panels, respectively. At each crystal momentum the square moduli of the four flat-band states have been summed up. The origin is located at the centre of the AA stacking region and $L_m$ denotes the moir\'e length scale.} 
\label{fig:tBLG_WF}
\end{figure*}

DFT calculations are performed using \textsc{ONETEP}, a linear-scaling DFT code~\cite{onetep_2020,Ratclif_PRB98_2018}. We use the Perdew-Burke-Ernzerhof (PBE) exchange-correlation functional~\cite{PBE_PRL77}. For the electron-ion interaction we employ the projector-augmented-wave (PAW) formalism~\cite{PAW_PRB50,JOLLET20141246} (where PAW pseudopotentials are generated from ultra-soft pseudopotentials~\cite{GARRITY2014446}). The kinetic energy cutoff is set to 800~eV. A basis consisting of four non-orthogonal generalized Wannier functions (NGWFs) per carbon atom is used, which are optimised \textit{in situ}, and the ensemble-DFT approach~\cite{Serrano_JCP139_2013,Marzari_PRL79_1997} is adopted for the minimisation of the ground state energy.

\section{Results and Discussion}

\subsection{Doping dependent band structure}
\label{sec:doping-dependence}

\begin{figure*}[ht]
\centering
\begin{subfigure}[b]{0.3\textwidth}
\centering
\includegraphics[width=\textwidth]{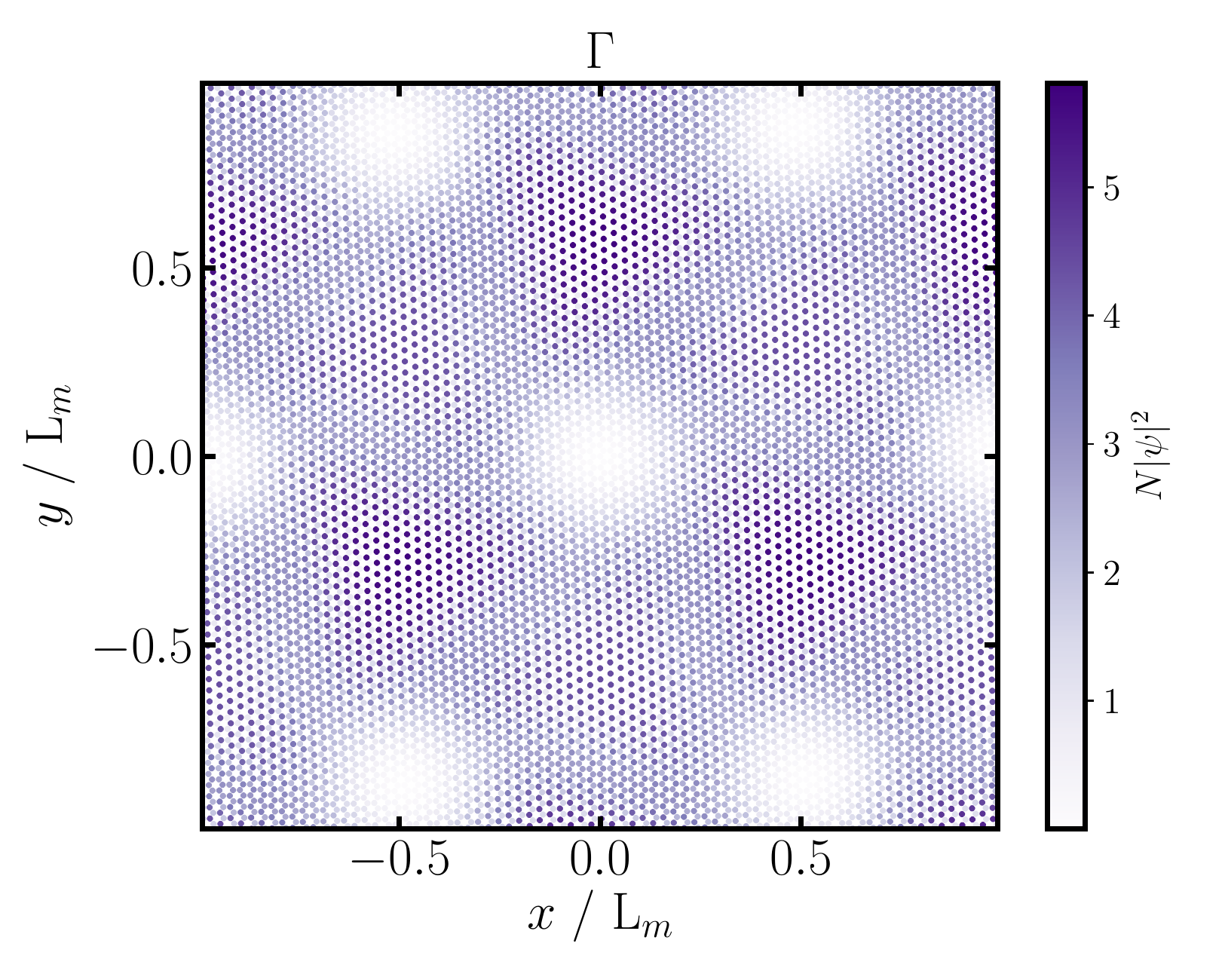}
\end{subfigure}
\centering
\begin{subfigure}[b]{0.3\textwidth}
\centering
\includegraphics[width=\textwidth]{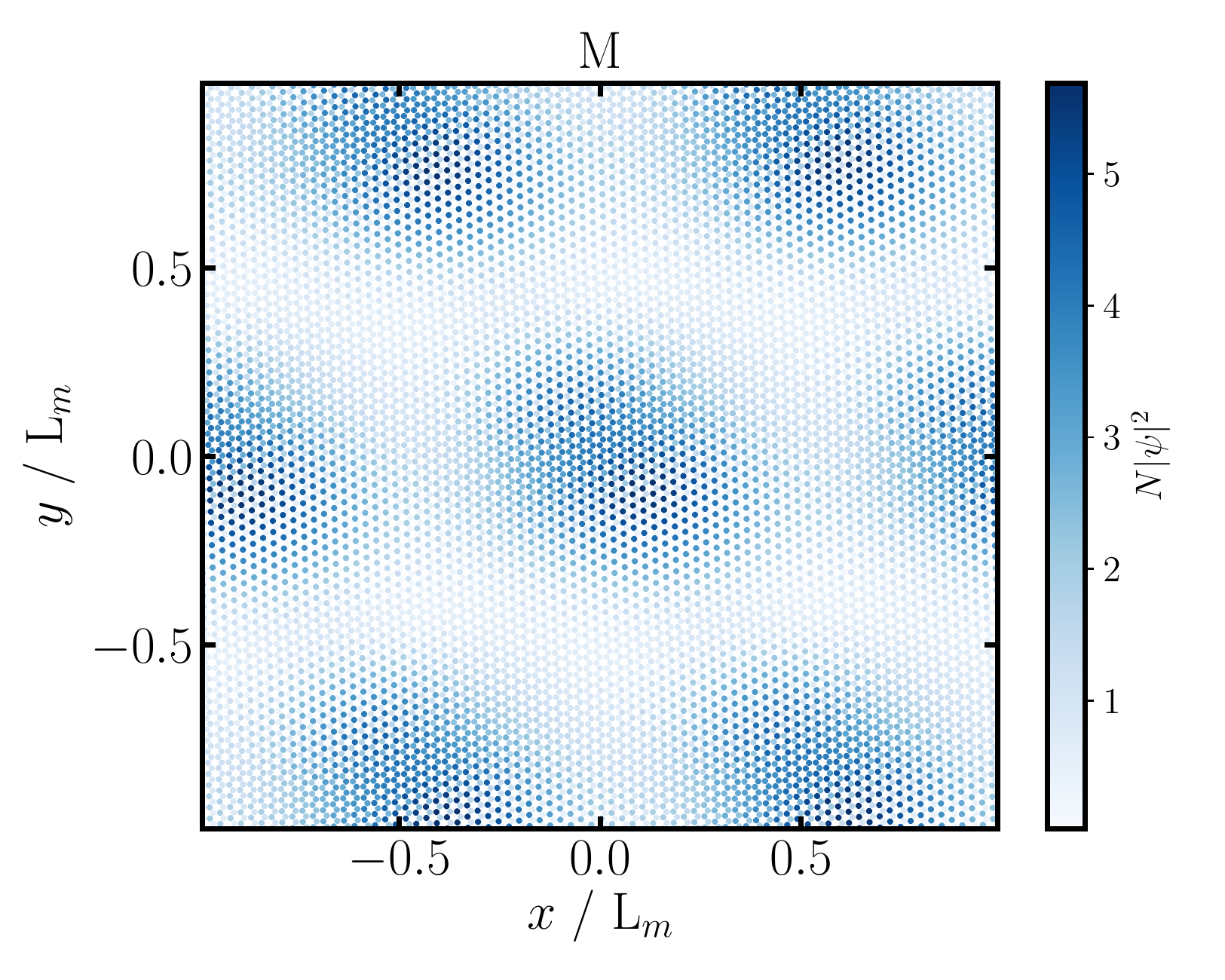}
\end{subfigure}
\centering
\begin{subfigure}[b]{0.3\textwidth}
\centering
\includegraphics[width=\textwidth]{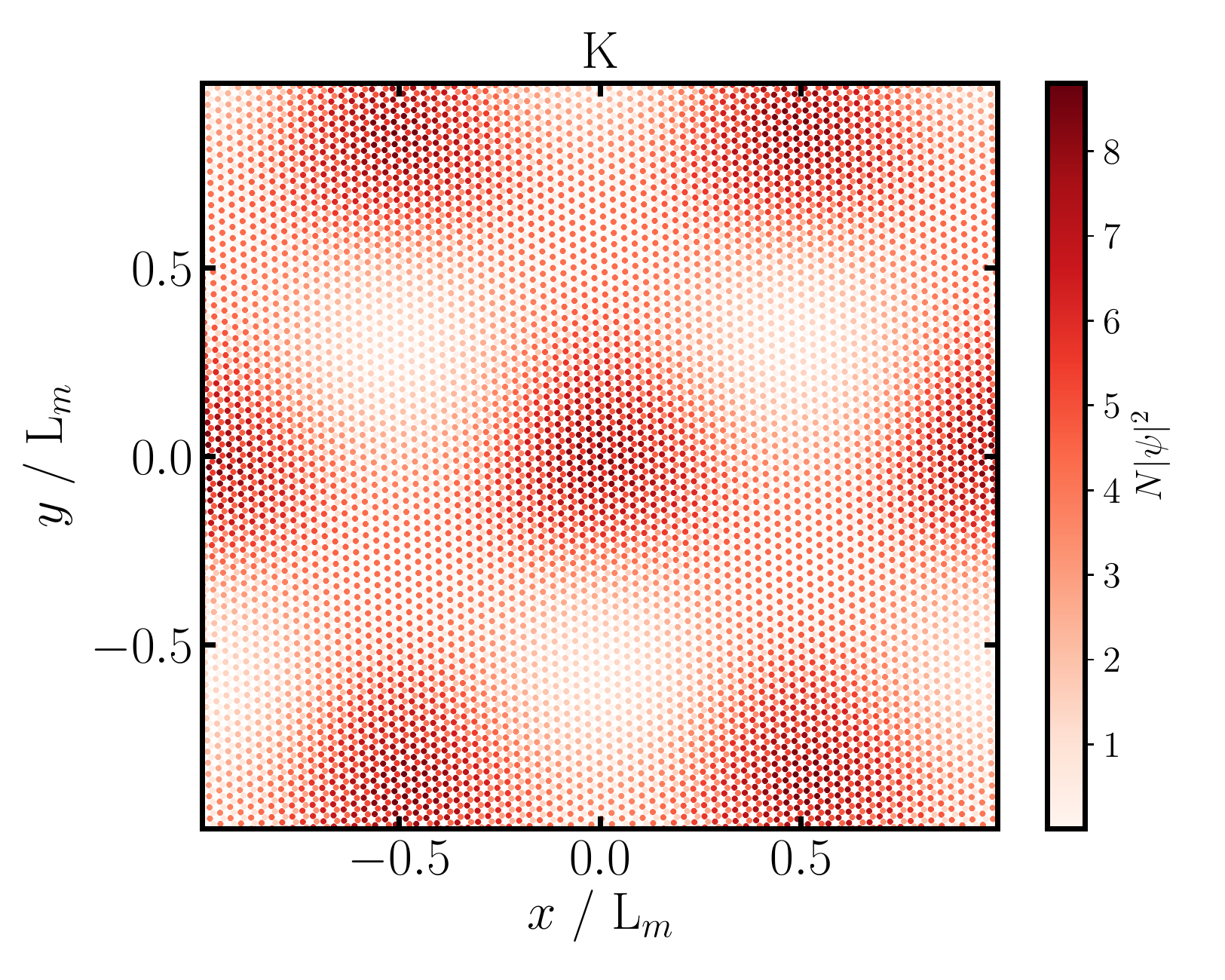}
\end{subfigure}
\centering
\begin{subfigure}[b]{0.3\textwidth}
\centering
\includegraphics[width=\textwidth]{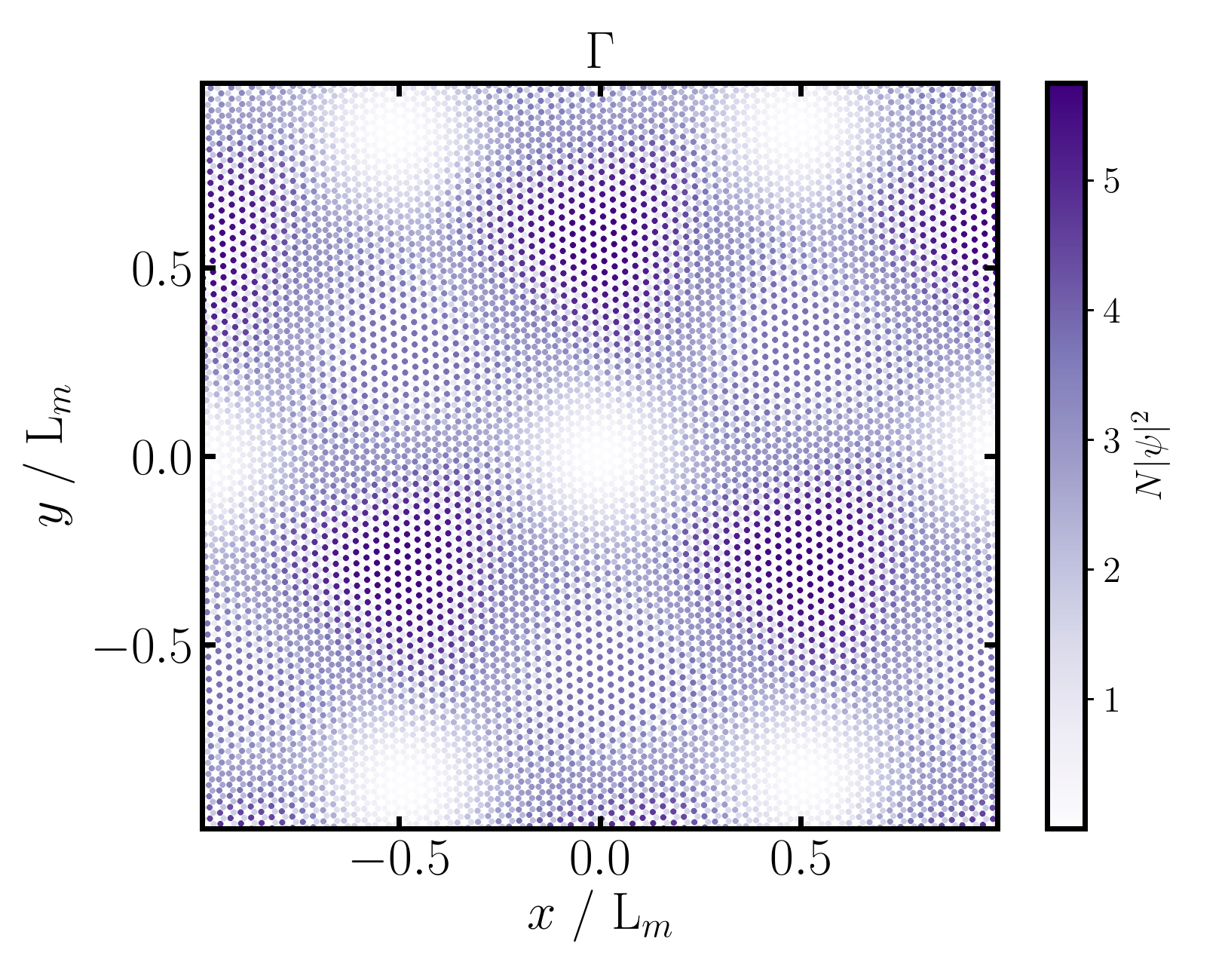}
\end{subfigure}
\centering
\begin{subfigure}[b]{0.3\textwidth}
\centering
\includegraphics[width=\textwidth]{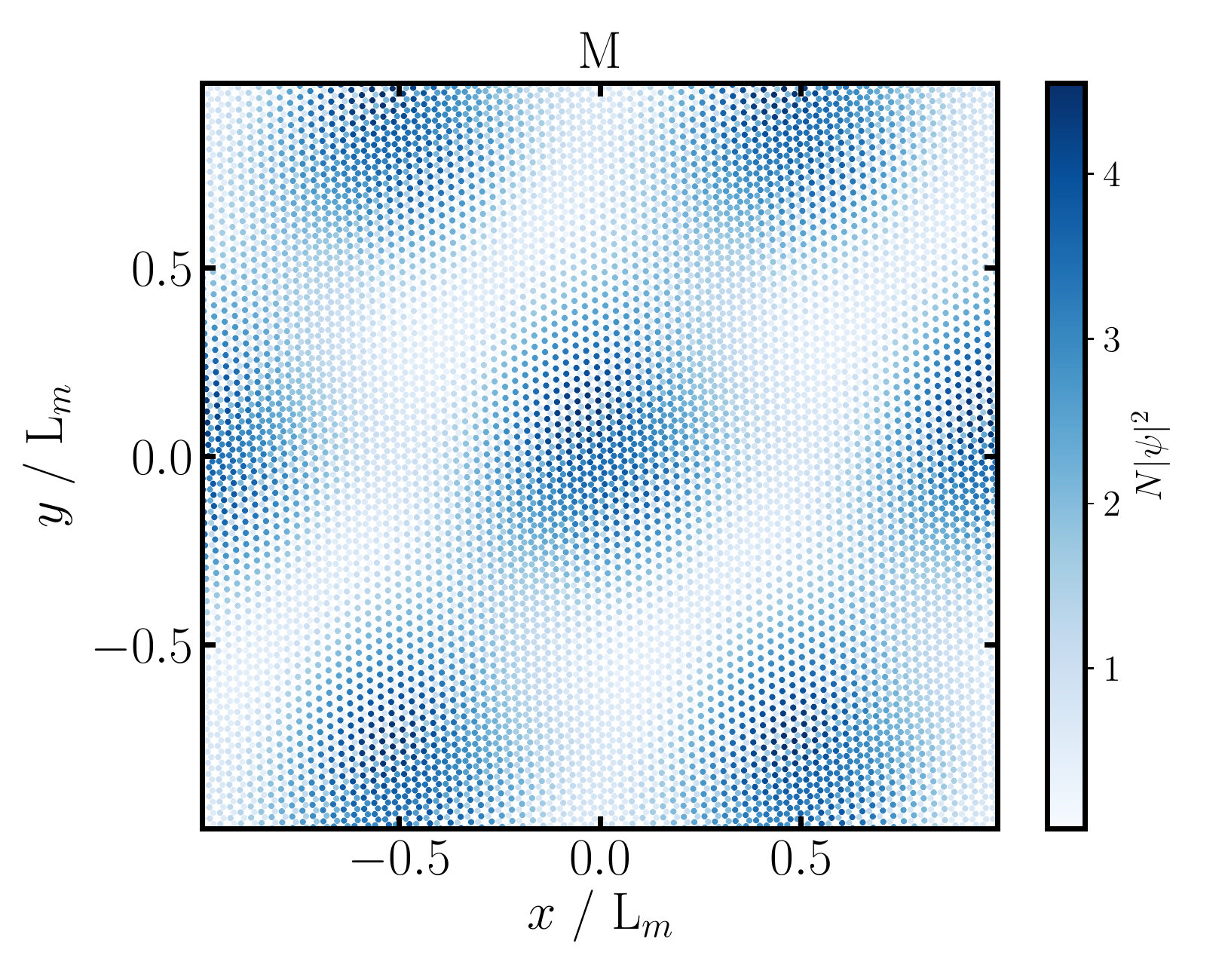}
\end{subfigure}
\centering
\begin{subfigure}[b]{0.3\textwidth}
\centering
\includegraphics[width=\textwidth]{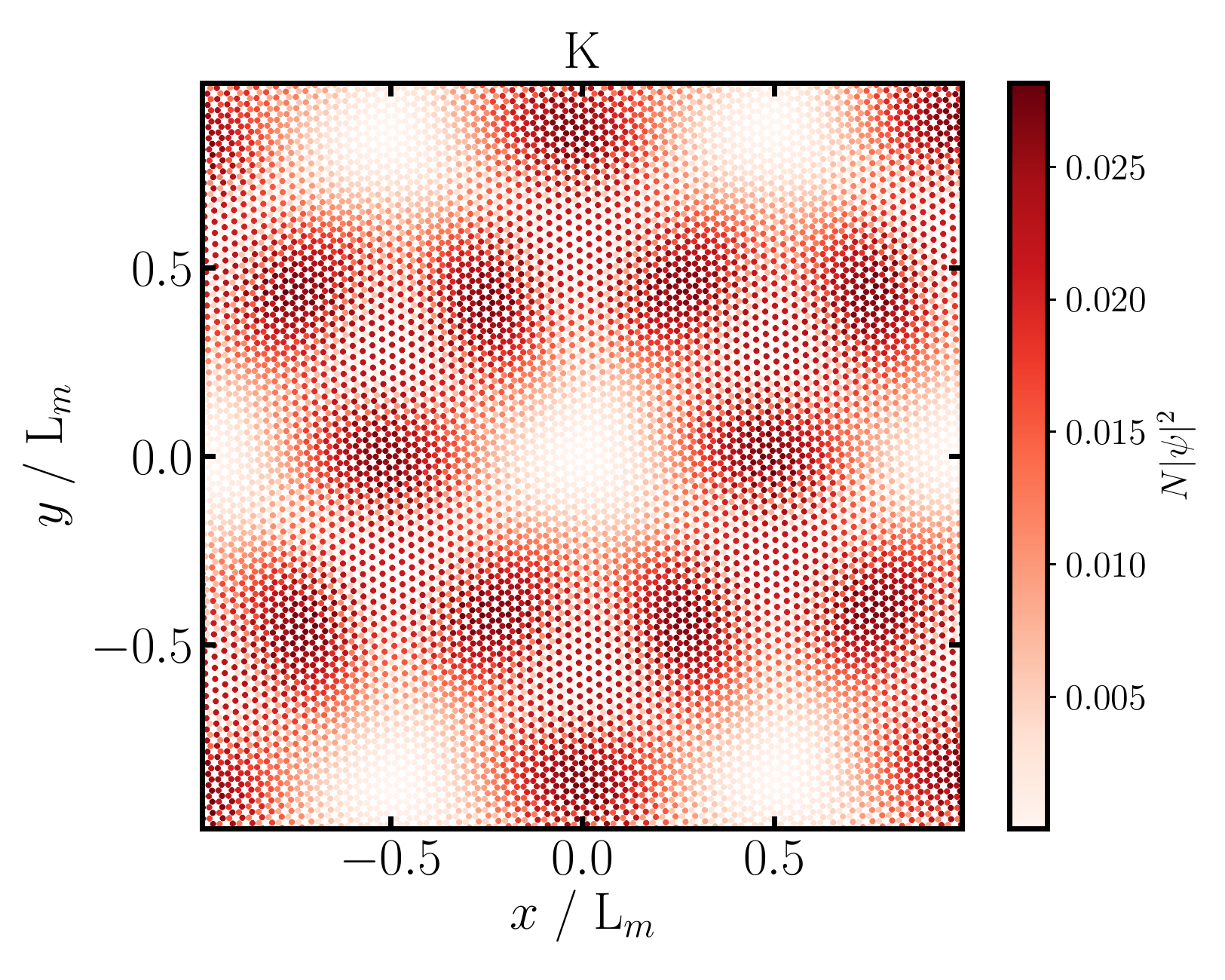}
\end{subfigure}
\caption{Square modulus of the flat-band wavefunctions on an inner layer of tDBLG at different crystal momenta. Top panels show results for the flat conduction band (whose square moduli have been summed), and the bottom panels show similar results for the two flat valence band states. See the caption of Fig.~\ref{fig:tBLG_WF} for further details.}
\label{fig:tDBLG_WF_inner}
\end{figure*}

\begin{figure*}[ht]
\centering
\begin{subfigure}[b]{0.3\textwidth}
\centering
\includegraphics[width=\textwidth]{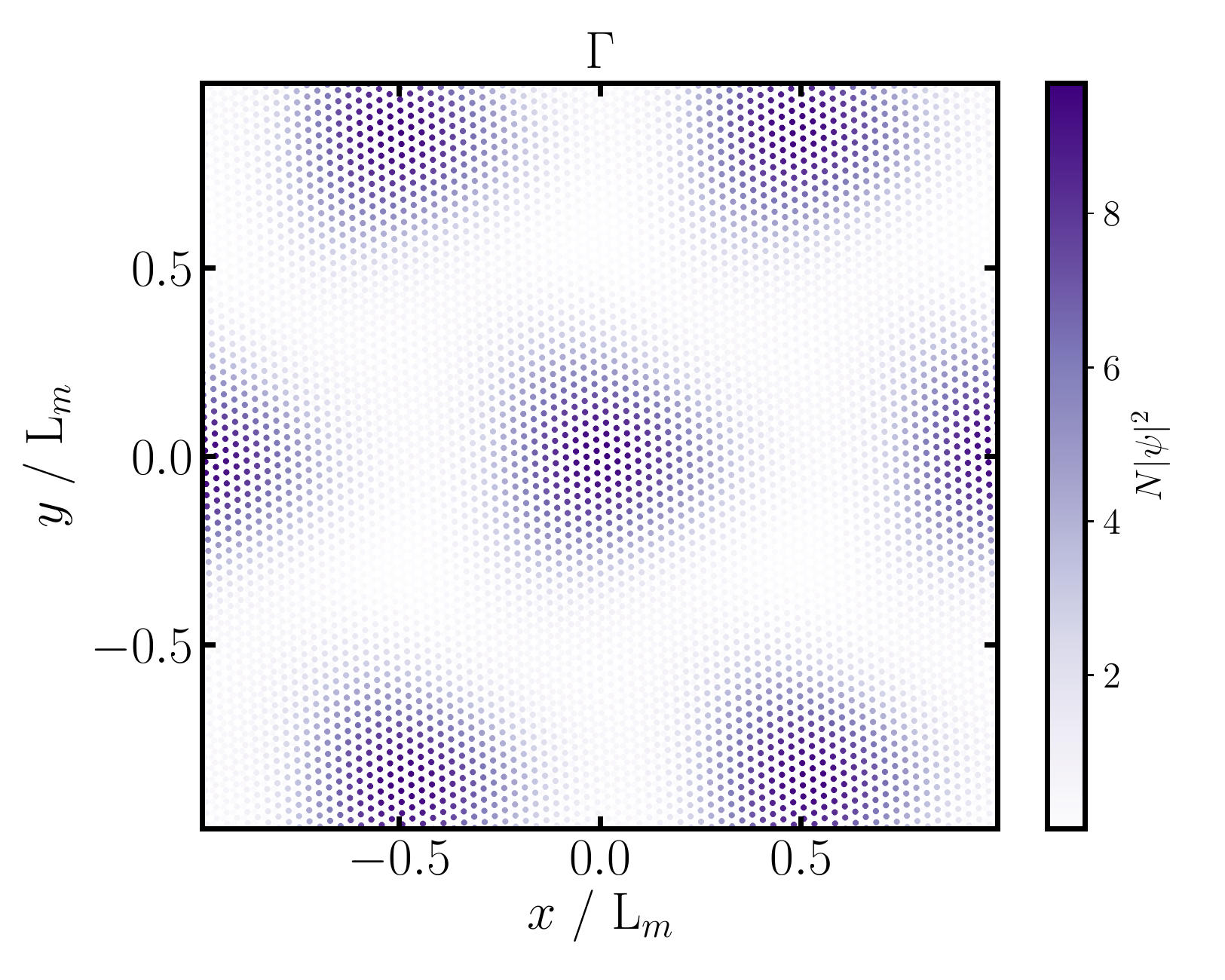}
\end{subfigure}
\centering
\begin{subfigure}[b]{0.3\textwidth}
\centering
\includegraphics[width=\textwidth]{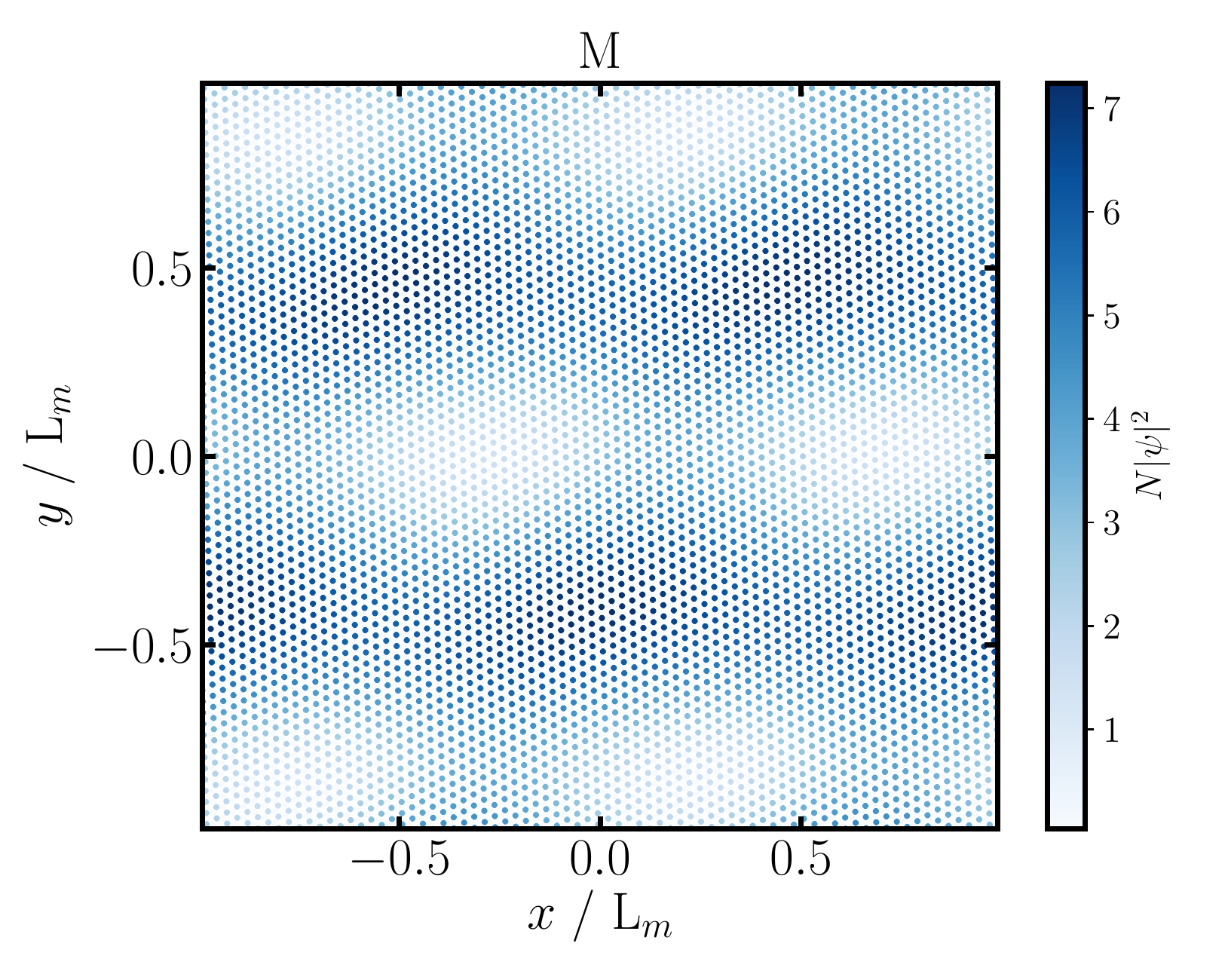}
\end{subfigure}
\centering
\begin{subfigure}[b]{0.3\textwidth}
\centering
\includegraphics[width=\textwidth]{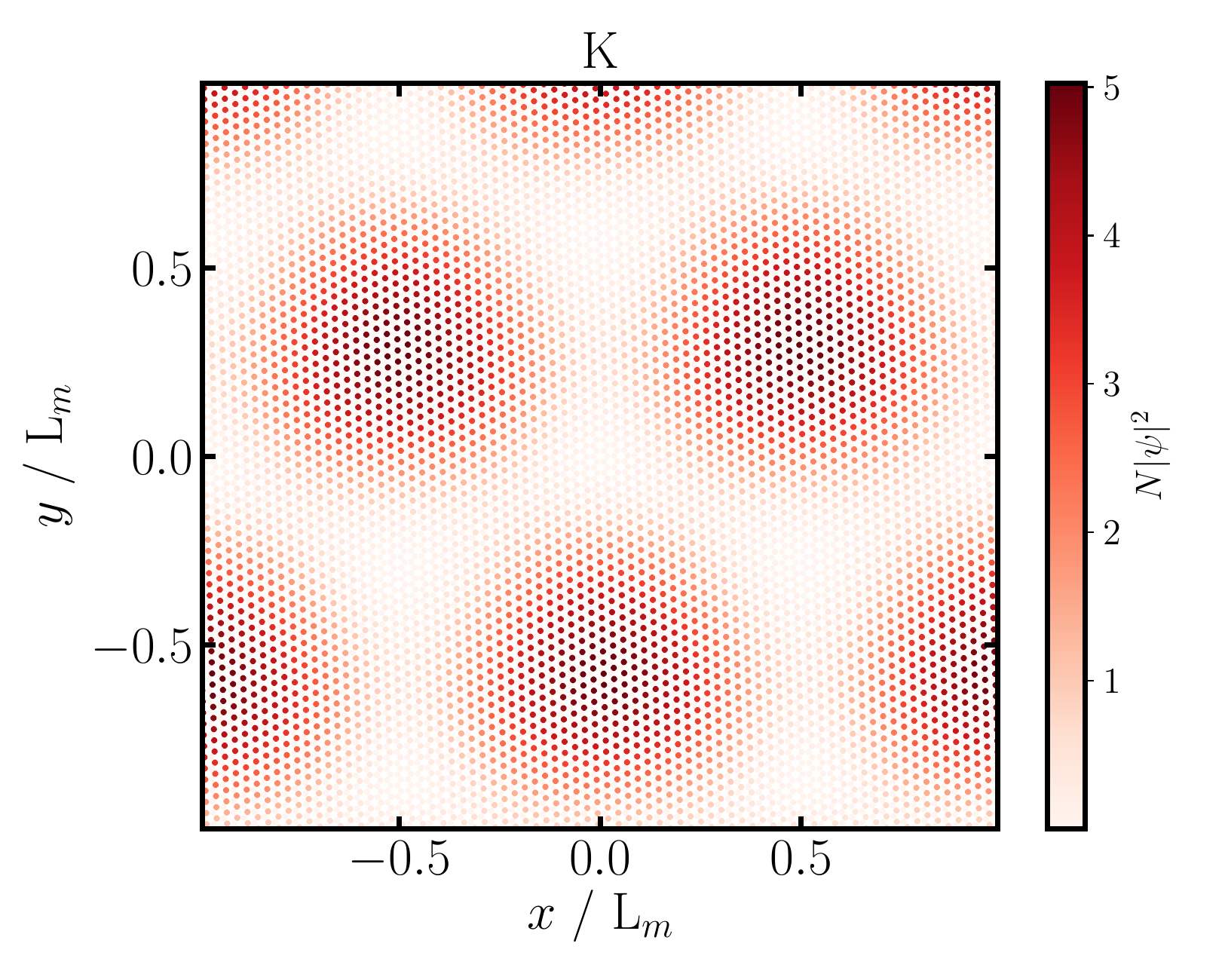}
\end{subfigure}
\centering
\begin{subfigure}[b]{0.3\textwidth}
\centering
\includegraphics[width=\textwidth]{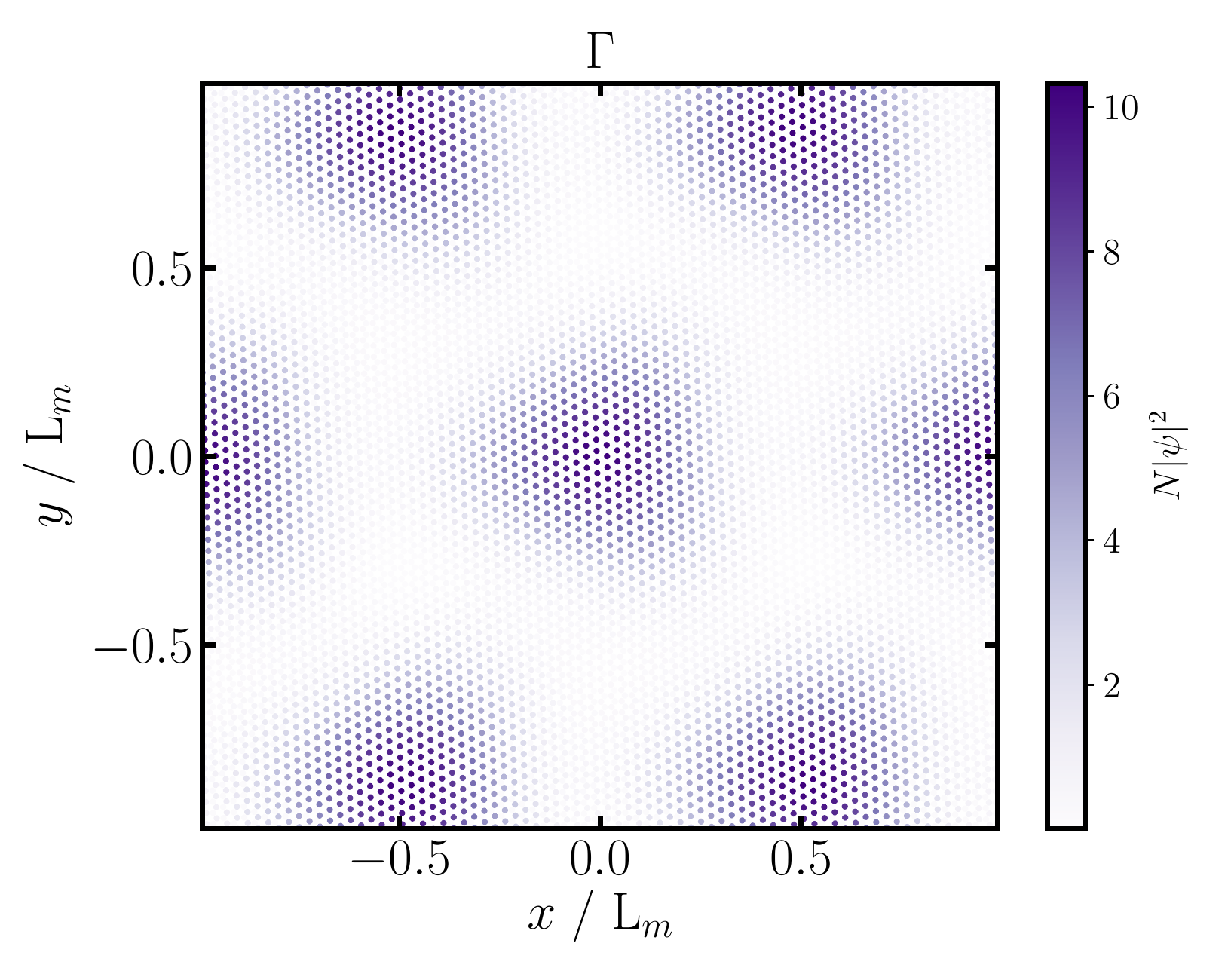}
\end{subfigure}
\centering
\begin{subfigure}[b]{0.3\textwidth}
\centering
\includegraphics[width=\textwidth]{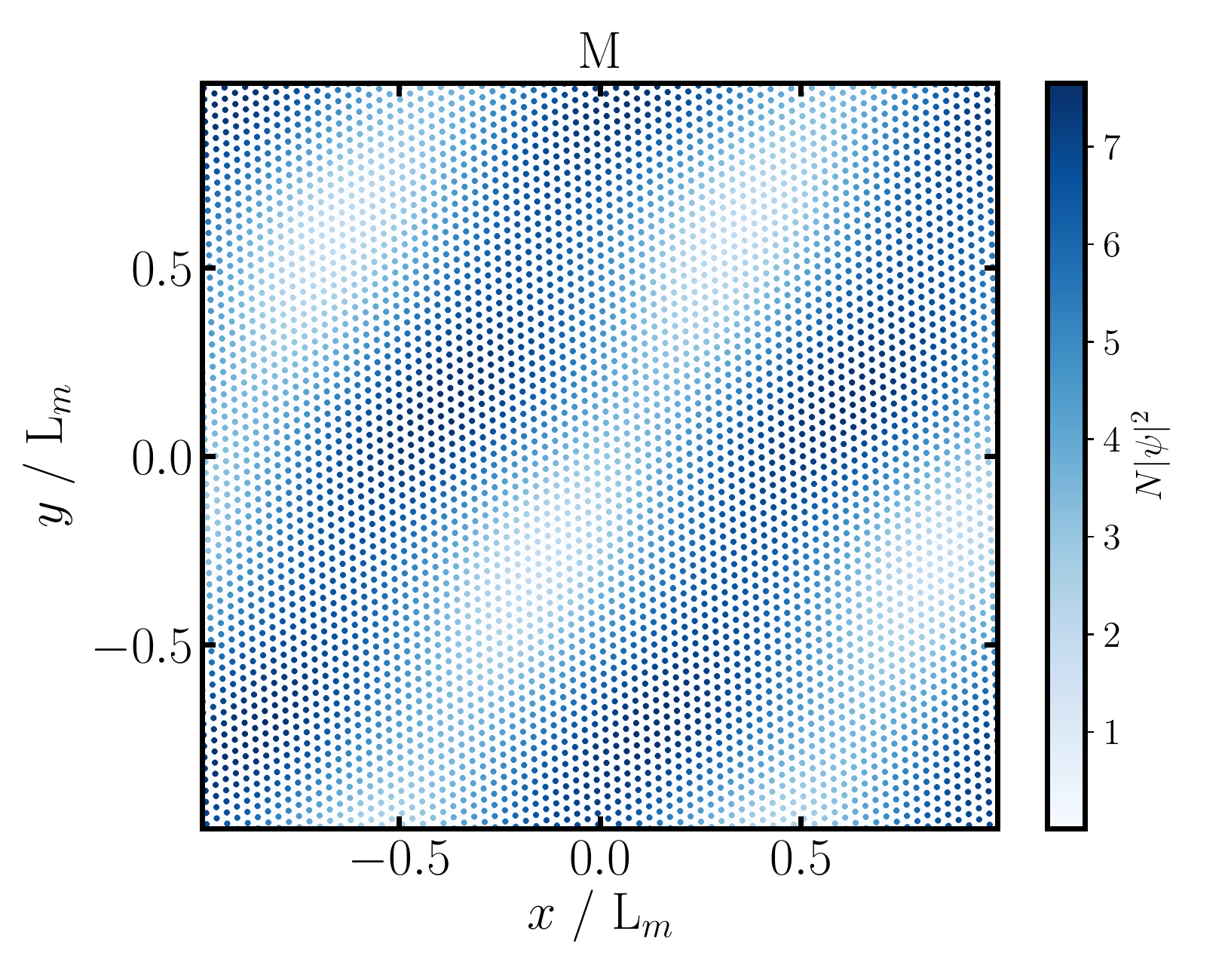}
\end{subfigure}
\centering
\begin{subfigure}[b]{0.3\textwidth}
\centering
\includegraphics[width=\textwidth]{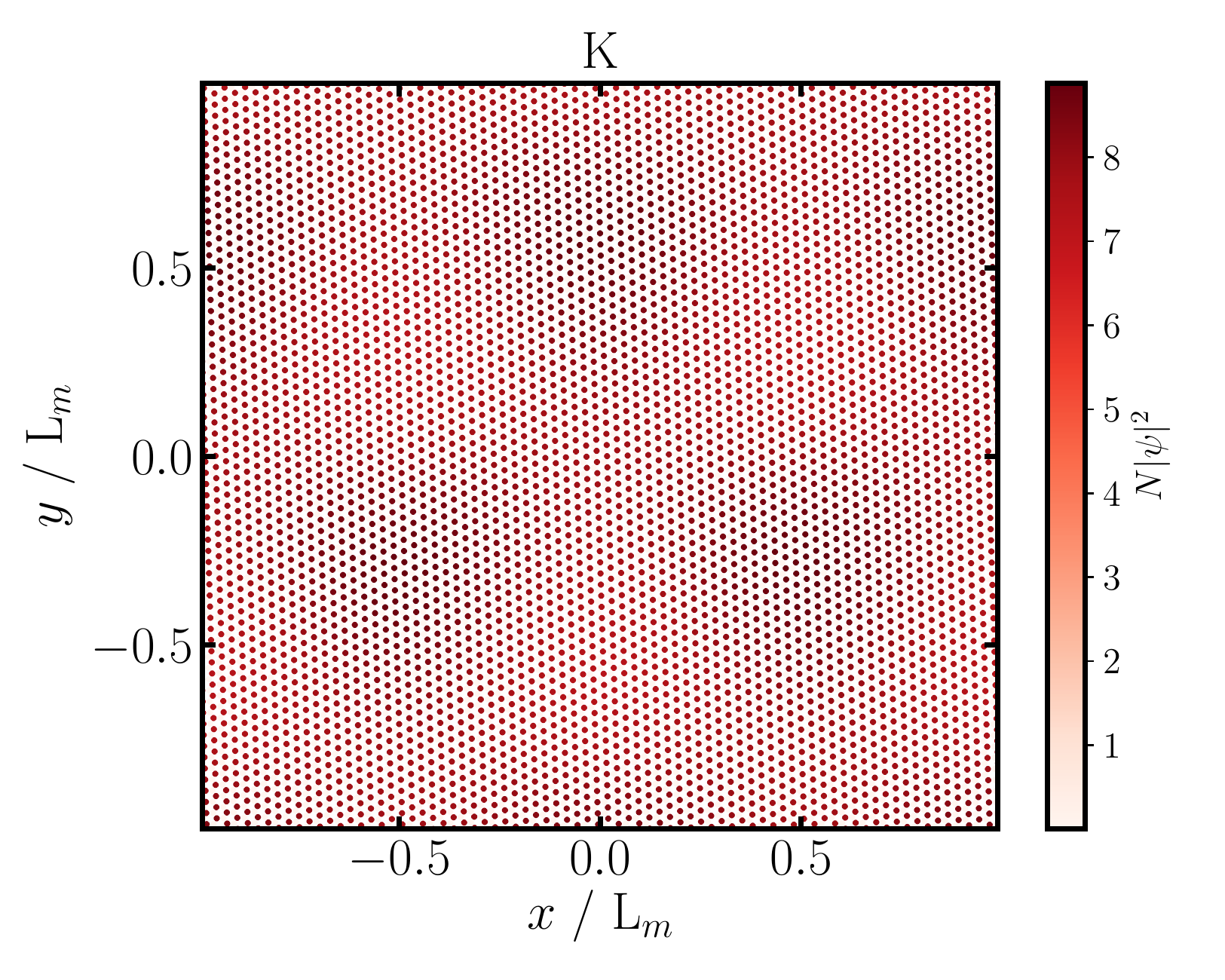}
\end{subfigure}
\caption{Square modulus of the flat-band wavefunctions on an outer layer of tDBLG at different crystal momenta. Top panels show results for the flat conduction band (whose square moduli have been summed), and the bottom panels show similar results for the two flat valence band states. See the caption of Fig.~\ref{fig:tBLG_WF} for further details.}
\label{fig:tDBLG_WF_outer}
\end{figure*}


In tDBLG, the low-energy electronic band structure is characterized by a set of four bands that are separated from all other bands. These bands become extremely flat close to the magic angle of $\sim$1.3$\degree$, see bottom left panel of Fig.~\ref{fig:BS_D} which shows the band structure at a twist angle of 1.41$\degree$. At angles slightly above the magic angle, see top left panel of Fig.~\ref{fig:BS_D} showing the band structure at a twist angle of 1.89$\degree$, the low-energy flat bands near the K and K$^\prime$ points exhibit a parabolic dispersion which is inherited from the parent AB-stacked bilayers. However, the two graphene sheets of the AB-stacked bilayers in tDBLG are no longer equivalent and this results in the opening of a band gap at the K and K$^{\prime}$ points - similar to the case when an electric field is applied perpendicular to an AB-stacked bilayer.

In Fig.~\ref{fig:BS_D} we show the Hartree theory band structure of doped tDBLG at a twist angle of $\theta = 1.89\degree$ (top left panel) and $\theta = 1.41\degree$ (bottom left panel), and compare the results to tBLG (right panels). Results are shown for $-3\le \nu \le 3$, where $\nu$ represents the number of electrons ($\nu>0$) or holes ($\nu<0$) added to each moir\'e unit cell. In stark contrast with tBLG~\cite{EE,Rademaker2019,Cea2019,PHD_4,Bascones2020,PHD_6,Lewandowski2021,Cea2020,Cea2021}, it can be seen that the band structure of tDBLG does not significantly change upon doping (all doping levels have been aligned such that the zero energy occurs at the mid-point between the upper and lower flat bands at the K-point). The flat bands distort only by a few meV relative to the charge neutral case (black lines), which is small compared to their bandwidths (129~meV at 1.89$\degree$ and 13~meV at 1.41$\degree$). By comparison, in tBLG the Hartree interactions induce band deformations of $\sim$25~meV. At the twist angle of 1.89$\degree$ (with a bandwidth of 260~meV), these deformations are modest in comparison to the bandwidth, but at the angle of 1.41$\degree$ (with a bandwidth of 111~meV), they are starting to become comparable to the bandwidth. Very close to the magic angle of tBLG (approximately 1.1$\degree$), the bandwidth shrinks to $\sim$5~meV and the Hartree deformations become the dominant energy scale~\cite{EE,Rademaker2019,Cea2019,PHD_4,Bascones2020,PHD_6,Lewandowski2021,Cea2020,Cea2021}.


To understand the absence of significant band deformations in tDBLG, we analyze the Hartree potential. Fig.~\ref{fig:pot_strc} shows the locally-averaged Hartree potential along the diagonal of the moir\'e unit cell on one of the outer layers (left panels) and one of the inner layers (right panels) for different doping levels $\nu$. We find that the Hartree potential varies by $\sim$25~meV on the inner layers, but only by $\sim$10~meV on the outer layers. In contrast, the Hartree potential of tBLG (and also of twisted trilayer graphene~\cite{Fischer_TTLG}) varies by $\sim$100~meV and is therefore the dominant energy scale for a wide range of twist angles near the magic angle~\cite{EE}. 


Upon electron doping tDBLG ($\nu>0$, top panels of Fig.~\ref{fig:pot_strc}), a positive peak in the Hartree potential emerges in the AA regions. When electrons are removed ($\nu<0$, bottom panels of Fig.~\ref{fig:pot_strc}) a negative trough is found instead in the AA regions. This is a consequence of the shape of the flat-band wavefunctions near the K and K$^\prime$ points. Specifically, these states are localized in the AA regions of the inner layers, similar to  tBLG~\cite{EE,Cea2019,Rademaker2019,PHD_4,Bascones2020,PHD_6} and tTLG~\cite{Fischer_TTLG}. 

\begin{figure*}[ht]
\centering
\begin{subfigure}[b]{0.4\textwidth}
\centering
\includegraphics[width=\textwidth]{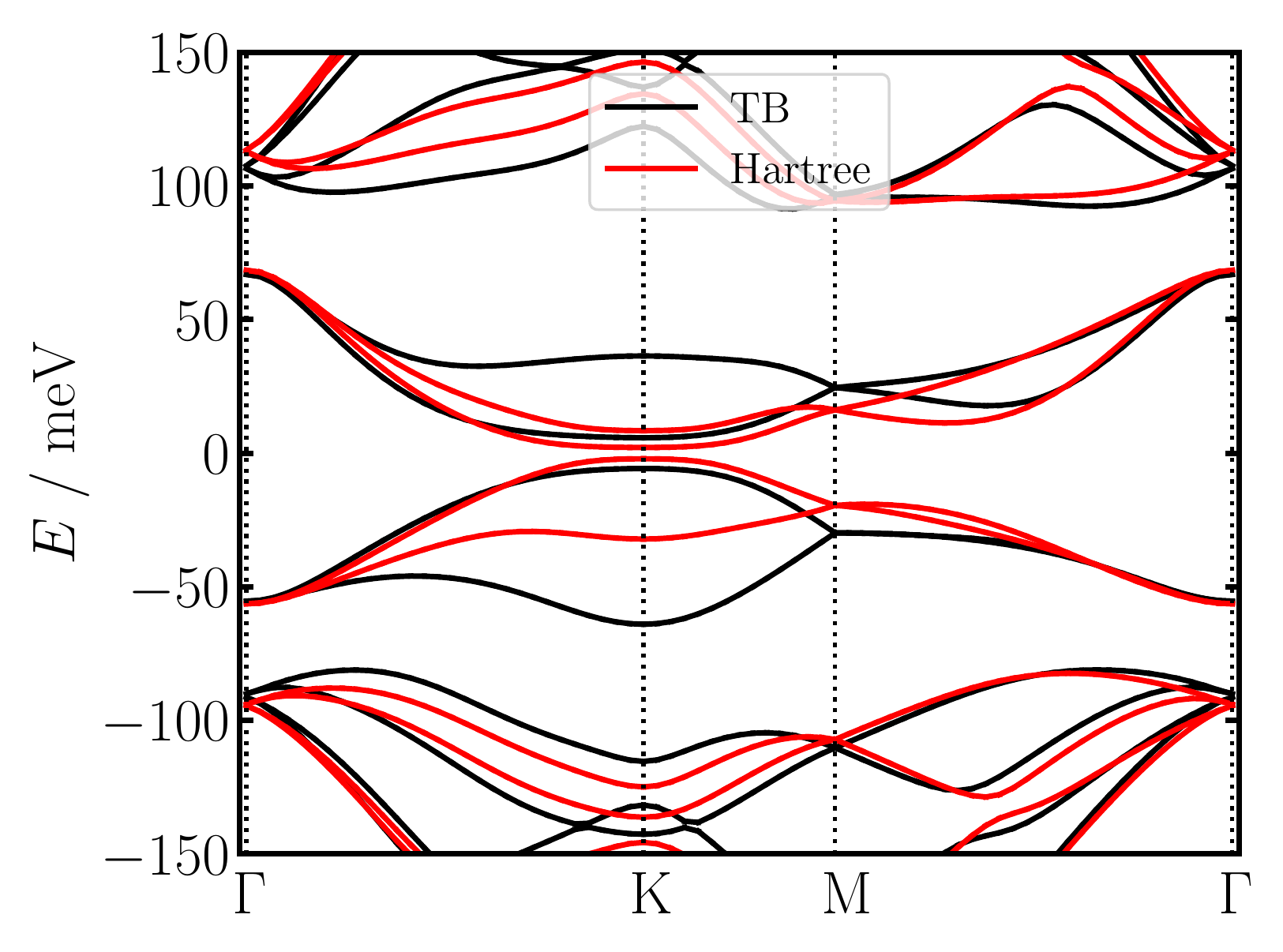}
\end{subfigure}
\centering
\begin{subfigure}[b]{0.4\textwidth}
\centering
\includegraphics[width=\textwidth]{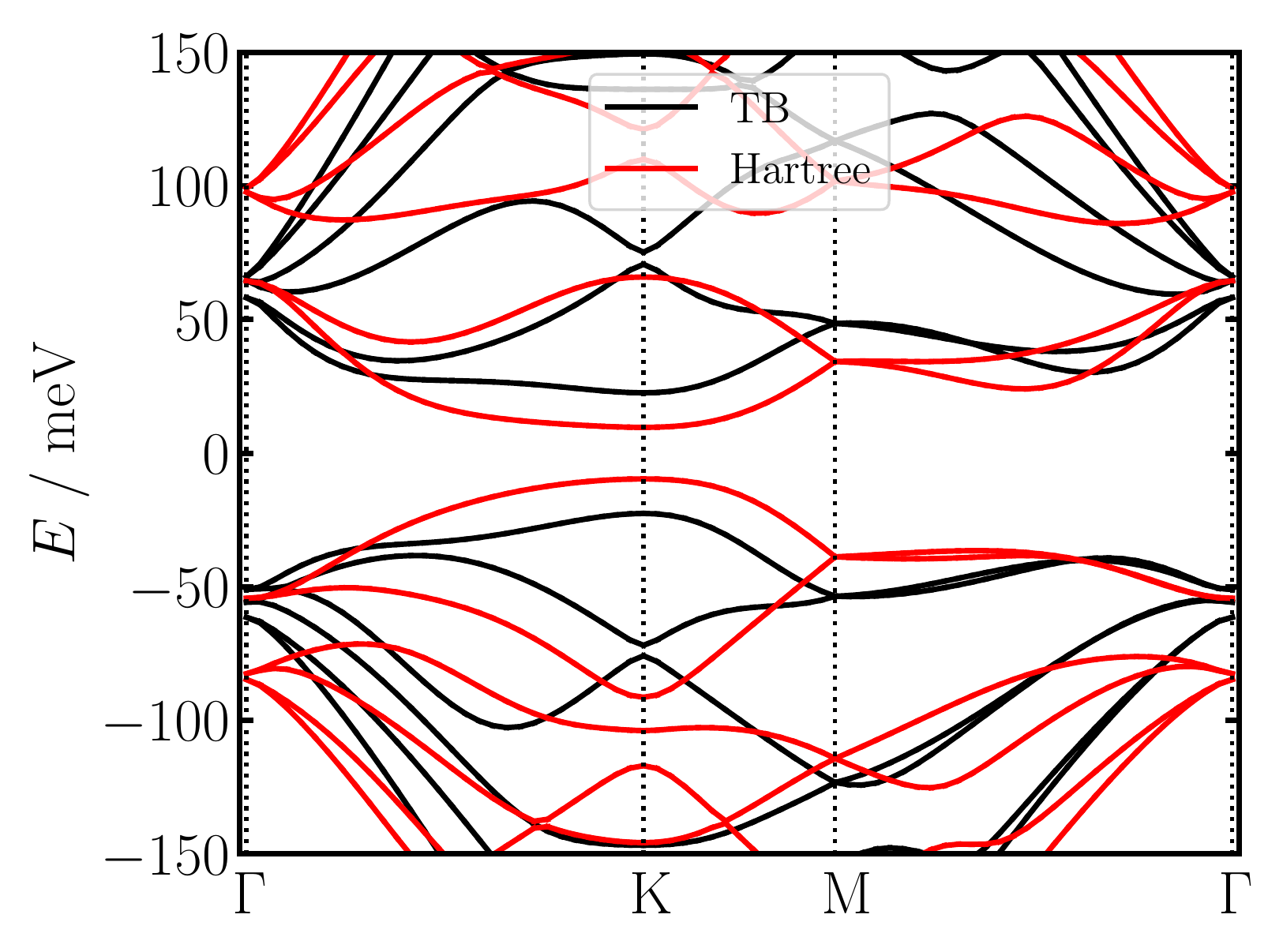}
\end{subfigure}
\caption{Comparison of Hartree theory (black) and tight-binding (red) band structures of undoped tDBLG at a twist angle of $\theta=$1.89$\degree$ in a perpendicular electric field with field strength 10~meV$\textrm{\AA}^{-1}$ (left) and 30~meV$\textrm{\AA}^{-1}$ (right).} 
\label{fig:BS_EF}
\end{figure*}

However, an important difference (besides the strength of the Hartree potential) between tDBLG and tBLG is the degree of real-space localization of flat-band states in different parts of the first Brillouin zone. Fig.~\ref{fig:tBLG_WF} shows that in tBLG that flat-band states at K and M are strongly localized in the AA regions, while the states at $\Gamma$ form rings around the AA regions. As a consequence, the Hartree potential gives rise to large energy shifts of the states at K and M relative to the states at $\Gamma$ (see Refs.~\citenum{EE,Cea2019,Rademaker2019,PHD_4,Bascones2020,PHD_6,Fischer_TTLG} for a more detailed description of these band deformations).

In contrast, Figs.~\ref{fig:tDBLG_WF_inner} and \ref{fig:tDBLG_WF_outer} show that flat-band states in tDBLG are much less localized. For example, the states at $\Gamma$ form rings around the AA regions in the inner layers (see Fig.~\ref{fig:tDBLG_WF_inner}), but are localized in the AA regions on the outer layers (see Fig.~\ref{fig:tDBLG_WF_outer}). The states at M are localized in the AA regions in the inner layers, but have delocalized stripe-like features on the outer layers. As a consequence, the Hartree potential does not give rise to significant relative shifts of these states. Finally, the valence states at the K-point are localised on the AA regions of the inner layers, but on the AB/BA regions of the outer layers; while the conduction states at K are almost entirely delocalized on the outer layers~\cite{CulchacF.J.2020Fbag}. This explains why significant band deformations are not observed in tDBLG even though the Hartree potential has a similar magnitude as the band width close to the magic angle. A similar explanation for the absence of strong band deformations was also offered in the continuum model of Ref.~\citenum{Pierre2020}. 

In the case of tBLG, the interaction-induced doping-dependent band distortions were shown to cause a pinning of the Fermi level at the van Hove singularities~\cite{EE,Cea2019,Rademaker2019,PHD_4,Bascones2020,PHD_6}, which was observed in tunneling experiments~\cite{NAT_SS,NAT_CO,NAT_MEI,Choi2021driven}. In tDBLG, as no significant band distortions are observed, we do not expect a similar Fermi level pinning as in tBLG. Indeed, recent tunnelling experiments did not observe Fermi level pinning at the van Hove singularity~\cite{Liu2021,Crommie2021,Carmen2021}. Note, however, that in these experiments some changes of the electronic structure were observed as function of doping. It was proposed~\cite{Carmen2021} that these changes are a consequence of the perpendicular electric field which accompanies doping in a device with a single metallic gate. We therefore study the effect of such electric fields in the next section.  


\subsection{Electric fields}

Figure~\ref{fig:BS_EF} compares the Hartree (red) and tight-binding (black) band structures of 1.89$\degree$ tDBLG at charge neutrality for two different electric field strengths. Application of a perpendicular electric field increases the gap between the valence and conduction bands. Without the field, the value of the gap is 9.9~meV, while its value is 19.2~meV for a field strength 30~meV\AA$^{-1}$. Moreover, the electric field lifts the valley degeneracy of both the valence and conduction bands (except at the $\Gamma$ and M points). This splitting of the conduction and valence bands increases substantially with the strength of the applied field with the valence bands undergoing more significant distortions than the conduction bands.

In the tight-binding approximation the band distortions are significantly more pronounced than in Hartree theory. The electric field causes the system to polarise such that in one of the bilayers there is an enrichment of electrons and in the other bilayer there is a depletion of electrons, with the outer layers exhibiting larger enrichment/depletion than the inner layers. When Hartree interactions are included, the Hartree potential opposes the external electric field to reduce its effect. 

\begin{figure*}[ht]
\centering
\begin{subfigure}[b]{0.405\textwidth}
\centering
\includegraphics[width=\textwidth]{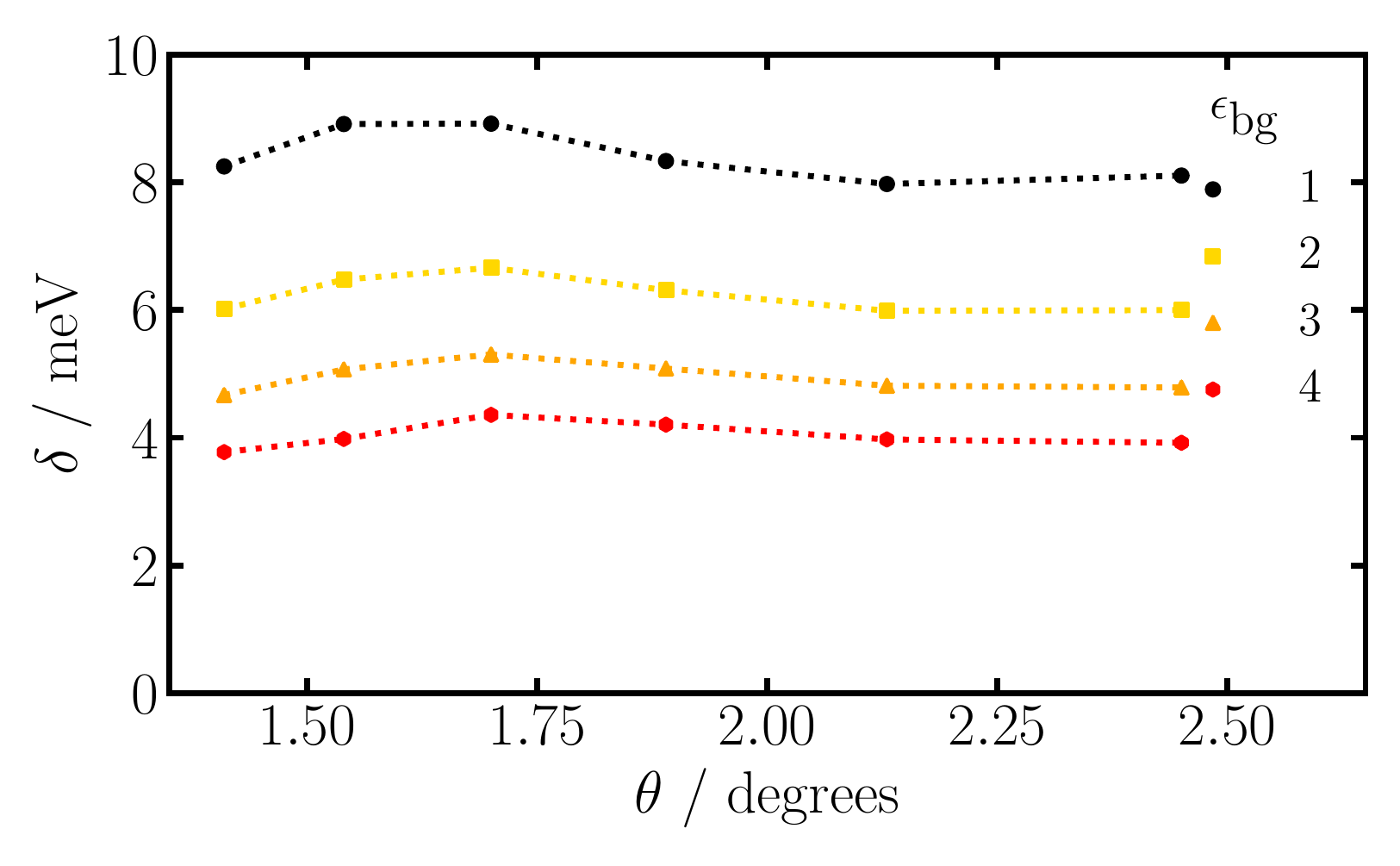}
\end{subfigure}
\begin{subfigure}[b]{0.4\textwidth}  
\centering 
\includegraphics[width=\textwidth]{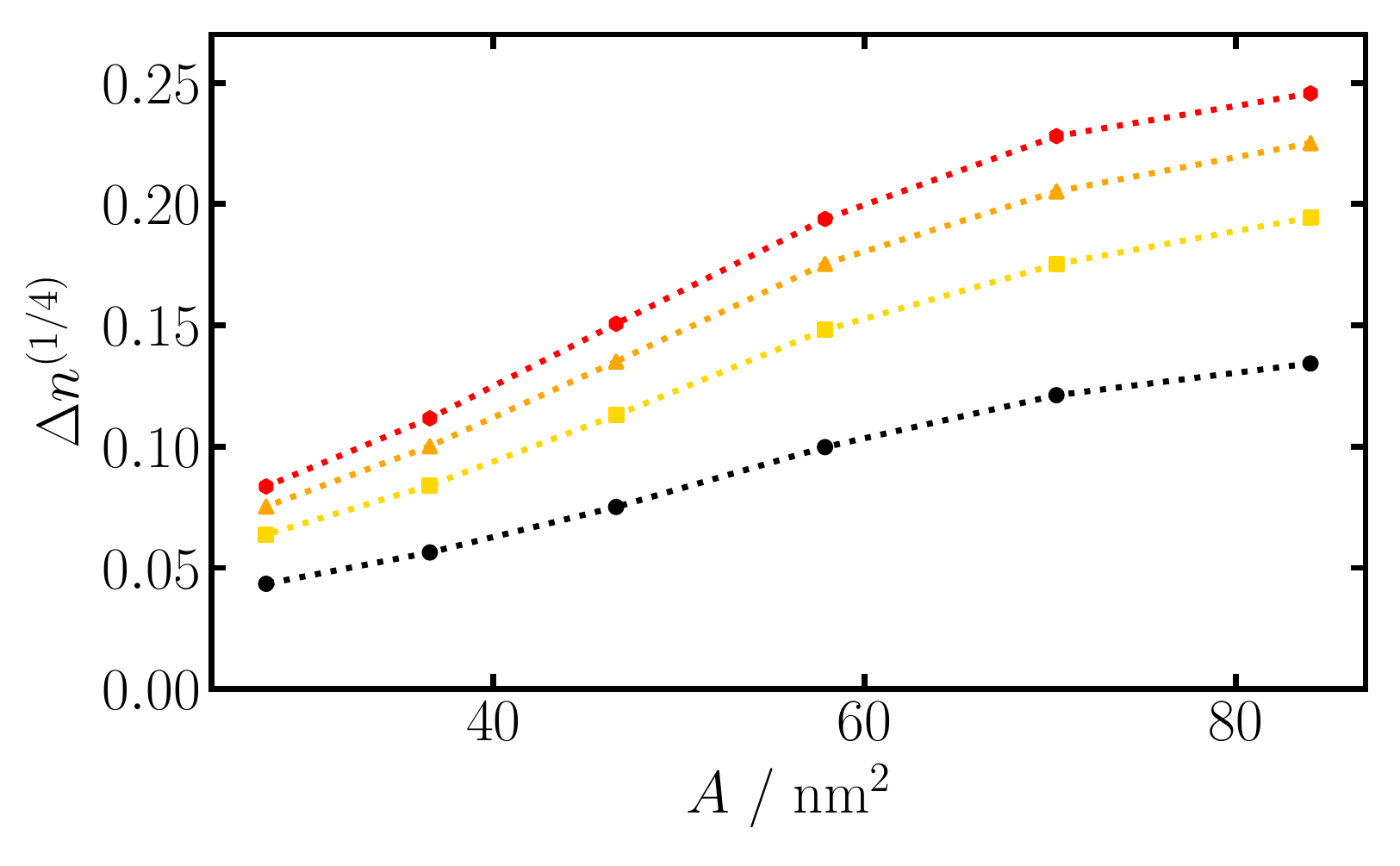}
\end{subfigure}
\caption{Left panel: The average Hartree potential difference between the outer and inner layers, $\delta$, for several dielectric constants as a function of twist angle $\theta$. Right panel: The total excess number of electrons $\Delta n^{(1/4)}$ on one of the outer layers (which we refer to as layers 1 and 4) within the moir\'e unit cell at various dielectric constants $\epsilon_\mathrm{bg}$ as a function of moir\'e unit cell area $A$.}
\label{fig:del_eps}
\end{figure*}

The extent to which the Hartree potential screens the electric field can be determined by computing an effective dielectric constant for each layer
\begin{equation}
    \epsilon^{(\alpha)} = \dfrac{V_{\textrm{ext}}^{(\alpha)}}{V_{\textrm{ext}}^{(\alpha)} + V_{\textrm{H}}^{(\alpha)}},
\end{equation}
\noindent where $V_{\textrm{ext}}^{(\alpha)} = \langle Ez_{i}^{(\alpha)} \rangle $ is the average potential due to the electric field in layer $\alpha$, with $E$ denoting the electric field strength,  $z_{i}^{(\alpha)}$ the $z$-coordinate of atom $i$ in layer $\alpha$, and $\langle\cdot\cdot\cdot\rangle$ an average over $i$ in layer $\alpha$. Also, $V_{\textrm{H}}^{(\alpha)} = \langle V_{\textrm{H}i}^{(\alpha)} \rangle $ is the averaged Hartree potential in each layer $\alpha$. For $E= 10$~meV$\textrm{\AA}^{-1}$, we find $\epsilon^{(\alpha)} = 2.60$ for all layers, while for $E= 30$~meV$\textrm{\AA}^{-1}$ the effective dielectric constant is reduced to 1.94. This reflects the fact that the electrons cannot screen larger electric fields as effectively. Note that these values were obtained with $\epsilon_\mathrm{bg}=4$. For free-standing tDBLG, ab initio DFT calculations have found an effective perpendicular dielectric constant of approximately 3~\cite{Santos2013,Nikita2021} for untwisted graphene multilayers in vacuum. These ab initio values are qualitatively similar to our findings.

\subsection{Crystal field}

Previous work established that an additional layer-dependent on-site potential (often referred to as the crystal field) must be included in tight-binding calculations of tBLG to achieve agreement with ab initio DFT band structures~\cite{haddadi2019moir,RickhausPeter2019GOiT,CulchacF.J.2020Fbag}. However, the origin of this on-site potential has remained unclear. In this section we evaluate the effective on-site energy generated by the Hartree potential and compare to previous work. 

At charge neutrality, the Hartree potential is more negative on the inner layers than on the outer layers, see Fig.~\ref{fig:pot_strc}, indicating that electrons have transferred from the inner to the outer layers. Adapting the definition of Ref.~\citenum{RickhausPeter2019GOiT} for the crystal field, we define $\delta$ as the difference between the layer-averaged on-site Hartree potential on the outer and inner layers according to
\begin{equation}
\delta = \langle \varepsilon_{\textrm{out}}^{\textrm{el}} \rangle - \langle \varepsilon_{\textrm{in}}^{\textrm{el}} \rangle ,
\label{eq:delta}
\end{equation}
where \( \langle \varepsilon_{\textrm{out}}^{\textrm{el}} \rangle \) and \( \langle \varepsilon_{\textrm{in}}^{\textrm{el}} \rangle \) are the on-site energies averaged over atoms in the outer and inner layer of a bilayer, respectively (note that the two bilayers of tDBLG are equivalent by symmetry). 

\begin{figure*}[ht]
\centering
\begin{subfigure}[b]{0.4\textwidth}
\centering
\includegraphics[width=\textwidth]{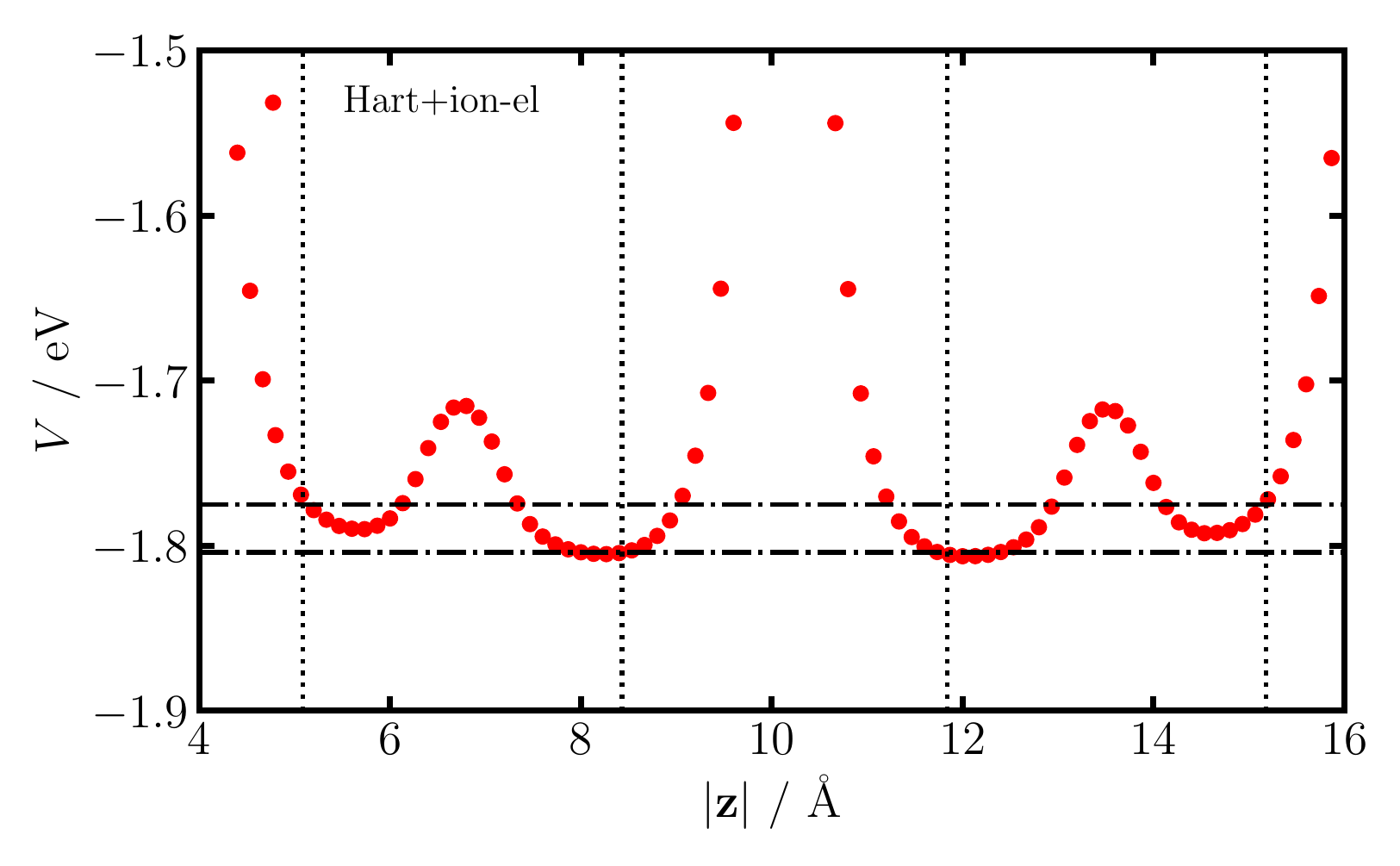}
\end{subfigure}
\centering
\begin{subfigure}[b]{0.4\textwidth}
\centering
\includegraphics[width=\textwidth]{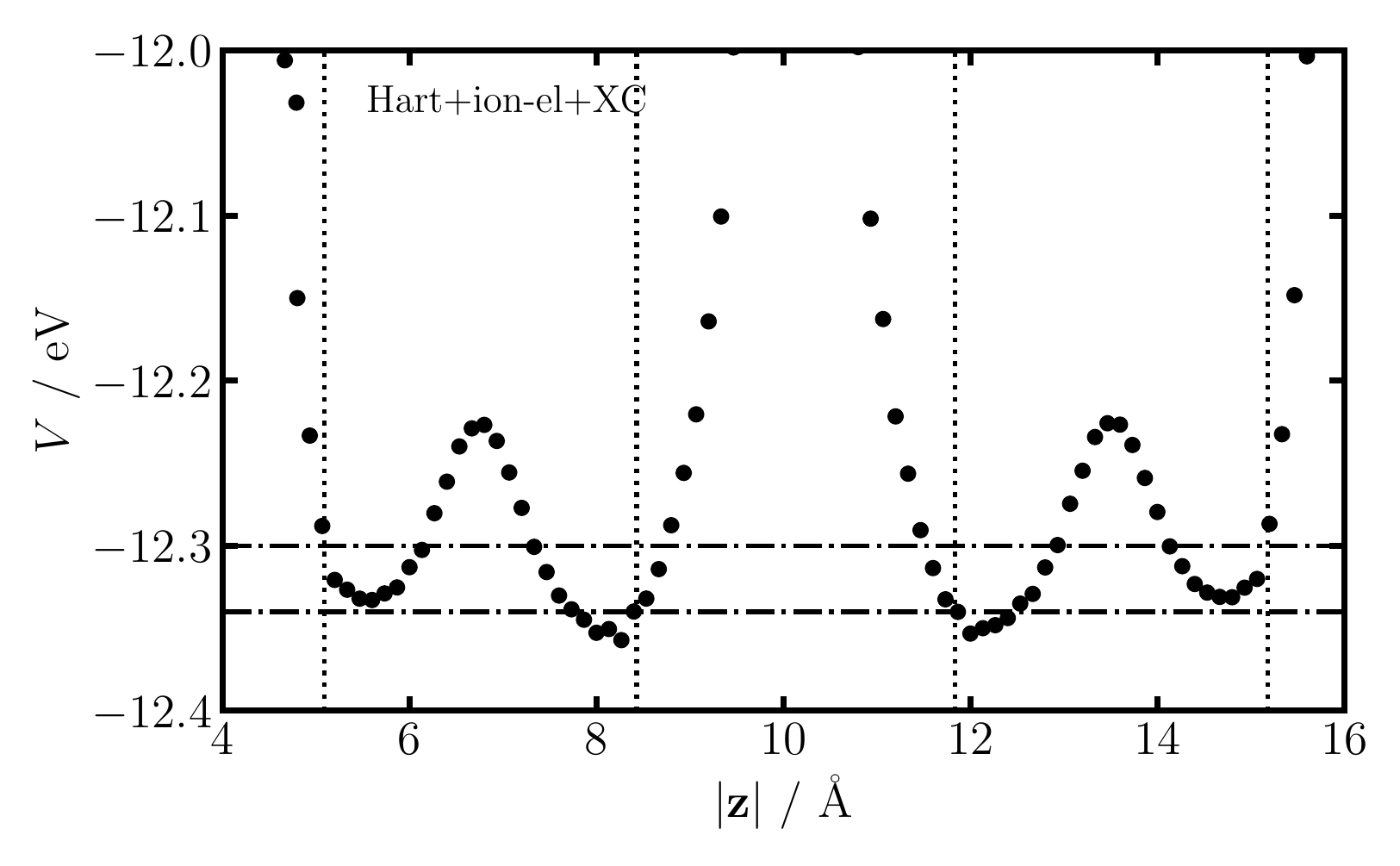}
\end{subfigure}
\caption{Left panel: Sum of Hartree potential and ion-electron potential as a function of $z$ (the coordinate perpendicular to the plane of the tDBLG system) obtained from an ab initio DFT calculation of tDBLG at a twist angle of 2.45$\degree$. Right panel: Full Kohn-Sham potential (including the exchange-correlation contribution) as a function of $z$ in 2.45$\degree$ tDBLG. The ab initio potentials are first averaged over the $x$ and $y$ coordinates, and the resulting function of $z$ is smoothed by taking its convolution with a rectangular function of width 3.20~$\textrm{\AA}$. The dotted vertical lines correspond to the $z$-averaged atomic positions of each layer. The horizontal dotted-dashed lines indicate where the potential crosses the $z$-averaged atomic positions of each layer.} 
\label{fig:DFT_CF}
\end{figure*}

The left panel of Fig.~\ref{fig:del_eps} shows the values of the Hartree crystal field $\delta$ as function of twist angle for different values of $\epsilon_\mathrm{bg}$. We find that $\delta$ does not sensitively depend on the twist angle, but increases as the background dielectric constant is reduced. For freestanding tDBLG (corresponding to $\epsilon_\mathrm{bg}=1$), we find $\delta \approx 8$~meV near the magic angle. We have verified that the band structure from a tight-binding calculation with a layer-dependent on-site potential of 8~meV agrees well with the full Hartree theory result indicating that the in-plane variations of the Hartree potential do not play an important role. Note that the value of $\delta$ has the same sign, but is somewhat smaller than the value proposed by Haddidi and coworkers~\cite{haddadi2019moir} who used a value of 30~meV. 

To understand the weak twist-angle dependence of $\delta$, we analyze the charge transfer between inner and outer layers. Within an idealized parallel plate capacitor model, $\delta$ should be proportional to the charge density per unit area of each graphene layer. The total number of polarized charges per moir\'e cell in layer $\alpha$ is given by 
\begin{equation}
\Delta n^{(\alpha)} = \sum_{j \in \alpha}(n_{j}-n_0),
\label{eq:Deln}
\end{equation}
where $j$ runs through all the atoms in layer $\alpha$. By symmetry $\Delta n$ has the same magnitude but opposite sign for the outer and inner layer of each bilayer. In the right panel of Fig.~\ref{fig:del_eps} we show the dependence of $\Delta n^{(1/4)}$ (i.e., the charge on the outer layers) on the area of the moir\'e unit cell. It can be seen that $\Delta n^{(1/4)}$ behaves approximately linearly resulting in a constant charge density per unit area which in turn gives rise to a constant $\delta$.

As the Hartree theory contribution to the crystal field from our atomistic model is significantly smaller than the value determined by Haddidi and coworkers~\cite{haddadi2019moir}, we also analyze the full Kohn-Sham potential obtained from an ab initio DFT calculation of tDBLG with a twist angle of 2.45$\degree$. The Kohn-Sham potential has three contributions: (1) the ion-electron (ion-el) potential which is often approximated with a pseudo-potential; (2) the Hartree contribution from electron-electron (el-el) interactions; and (3) the exchange-correlation contribution.

Figure~\ref{fig:DFT_CF} shows the averaged Kohn-Sham potential of tDBLG as function of $z$. The potential has been averaged over the $x$ and $y$ directions, and the resulting function of $z$ has been smoothed by taking its convolution with a rectangular function of width 3.20~$\textrm{\AA}$ (we have verified that our results do not depend sensitively on the value of this width). In the left panel, we subtract the exchange-correlation contribution from the Kohn-Sham potential which yields an ab initio Hartree theory potential (note, however, that the charge density was obtained from the full potential including exchange-correlation effects). We find that the ab initio Hartree potential is approximately 30~meV smaller on the inner layers than the outer layers, as indicated with the horizontal lines. When the exchange-correlation potential is included (right panel of Fig.~\ref{fig:del_eps}), the potential difference between inner and outer layers increases to approximately 40~meV, in good agreement with the findings in Refs.~\citenum{haddadi2019moir},~\citenum{RickhausPeter2019GOiT} and~\citenum{CulchacF.J.2020Fbag}. These results demonstrate that an ab initio description of the potential is required to obtain a quantitatively accurate description of the crystal field.

\section{Conclusion}

We have used atomistic Hartree theory calculation to investigate the role of electron-electron interactions in tDBLG and studied the effects of changes in twist angle, dielectric environment, doping and applied electric fields. Our calculations reveal that the band structure of tDBLG is largely insensitive to electron or hole doping in stark contrast to tBLG. Application of a perpendicular electric field changes the band gap and lifts the valley degeneracy of the bands. Electron-electron interactions screen the electric field and we obtain an effective dielectric constant that is quantitatively similar to ab initio results for untwisted graphene multilayers. Finally, we analyze the contribution of Hartree interactions to the crystal field which is defined as the difference between the on-site energies of the inner and outer layers. We find that the difference of the average Hartree potentials in the inner and outer layers has the same sign, but a smaller magnitude compared to previous studies which determined the on-site potential by fitting tight-binding band structures to ab initio DFT results. To understand this difference, we carry out ab initio DFT calculations of tDBLG and analyze the Kohn-Sham potential. We find that the locally averaged Kohn-Sham potential difference between the inner and outer layers agrees well with the previously reported value
of the crystal field, indicating that a quantitative description of this effect requires an ab initio description of the subtle interplay between electron-ion and electron-electron interactions.

\section{Acknowledgements}

CC was supported through a UROP Bursary from Imperial College. ZG was supported through a studentship in the Centre for Doctoral Training on Theory and Simulation of Materials at Imperial College London funded by the EPSRC (EP/L015579/1). We acknowledge funding from EPSRC grant EP/S025324/1 and the Thomas Young Centre under grant number TYC-101. We acknowledge the Imperial College London Research Computing Service (DOI:10.14469/hpc/2232) for the computational resources used in carrying out this work. Via our membership of the UK's HEC Materials Chemistry Consortium, which is funded by EPSRC (EP/L000202, EP/R029431), this work used the ARCHER UK National Supercomputing Service.

\bibliography{REF}

\begin{thebibliography}{73}%
\makeatletter
\providecommand \@ifxundefined [1]{%
 \@ifx{#1\undefined}
}%
\providecommand \@ifnum [1]{%
 \ifnum #1\expandafter \@firstoftwo
 \else \expandafter \@secondoftwo
 \fi
}%
\providecommand \@ifx [1]{%
 \ifx #1\expandafter \@firstoftwo
 \else \expandafter \@secondoftwo
 \fi
}%
\providecommand \natexlab [1]{#1}%
\providecommand \enquote  [1]{``#1''}%
\providecommand \bibnamefont  [1]{#1}%
\providecommand \bibfnamefont [1]{#1}%
\providecommand \citenamefont [1]{#1}%
\providecommand \href@noop [0]{\@secondoftwo}%
\providecommand \href [0]{\begingroup \@sanitize@url \@href}%
\providecommand \@href[1]{\@@startlink{#1}\@@href}%
\providecommand \@@href[1]{\endgroup#1\@@endlink}%
\providecommand \@sanitize@url [0]{\catcode `\\12\catcode `\$12\catcode
  `\&12\catcode `\#12\catcode `\^12\catcode `\_12\catcode `\%12\relax}%
\providecommand \@@startlink[1]{}%
\providecommand \@@endlink[0]{}%
\providecommand \url  [0]{\begingroup\@sanitize@url \@url }%
\providecommand \@url [1]{\endgroup\@href {#1}{\urlprefix }}%
\providecommand \urlprefix  [0]{URL }%
\providecommand \Eprint [0]{\href }%
\providecommand \doibase [0]{http://dx.doi.org/}%
\providecommand \selectlanguage [0]{\@gobble}%
\providecommand \bibinfo  [0]{\@secondoftwo}%
\providecommand \bibfield  [0]{\@secondoftwo}%
\providecommand \translation [1]{[#1]}%
\providecommand \BibitemOpen [0]{}%
\providecommand \bibitemStop [0]{}%
\providecommand \bibitemNoStop [0]{.\EOS\space}%
\providecommand \EOS [0]{\spacefactor3000\relax}%
\providecommand \BibitemShut  [1]{\csname bibitem#1\endcsname}%
\let\auto@bib@innerbib\@empty
\bibitem [{\citenamefont {Carr}\ \emph {et~al.}(2017)\citenamefont {Carr},
  \citenamefont {Massatt}, \citenamefont {Fang}, \citenamefont {Cazeaux},
  \citenamefont {Luskin},\ and\ \citenamefont {Kaxiras}}]{Carr_2017}%
  \BibitemOpen
  \bibfield  {author} {\bibinfo {author} {\bibfnamefont {S.}~\bibnamefont
  {Carr}}, \bibinfo {author} {\bibfnamefont {D.}~\bibnamefont {Massatt}},
  \bibinfo {author} {\bibfnamefont {S.}~\bibnamefont {Fang}}, \bibinfo {author}
  {\bibfnamefont {P.}~\bibnamefont {Cazeaux}}, \bibinfo {author} {\bibfnamefont
  {M.}~\bibnamefont {Luskin}}, \ and\ \bibinfo {author} {\bibfnamefont
  {E.}~\bibnamefont {Kaxiras}},\ }\href@noop {} {\bibfield  {journal} {\bibinfo
   {journal} {Phys. Rev. B}\ }\textbf {\bibinfo {volume} {95}},\ \bibinfo
  {pages} {075420} (\bibinfo {year} {2017})}\BibitemShut {NoStop}%
\bibitem [{\citenamefont {Kennes}\ \emph {et~al.}(2021)\citenamefont {Kennes},
  \citenamefont {Claassen}, \citenamefont {Xian}, \citenamefont {Georges},
  \citenamefont {Millis}, \citenamefont {Hone}, \citenamefont {Dean},
  \citenamefont {Basov}, \citenamefont {Pasupathy},\ and\ \citenamefont
  {Rubio}}]{moiresim}%
  \BibitemOpen
  \bibfield  {author} {\bibinfo {author} {\bibfnamefont {D.~M.}\ \bibnamefont
  {Kennes}}, \bibinfo {author} {\bibfnamefont {M.}~\bibnamefont {Claassen}},
  \bibinfo {author} {\bibfnamefont {L.}~\bibnamefont {Xian}}, \bibinfo {author}
  {\bibfnamefont {A.}~\bibnamefont {Georges}}, \bibinfo {author} {\bibfnamefont
  {A.~J.}\ \bibnamefont {Millis}}, \bibinfo {author} {\bibfnamefont
  {J.}~\bibnamefont {Hone}}, \bibinfo {author} {\bibfnamefont {C.~R.}\
  \bibnamefont {Dean}}, \bibinfo {author} {\bibfnamefont {D.~N.}\ \bibnamefont
  {Basov}}, \bibinfo {author} {\bibfnamefont {A.}~\bibnamefont {Pasupathy}}, \
  and\ \bibinfo {author} {\bibfnamefont {A.}~\bibnamefont {Rubio}},\
  }\href@noop {} {\bibfield  {journal} {\bibinfo  {journal} {Nat. Phys.}\
  }\textbf {\bibinfo {volume} {17}},\ \bibinfo {pages} {155–163} (\bibinfo
  {year} {2021})}\BibitemShut {NoStop}%
\bibitem [{\citenamefont {Carr}\ \emph {et~al.}(2020)\citenamefont {Carr},
  \citenamefont {Fang},\ and\ \citenamefont {Kaxiras}}]{CarrRev2020}%
  \BibitemOpen
  \bibfield  {author} {\bibinfo {author} {\bibfnamefont {S.}~\bibnamefont
  {Carr}}, \bibinfo {author} {\bibfnamefont {S.}~\bibnamefont {Fang}}, \ and\
  \bibinfo {author} {\bibfnamefont {E.}~\bibnamefont {Kaxiras}},\ }\href@noop
  {} {\bibfield  {journal} {\bibinfo  {journal} {Nat. Rev. Mater}\ }\textbf
  {\bibinfo {volume} {5}},\ \bibinfo {pages} {748–763} (\bibinfo {year}
  {2020})}\BibitemShut {NoStop}%
\bibitem [{\citenamefont {dos Santos}\ \emph {et~al.}(2007)\citenamefont {dos
  Santos}, \citenamefont {Peres},\ and\ \citenamefont {Neto}}]{GBWT}%
  \BibitemOpen
  \bibfield  {author} {\bibinfo {author} {\bibfnamefont {J.~M. B.~L.}\
  \bibnamefont {dos Santos}}, \bibinfo {author} {\bibfnamefont {N.~M.~R.}\
  \bibnamefont {Peres}}, \ and\ \bibinfo {author} {\bibfnamefont {A.~H.~C.}\
  \bibnamefont {Neto}},\ }\href@noop {} {\bibfield  {journal} {\bibinfo
  {journal} {Phys. Rev. Lett.}\ }\textbf {\bibinfo {volume} {99}},\ \bibinfo
  {pages} {256802} (\bibinfo {year} {2007})}\BibitemShut {NoStop}%
\bibitem [{\citenamefont {Bistritzer}\ and\ \citenamefont
  {MacDonald}(2011)}]{Bistritzer12233}%
  \BibitemOpen
  \bibfield  {author} {\bibinfo {author} {\bibfnamefont {R.}~\bibnamefont
  {Bistritzer}}\ and\ \bibinfo {author} {\bibfnamefont {A.~H.}\ \bibnamefont
  {MacDonald}},\ }\href@noop {} {\bibfield  {journal} {\bibinfo  {journal}
  {PNAS}\ }\textbf {\bibinfo {volume} {108}},\ \bibinfo {pages} {12233}
  (\bibinfo {year} {2011})}\BibitemShut {NoStop}%
\bibitem [{\citenamefont {de~Laissardi\`ere}\ \emph {et~al.}(2010)\citenamefont
  {de~Laissardi\`ere}, \citenamefont {Mayou},\ and\ \citenamefont
  {Magaud}}]{LDE}%
  \BibitemOpen
  \bibfield  {author} {\bibinfo {author} {\bibfnamefont {G.~T.}\ \bibnamefont
  {de~Laissardi\`ere}}, \bibinfo {author} {\bibfnamefont {D.}~\bibnamefont
  {Mayou}}, \ and\ \bibinfo {author} {\bibfnamefont {L.}~\bibnamefont
  {Magaud}},\ }\href@noop {} {\bibfield  {journal} {\bibinfo  {journal} {Nano
  Lett.}\ }\textbf {\bibinfo {volume} {10}},\ \bibinfo {pages} {804} (\bibinfo
  {year} {2010})}\BibitemShut {NoStop}%
\bibitem [{\citenamefont {de~Laissardi\`ere}\ \emph {et~al.}(2012)\citenamefont
  {de~Laissardi\`ere}, \citenamefont {Mayou},\ and\ \citenamefont
  {Magaud}}]{NSCS}%
  \BibitemOpen
  \bibfield  {author} {\bibinfo {author} {\bibfnamefont {G.~T.}\ \bibnamefont
  {de~Laissardi\`ere}}, \bibinfo {author} {\bibfnamefont {D.}~\bibnamefont
  {Mayou}}, \ and\ \bibinfo {author} {\bibfnamefont {L.}~\bibnamefont
  {Magaud}},\ }\href@noop {} {\bibfield  {journal} {\bibinfo  {journal} {Phys.
  Rev. B}\ }\textbf {\bibinfo {volume} {86}},\ \bibinfo {pages} {125413}
  (\bibinfo {year} {2012})}\BibitemShut {NoStop}%
\bibitem [{\citenamefont {Su\'arez~Morell}\ \emph {et~al.}(2010)\citenamefont
  {Su\'arez~Morell}, \citenamefont {Correa}, \citenamefont {Vargas},
  \citenamefont {Pacheco},\ and\ \citenamefont
  {Barticevic}}]{PhysRevB.82.121407}%
  \BibitemOpen
  \bibfield  {author} {\bibinfo {author} {\bibfnamefont {E.}~\bibnamefont
  {Su\'arez~Morell}}, \bibinfo {author} {\bibfnamefont {J.~D.}\ \bibnamefont
  {Correa}}, \bibinfo {author} {\bibfnamefont {P.}~\bibnamefont {Vargas}},
  \bibinfo {author} {\bibfnamefont {M.}~\bibnamefont {Pacheco}}, \ and\
  \bibinfo {author} {\bibfnamefont {Z.}~\bibnamefont {Barticevic}},\
  }\href@noop {} {\bibfield  {journal} {\bibinfo  {journal} {Phys. Rev. B}\
  }\textbf {\bibinfo {volume} {82}},\ \bibinfo {pages} {121407} (\bibinfo
  {year} {2010})}\BibitemShut {NoStop}%
\bibitem [{\citenamefont {Tritsaris}\ \emph {et~al.}(2020)\citenamefont
  {Tritsaris}, \citenamefont {Carr}, \citenamefont {Zhu}, \citenamefont {Xie},
  \citenamefont {Torrisi}, \citenamefont {Tang}, \citenamefont {Mattheakis},
  \citenamefont {Larson},\ and\ \citenamefont {Kaxiras}}]{Tritsaris_2020}%
  \BibitemOpen
  \bibfield  {author} {\bibinfo {author} {\bibfnamefont {G.~A.}\ \bibnamefont
  {Tritsaris}}, \bibinfo {author} {\bibfnamefont {S.}~\bibnamefont {Carr}},
  \bibinfo {author} {\bibfnamefont {Z.}~\bibnamefont {Zhu}}, \bibinfo {author}
  {\bibfnamefont {Y.}~\bibnamefont {Xie}}, \bibinfo {author} {\bibfnamefont
  {S.~B.}\ \bibnamefont {Torrisi}}, \bibinfo {author} {\bibfnamefont
  {J.}~\bibnamefont {Tang}}, \bibinfo {author} {\bibfnamefont {M.}~\bibnamefont
  {Mattheakis}}, \bibinfo {author} {\bibfnamefont {D.~T.}\ \bibnamefont
  {Larson}}, \ and\ \bibinfo {author} {\bibfnamefont {E.}~\bibnamefont
  {Kaxiras}},\ }\href {\doibase 10.1088/2053-1583/ab8f62} {\bibfield  {journal}
  {\bibinfo  {journal} {2D Materials}\ }\textbf {\bibinfo {volume} {7}},\
  \bibinfo {pages} {035028} (\bibinfo {year} {2020})}\BibitemShut {NoStop}%
\bibitem [{\citenamefont {Cao}\ \emph {et~al.}(2018{\natexlab{a}})\citenamefont
  {Cao}, \citenamefont {Fatemi}, \citenamefont {Demir}, \citenamefont {Fang},
  \citenamefont {Tomarken}, \citenamefont {Luo}, \citenamefont
  {Sanchez-Yamagishi}, \citenamefont {Watanabe}, \citenamefont {Taniguchi},
  \citenamefont {Kaxiras}, \citenamefont {Ashoori},\ and\ \citenamefont
  {Jarillo-Herrero}}]{NAT_I}%
  \BibitemOpen
  \bibfield  {author} {\bibinfo {author} {\bibfnamefont {Y.}~\bibnamefont
  {Cao}}, \bibinfo {author} {\bibfnamefont {V.}~\bibnamefont {Fatemi}},
  \bibinfo {author} {\bibfnamefont {A.}~\bibnamefont {Demir}}, \bibinfo
  {author} {\bibfnamefont {S.}~\bibnamefont {Fang}}, \bibinfo {author}
  {\bibfnamefont {S.~L.}\ \bibnamefont {Tomarken}}, \bibinfo {author}
  {\bibfnamefont {J.~Y.}\ \bibnamefont {Luo}}, \bibinfo {author} {\bibfnamefont
  {J.~D.}\ \bibnamefont {Sanchez-Yamagishi}}, \bibinfo {author} {\bibfnamefont
  {K.}~\bibnamefont {Watanabe}}, \bibinfo {author} {\bibfnamefont
  {T.}~\bibnamefont {Taniguchi}}, \bibinfo {author} {\bibfnamefont
  {E.}~\bibnamefont {Kaxiras}}, \bibinfo {author} {\bibfnamefont {R.~C.}\
  \bibnamefont {Ashoori}}, \ and\ \bibinfo {author} {\bibfnamefont
  {P.}~\bibnamefont {Jarillo-Herrero}},\ }\href@noop {} {\bibfield  {journal}
  {\bibinfo  {journal} {Nature}\ }\textbf {\bibinfo {volume} {556}},\ \bibinfo
  {pages} {80} (\bibinfo {year} {2018}{\natexlab{a}})}\BibitemShut {NoStop}%
\bibitem [{\citenamefont {Cao}\ \emph {et~al.}(2018{\natexlab{b}})\citenamefont
  {Cao}, \citenamefont {Fatemi}, \citenamefont {Fang}, \citenamefont
  {Watanabe}, \citenamefont {Taniguchi}, \citenamefont {Kaxiras},\ and\
  \citenamefont {Jarillo-Herrero}}]{NAT_S}%
  \BibitemOpen
  \bibfield  {author} {\bibinfo {author} {\bibfnamefont {Y.}~\bibnamefont
  {Cao}}, \bibinfo {author} {\bibfnamefont {V.}~\bibnamefont {Fatemi}},
  \bibinfo {author} {\bibfnamefont {S.}~\bibnamefont {Fang}}, \bibinfo {author}
  {\bibfnamefont {K.}~\bibnamefont {Watanabe}}, \bibinfo {author}
  {\bibfnamefont {T.}~\bibnamefont {Taniguchi}}, \bibinfo {author}
  {\bibfnamefont {E.}~\bibnamefont {Kaxiras}}, \ and\ \bibinfo {author}
  {\bibfnamefont {P.}~\bibnamefont {Jarillo-Herrero}},\ }\href@noop {}
  {\bibfield  {journal} {\bibinfo  {journal} {Nature}\ }\textbf {\bibinfo
  {volume} {556}},\ \bibinfo {pages} {43} (\bibinfo {year}
  {2018}{\natexlab{b}})}\BibitemShut {NoStop}%
\bibitem [{\citenamefont {Yankowitz}\ \emph {et~al.}(2019)\citenamefont
  {Yankowitz}, \citenamefont {Chen}, \citenamefont {Polshyn}, \citenamefont
  {Zhang}, \citenamefont {Watanabe}, \citenamefont {Taniguchi}, \citenamefont
  {Graf}, \citenamefont {Young},\ and\ \citenamefont {Dean}}]{TSTBLG}%
  \BibitemOpen
  \bibfield  {author} {\bibinfo {author} {\bibfnamefont {M.}~\bibnamefont
  {Yankowitz}}, \bibinfo {author} {\bibfnamefont {S.}~\bibnamefont {Chen}},
  \bibinfo {author} {\bibfnamefont {H.}~\bibnamefont {Polshyn}}, \bibinfo
  {author} {\bibfnamefont {Y.}~\bibnamefont {Zhang}}, \bibinfo {author}
  {\bibfnamefont {K.}~\bibnamefont {Watanabe}}, \bibinfo {author}
  {\bibfnamefont {T.}~\bibnamefont {Taniguchi}}, \bibinfo {author}
  {\bibfnamefont {D.}~\bibnamefont {Graf}}, \bibinfo {author} {\bibfnamefont
  {A.~F.}\ \bibnamefont {Young}}, \ and\ \bibinfo {author} {\bibfnamefont
  {C.~R.}\ \bibnamefont {Dean}},\ }\href@noop {} {\bibfield  {journal}
  {\bibinfo  {journal} {Science}\ }\textbf {\bibinfo {volume} {363}},\ \bibinfo
  {pages} {1059} (\bibinfo {year} {2019})}\BibitemShut {NoStop}%
\bibitem [{\citenamefont {Lu}\ \emph {et~al.}(2019)\citenamefont {Lu},
  \citenamefont {Stepanov}, \citenamefont {Yang}, \citenamefont {Xie},
  \citenamefont {Aamir}, \citenamefont {Das}, \citenamefont {Urgell},
  \citenamefont {Watanabe}, \citenamefont {Taniguchi}, \citenamefont {Zhang},
  \citenamefont {Bachtold}, \citenamefont {MacDonald},\ and\ \citenamefont
  {Efetov}}]{SOM}%
  \BibitemOpen
  \bibfield  {author} {\bibinfo {author} {\bibfnamefont {X.}~\bibnamefont
  {Lu}}, \bibinfo {author} {\bibfnamefont {P.}~\bibnamefont {Stepanov}},
  \bibinfo {author} {\bibfnamefont {W.}~\bibnamefont {Yang}}, \bibinfo {author}
  {\bibfnamefont {M.}~\bibnamefont {Xie}}, \bibinfo {author} {\bibfnamefont
  {M.~A.}\ \bibnamefont {Aamir}}, \bibinfo {author} {\bibfnamefont
  {I.}~\bibnamefont {Das}}, \bibinfo {author} {\bibfnamefont {C.}~\bibnamefont
  {Urgell}}, \bibinfo {author} {\bibfnamefont {K.}~\bibnamefont {Watanabe}},
  \bibinfo {author} {\bibfnamefont {T.}~\bibnamefont {Taniguchi}}, \bibinfo
  {author} {\bibfnamefont {G.}~\bibnamefont {Zhang}}, \bibinfo {author}
  {\bibfnamefont {A.}~\bibnamefont {Bachtold}}, \bibinfo {author}
  {\bibfnamefont {A.~H.}\ \bibnamefont {MacDonald}}, \ and\ \bibinfo {author}
  {\bibfnamefont {D.~K.}\ \bibnamefont {Efetov}},\ }\href@noop {} {\bibfield
  {journal} {\bibinfo  {journal} {Nature}\ }\textbf {\bibinfo {volume} {574}},\
  \bibinfo {pages} {653–657} (\bibinfo {year} {2019})}\BibitemShut {NoStop}%
\bibitem [{\citenamefont {Jiang}\ \emph {et~al.}(2019)\citenamefont {Jiang},
  \citenamefont {Lai}, \citenamefont {Watanabe}, \citenamefont {Taniguchi},
  \citenamefont {Haule}, \citenamefont {Mao},\ and\ \citenamefont
  {Andrei}}]{NAT_CO}%
  \BibitemOpen
  \bibfield  {author} {\bibinfo {author} {\bibfnamefont {Y.}~\bibnamefont
  {Jiang}}, \bibinfo {author} {\bibfnamefont {X.}~\bibnamefont {Lai}}, \bibinfo
  {author} {\bibfnamefont {K.}~\bibnamefont {Watanabe}}, \bibinfo {author}
  {\bibfnamefont {T.}~\bibnamefont {Taniguchi}}, \bibinfo {author}
  {\bibfnamefont {K.}~\bibnamefont {Haule}}, \bibinfo {author} {\bibfnamefont
  {J.}~\bibnamefont {Mao}}, \ and\ \bibinfo {author} {\bibfnamefont {E.~Y.}\
  \bibnamefont {Andrei}},\ }\href@noop {} {\bibfield  {journal} {\bibinfo
  {journal} {Nature}\ }\textbf {\bibinfo {volume} {573}},\ \bibinfo {pages}
  {91} (\bibinfo {year} {2019})}\BibitemShut {NoStop}%
\bibitem [{\citenamefont {Xie}\ \emph {et~al.}(2019)\citenamefont {Xie},
  \citenamefont {Lian}, \citenamefont {J\"{a}ck}, \citenamefont {Liu},
  \citenamefont {Chiu}, \citenamefont {Watanabe}, \citenamefont {Taniguchi},
  \citenamefont {Bernevig},\ and\ \citenamefont {Yazdani}}]{NAT_SS}%
  \BibitemOpen
  \bibfield  {author} {\bibinfo {author} {\bibfnamefont {Y.}~\bibnamefont
  {Xie}}, \bibinfo {author} {\bibfnamefont {B.}~\bibnamefont {Lian}}, \bibinfo
  {author} {\bibfnamefont {B.}~\bibnamefont {J\"{a}ck}}, \bibinfo {author}
  {\bibfnamefont {X.}~\bibnamefont {Liu}}, \bibinfo {author} {\bibfnamefont
  {C.-L.}\ \bibnamefont {Chiu}}, \bibinfo {author} {\bibfnamefont
  {K.}~\bibnamefont {Watanabe}}, \bibinfo {author} {\bibfnamefont
  {T.}~\bibnamefont {Taniguchi}}, \bibinfo {author} {\bibfnamefont {B.~A.}\
  \bibnamefont {Bernevig}}, \ and\ \bibinfo {author} {\bibfnamefont
  {A.}~\bibnamefont {Yazdani}},\ }\href@noop {} {\bibfield  {journal} {\bibinfo
   {journal} {Nature}\ }\textbf {\bibinfo {volume} {572}},\ \bibinfo {pages}
  {101} (\bibinfo {year} {2019})}\BibitemShut {NoStop}%
\bibitem [{\citenamefont {Kerelsky}\ \emph {et~al.}(2019)\citenamefont
  {Kerelsky}, \citenamefont {McGilly}, \citenamefont {Kennes}, \citenamefont
  {Xian}, \citenamefont {Yankowitz}, \citenamefont {Chen}, \citenamefont
  {Watanabe}, \citenamefont {Taniguchi}, \citenamefont {Hone}, \citenamefont
  {Dean}, \citenamefont {Rubio},\ and\ \citenamefont {Pasupathy}}]{NAT_MEI}%
  \BibitemOpen
  \bibfield  {author} {\bibinfo {author} {\bibfnamefont {A.}~\bibnamefont
  {Kerelsky}}, \bibinfo {author} {\bibfnamefont {L.~J.}\ \bibnamefont
  {McGilly}}, \bibinfo {author} {\bibfnamefont {D.~M.}\ \bibnamefont {Kennes}},
  \bibinfo {author} {\bibfnamefont {L.}~\bibnamefont {Xian}}, \bibinfo {author}
  {\bibfnamefont {M.}~\bibnamefont {Yankowitz}}, \bibinfo {author}
  {\bibfnamefont {S.}~\bibnamefont {Chen}}, \bibinfo {author} {\bibfnamefont
  {K.}~\bibnamefont {Watanabe}}, \bibinfo {author} {\bibfnamefont
  {T.}~\bibnamefont {Taniguchi}}, \bibinfo {author} {\bibfnamefont
  {J.}~\bibnamefont {Hone}}, \bibinfo {author} {\bibfnamefont {C.}~\bibnamefont
  {Dean}}, \bibinfo {author} {\bibfnamefont {A.}~\bibnamefont {Rubio}}, \ and\
  \bibinfo {author} {\bibfnamefont {A.~N.}\ \bibnamefont {Pasupathy}},\
  }\href@noop {} {\bibfield  {journal} {\bibinfo  {journal} {Nature}\ }\textbf
  {\bibinfo {volume} {572}},\ \bibinfo {pages} {95} (\bibinfo {year}
  {2019})}\BibitemShut {NoStop}%
\bibitem [{\citenamefont {Wong}\ \emph {et~al.}(2020)\citenamefont {Wong},
  \citenamefont {Nuckolls}, \citenamefont {Oh}, \citenamefont {Lian},
  \citenamefont {Yonglong~Xie}, \citenamefont {Watanabe}, \citenamefont
  {Taniguchi}, \citenamefont {Bernevig},\ and\ \citenamefont
  {Yazdani}}]{Wong2020}%
  \BibitemOpen
  \bibfield  {author} {\bibinfo {author} {\bibfnamefont {D.}~\bibnamefont
  {Wong}}, \bibinfo {author} {\bibfnamefont {K.~P.}\ \bibnamefont {Nuckolls}},
  \bibinfo {author} {\bibfnamefont {M.}~\bibnamefont {Oh}}, \bibinfo {author}
  {\bibfnamefont {B.}~\bibnamefont {Lian}}, \bibinfo {author} {\bibfnamefont
  {S.~J.}\ \bibnamefont {Yonglong~Xie}}, \bibinfo {author} {\bibfnamefont
  {K.}~\bibnamefont {Watanabe}}, \bibinfo {author} {\bibfnamefont
  {T.}~\bibnamefont {Taniguchi}}, \bibinfo {author} {\bibfnamefont {B.~A.}\
  \bibnamefont {Bernevig}}, \ and\ \bibinfo {author} {\bibfnamefont
  {A.}~\bibnamefont {Yazdani}},\ }\href@noop {} {\bibfield  {journal} {\bibinfo
   {journal} {Nature}\ }\textbf {\bibinfo {volume} {582}},\ \bibinfo {pages}
  {198–202} (\bibinfo {year} {2020})}\BibitemShut {NoStop}%
\bibitem [{\citenamefont {Zondiner}\ \emph {et~al.}(2020)\citenamefont
  {Zondiner}, \citenamefont {Rozen}, \citenamefont {Rodan-Legrain},
  \citenamefont {Cao}, \citenamefont {Queiroz}, \citenamefont {Taniguchi},
  \citenamefont {Watanabe}, \citenamefont {Oreg}, \citenamefont {von Oppen},
  \citenamefont {Stern}, \citenamefont {Berg}, \citenamefont
  {Jarillo-Herrero},\ and\ \citenamefont {Ilani}}]{Zondiner2020}%
  \BibitemOpen
  \bibfield  {author} {\bibinfo {author} {\bibfnamefont {U.}~\bibnamefont
  {Zondiner}}, \bibinfo {author} {\bibfnamefont {A.}~\bibnamefont {Rozen}},
  \bibinfo {author} {\bibfnamefont {D.}~\bibnamefont {Rodan-Legrain}}, \bibinfo
  {author} {\bibfnamefont {Y.}~\bibnamefont {Cao}}, \bibinfo {author}
  {\bibfnamefont {R.}~\bibnamefont {Queiroz}}, \bibinfo {author} {\bibfnamefont
  {T.}~\bibnamefont {Taniguchi}}, \bibinfo {author} {\bibfnamefont
  {K.}~\bibnamefont {Watanabe}}, \bibinfo {author} {\bibfnamefont
  {Y.}~\bibnamefont {Oreg}}, \bibinfo {author} {\bibfnamefont {F.}~\bibnamefont
  {von Oppen}}, \bibinfo {author} {\bibfnamefont {A.}~\bibnamefont {Stern}},
  \bibinfo {author} {\bibfnamefont {E.}~\bibnamefont {Berg}}, \bibinfo {author}
  {\bibfnamefont {P.}~\bibnamefont {Jarillo-Herrero}}, \ and\ \bibinfo {author}
  {\bibfnamefont {S.}~\bibnamefont {Ilani}},\ }\href@noop {} {\bibfield
  {journal} {\bibinfo  {journal} {Nature}\ }\textbf {\bibinfo {volume} {582}},\
  \bibinfo {pages} {203} (\bibinfo {year} {2020})}\BibitemShut {NoStop}%
\bibitem [{\citenamefont {Cao}\ \emph {et~al.}(2020{\natexlab{a}})\citenamefont
  {Cao}, \citenamefont {Chowdhury}, \citenamefont {Rodan-Legrain},
  \citenamefont {Rubies-Bigord\`a}, \citenamefont {Watanabe}, \citenamefont
  {Taniguchi}, \citenamefont {Senthil},\ and\ \citenamefont
  {Jarillo-Herrero}}]{SMTBLG}%
  \BibitemOpen
  \bibfield  {author} {\bibinfo {author} {\bibfnamefont {Y.}~\bibnamefont
  {Cao}}, \bibinfo {author} {\bibfnamefont {D.}~\bibnamefont {Chowdhury}},
  \bibinfo {author} {\bibfnamefont {D.}~\bibnamefont {Rodan-Legrain}}, \bibinfo
  {author} {\bibfnamefont {O.}~\bibnamefont {Rubies-Bigord\`a}}, \bibinfo
  {author} {\bibfnamefont {K.}~\bibnamefont {Watanabe}}, \bibinfo {author}
  {\bibfnamefont {T.}~\bibnamefont {Taniguchi}}, \bibinfo {author}
  {\bibfnamefont {T.}~\bibnamefont {Senthil}}, \ and\ \bibinfo {author}
  {\bibfnamefont {P.}~\bibnamefont {Jarillo-Herrero}},\ }\href@noop {}
  {\bibfield  {journal} {\bibinfo  {journal} {Phys. Rev. Lett.}\ }\textbf
  {\bibinfo {volume} {124}},\ \bibinfo {pages} {076801} (\bibinfo {year}
  {2020}{\natexlab{a}})}\BibitemShut {NoStop}%
\bibitem [{\citenamefont {Polshyn}\ \emph {et~al.}(2019)\citenamefont
  {Polshyn}, \citenamefont {Yankowitz}, \citenamefont {Chen}, \citenamefont
  {Zhang}, \citenamefont {Watanabe}, \citenamefont {Taniguchi}, \citenamefont
  {Dean},\ and\ \citenamefont {Young}}]{Polshyn2019strange}%
  \BibitemOpen
  \bibfield  {author} {\bibinfo {author} {\bibfnamefont {H.}~\bibnamefont
  {Polshyn}}, \bibinfo {author} {\bibfnamefont {M.}~\bibnamefont {Yankowitz}},
  \bibinfo {author} {\bibfnamefont {S.}~\bibnamefont {Chen}}, \bibinfo {author}
  {\bibfnamefont {Y.}~\bibnamefont {Zhang}}, \bibinfo {author} {\bibfnamefont
  {K.}~\bibnamefont {Watanabe}}, \bibinfo {author} {\bibfnamefont
  {T.}~\bibnamefont {Taniguchi}}, \bibinfo {author} {\bibfnamefont {C.~R.}\
  \bibnamefont {Dean}}, \ and\ \bibinfo {author} {\bibfnamefont {A.~F.}\
  \bibnamefont {Young}},\ }\href@noop {} {\bibfield  {journal} {\bibinfo
  {journal} {Nat. Phys.}\ }\textbf {\bibinfo {volume} {15}},\ \bibinfo {pages}
  {1011–1016} (\bibinfo {year} {2019})}\BibitemShut {NoStop}%
\bibitem [{\citenamefont {Cao}\ \emph {et~al.}(2021)\citenamefont {Cao},
  \citenamefont {Rodan-Legrain}, \citenamefont {Park}, \citenamefont {Yuan},
  \citenamefont {Watanabe}, \citenamefont {Taniguchi}, \citenamefont
  {Fernandes}, \citenamefont {Fu},\ and\ \citenamefont
  {Jarillo-Herrero}}]{Caonematicity2020}%
  \BibitemOpen
  \bibfield  {author} {\bibinfo {author} {\bibfnamefont {Y.}~\bibnamefont
  {Cao}}, \bibinfo {author} {\bibfnamefont {D.}~\bibnamefont {Rodan-Legrain}},
  \bibinfo {author} {\bibfnamefont {J.~M.}\ \bibnamefont {Park}}, \bibinfo
  {author} {\bibfnamefont {F.~N.}\ \bibnamefont {Yuan}}, \bibinfo {author}
  {\bibfnamefont {K.}~\bibnamefont {Watanabe}}, \bibinfo {author}
  {\bibfnamefont {T.}~\bibnamefont {Taniguchi}}, \bibinfo {author}
  {\bibfnamefont {R.~M.}\ \bibnamefont {Fernandes}}, \bibinfo {author}
  {\bibfnamefont {L.}~\bibnamefont {Fu}}, \ and\ \bibinfo {author}
  {\bibfnamefont {P.}~\bibnamefont {Jarillo-Herrero}},\ }\href@noop {}
  {\bibfield  {journal} {\bibinfo  {journal} {Science}\ }\textbf {\bibinfo
  {volume} {372}},\ \bibinfo {pages} {264} (\bibinfo {year}
  {2021})}\BibitemShut {NoStop}%
\bibitem [{\citenamefont {Koshino}(2019)}]{KoshinoMikito2019Bsat}%
  \BibitemOpen
  \bibfield  {author} {\bibinfo {author} {\bibfnamefont {M.}~\bibnamefont
  {Koshino}},\ }\href@noop {} {\bibfield  {journal} {\bibinfo  {journal} {Phys.
  Rev. B}\ }\textbf {\bibinfo {volume} {99}},\ \bibinfo {pages} {235406}
  (\bibinfo {year} {2019})}\BibitemShut {NoStop}%
\bibitem [{\citenamefont {Liu}\ \emph {et~al.}(2020)\citenamefont {Liu},
  \citenamefont {Hao}, \citenamefont {Khalaf}, \citenamefont {Lee},
  \citenamefont {Watanabe}, \citenamefont {Taniguchi}, \citenamefont
  {Vishwanath},\ and\ \citenamefont {Kim}}]{BIBI}%
  \BibitemOpen
  \bibfield  {author} {\bibinfo {author} {\bibfnamefont {X.}~\bibnamefont
  {Liu}}, \bibinfo {author} {\bibfnamefont {Z.}~\bibnamefont {Hao}}, \bibinfo
  {author} {\bibfnamefont {E.}~\bibnamefont {Khalaf}}, \bibinfo {author}
  {\bibfnamefont {J.~Y.}\ \bibnamefont {Lee}}, \bibinfo {author} {\bibfnamefont
  {K.}~\bibnamefont {Watanabe}}, \bibinfo {author} {\bibfnamefont
  {T.}~\bibnamefont {Taniguchi}}, \bibinfo {author} {\bibfnamefont
  {A.}~\bibnamefont {Vishwanath}}, \ and\ \bibinfo {author} {\bibfnamefont
  {P.}~\bibnamefont {Kim}},\ }\href@noop {} {\bibfield  {journal} {\bibinfo
  {journal} {Nature}\ }\textbf {\bibinfo {volume} {583}},\ \bibinfo {pages}
  {221} (\bibinfo {year} {2020})}\BibitemShut {NoStop}%
\bibitem [{\citenamefont {Haddadi}\ \emph {et~al.}(2020)\citenamefont
  {Haddadi}, \citenamefont {Wu}, \citenamefont {Kruchkov},\ and\ \citenamefont
  {Yazyev}}]{haddadi2019moir}%
  \BibitemOpen
  \bibfield  {author} {\bibinfo {author} {\bibfnamefont {F.}~\bibnamefont
  {Haddadi}}, \bibinfo {author} {\bibfnamefont {Q.}~\bibnamefont {Wu}},
  \bibinfo {author} {\bibfnamefont {A.~J.}\ \bibnamefont {Kruchkov}}, \ and\
  \bibinfo {author} {\bibfnamefont {O.~V.}\ \bibnamefont {Yazyev}},\
  }\href@noop {} {\bibfield  {journal} {\bibinfo  {journal} {Nano Lett.}\
  }\textbf {\bibinfo {volume} {20}},\ \bibinfo {pages} {2410–2415} (\bibinfo
  {year} {2020})}\BibitemShut {NoStop}%
\bibitem [{\citenamefont {Burg}\ \emph {et~al.}(2019)\citenamefont {Burg},
  \citenamefont {Zhu}, \citenamefont {Taniguchi}, \citenamefont {Watanabe},
  \citenamefont {MacDonald},\ and\ \citenamefont
  {Tutuc}}]{PhysRevLett.123.197702}%
  \BibitemOpen
  \bibfield  {author} {\bibinfo {author} {\bibfnamefont {G.~W.}\ \bibnamefont
  {Burg}}, \bibinfo {author} {\bibfnamefont {J.}~\bibnamefont {Zhu}}, \bibinfo
  {author} {\bibfnamefont {T.}~\bibnamefont {Taniguchi}}, \bibinfo {author}
  {\bibfnamefont {K.}~\bibnamefont {Watanabe}}, \bibinfo {author}
  {\bibfnamefont {A.~H.}\ \bibnamefont {MacDonald}}, \ and\ \bibinfo {author}
  {\bibfnamefont {E.}~\bibnamefont {Tutuc}},\ }\href@noop {} {\bibfield
  {journal} {\bibinfo  {journal} {Phys. Rev. Lett.}\ }\textbf {\bibinfo
  {volume} {123}},\ \bibinfo {pages} {197702} (\bibinfo {year}
  {2019})}\BibitemShut {NoStop}%
\bibitem [{\citenamefont {Leey}\ \emph {et~al.}(2019)\citenamefont {Leey},
  \citenamefont {Khalafy}, \citenamefont {Liu}, \citenamefont {Liu},
  \citenamefont {Hao}, \citenamefont {Kim},\ and\ \citenamefont
  {Vishwanath}}]{STSCTDB}%
  \BibitemOpen
  \bibfield  {author} {\bibinfo {author} {\bibfnamefont {J.~Y.}\ \bibnamefont
  {Leey}}, \bibinfo {author} {\bibfnamefont {E.}~\bibnamefont {Khalafy}},
  \bibinfo {author} {\bibfnamefont {S.}~\bibnamefont {Liu}}, \bibinfo {author}
  {\bibfnamefont {X.}~\bibnamefont {Liu}}, \bibinfo {author} {\bibfnamefont
  {Z.}~\bibnamefont {Hao}}, \bibinfo {author} {\bibfnamefont {P.}~\bibnamefont
  {Kim}}, \ and\ \bibinfo {author} {\bibfnamefont {A.}~\bibnamefont
  {Vishwanath}},\ }\href@noop {} {\bibfield  {journal} {\bibinfo  {journal}
  {Nat. Commun.}\ }\textbf {\bibinfo {volume} {10}},\ \bibinfo {pages} {5333}
  (\bibinfo {year} {2019})}\BibitemShut {NoStop}%
\bibitem [{\citenamefont {Samajdar}\ and\ \citenamefont
  {Scheurer}(2020)}]{Samajdar2020}%
  \BibitemOpen
  \bibfield  {author} {\bibinfo {author} {\bibfnamefont {R.}~\bibnamefont
  {Samajdar}}\ and\ \bibinfo {author} {\bibfnamefont {M.~S.}\ \bibnamefont
  {Scheurer}},\ }\href@noop {} {\bibfield  {journal} {\bibinfo  {journal}
  {Phys. Rev. B}\ }\textbf {\bibinfo {volume} {102}},\ \bibinfo {pages}
  {064501} (\bibinfo {year} {2020})}\BibitemShut {NoStop}%
\bibitem [{\citenamefont {Cao}\ \emph {et~al.}(2020{\natexlab{b}})\citenamefont
  {Cao}, \citenamefont {Rodan-Legrain}, \citenamefont {Rubies-Bigorda},
  \citenamefont {Park}, \citenamefont {Watanabe}, \citenamefont {Taniguchi},\
  and\ \citenamefont {Jarillo-Herrero}}]{cao2019electric}%
  \BibitemOpen
  \bibfield  {author} {\bibinfo {author} {\bibfnamefont {Y.}~\bibnamefont
  {Cao}}, \bibinfo {author} {\bibfnamefont {D.}~\bibnamefont {Rodan-Legrain}},
  \bibinfo {author} {\bibfnamefont {O.}~\bibnamefont {Rubies-Bigorda}},
  \bibinfo {author} {\bibfnamefont {J.~M.}\ \bibnamefont {Park}}, \bibinfo
  {author} {\bibfnamefont {K.}~\bibnamefont {Watanabe}}, \bibinfo {author}
  {\bibfnamefont {T.}~\bibnamefont {Taniguchi}}, \ and\ \bibinfo {author}
  {\bibfnamefont {P.}~\bibnamefont {Jarillo-Herrero}},\ }\href@noop {}
  {\bibfield  {journal} {\bibinfo  {journal} {Nature}\ }\textbf {\bibinfo
  {volume} {583}},\ \bibinfo {pages} {215} (\bibinfo {year}
  {2020}{\natexlab{b}})}\BibitemShut {NoStop}%
\bibitem [{\citenamefont {Shen}\ \emph {et~al.}(2020)\citenamefont {Shen},
  \citenamefont {Chu}, \citenamefont {Wu}, \citenamefont {Li}, \citenamefont
  {Wang}, \citenamefont {Zhao}, \citenamefont {Tang}, \citenamefont {Liu},
  \citenamefont {Tian}, \citenamefont {Watanabe}, \citenamefont {Taniguchi},
  \citenamefont {Yang}, \citenamefont {Meng}, \citenamefont {Shi},
  \citenamefont {Yazyev},\ and\ \citenamefont {Zhang}}]{TCT}%
  \BibitemOpen
  \bibfield  {author} {\bibinfo {author} {\bibfnamefont {C.}~\bibnamefont
  {Shen}}, \bibinfo {author} {\bibfnamefont {Y.}~\bibnamefont {Chu}}, \bibinfo
  {author} {\bibfnamefont {Q.}~\bibnamefont {Wu}}, \bibinfo {author}
  {\bibfnamefont {N.}~\bibnamefont {Li}}, \bibinfo {author} {\bibfnamefont
  {S.}~\bibnamefont {Wang}}, \bibinfo {author} {\bibfnamefont {Y.}~\bibnamefont
  {Zhao}}, \bibinfo {author} {\bibfnamefont {J.}~\bibnamefont {Tang}}, \bibinfo
  {author} {\bibfnamefont {J.}~\bibnamefont {Liu}}, \bibinfo {author}
  {\bibfnamefont {J.}~\bibnamefont {Tian}}, \bibinfo {author} {\bibfnamefont
  {K.}~\bibnamefont {Watanabe}}, \bibinfo {author} {\bibfnamefont
  {T.}~\bibnamefont {Taniguchi}}, \bibinfo {author} {\bibfnamefont
  {R.}~\bibnamefont {Yang}}, \bibinfo {author} {\bibfnamefont {Z.~Y.}\
  \bibnamefont {Meng}}, \bibinfo {author} {\bibfnamefont {D.}~\bibnamefont
  {Shi}}, \bibinfo {author} {\bibfnamefont {O.~V.}\ \bibnamefont {Yazyev}}, \
  and\ \bibinfo {author} {\bibfnamefont {G.}~\bibnamefont {Zhang}},\
  }\href@noop {} {\bibfield  {journal} {\bibinfo  {journal} {Nat. Phys.}\
  }\textbf {\bibinfo {volume} {16}},\ \bibinfo {pages} {520} (\bibinfo {year}
  {2020})}\BibitemShut {NoStop}%
\bibitem [{\citenamefont {He}\ \emph {et~al.}(2021)\citenamefont {He},
  \citenamefont {Li}, \citenamefont {Cai}, \citenamefont {Liu}, \citenamefont
  {Watanabe}, \citenamefont {Taniguchi}, \citenamefont {Xu},\ and\
  \citenamefont {Yankowitz}}]{He2021}%
  \BibitemOpen
  \bibfield  {author} {\bibinfo {author} {\bibfnamefont {M.}~\bibnamefont
  {He}}, \bibinfo {author} {\bibfnamefont {Y.}~\bibnamefont {Li}}, \bibinfo
  {author} {\bibfnamefont {J.}~\bibnamefont {Cai}}, \bibinfo {author}
  {\bibfnamefont {Y.}~\bibnamefont {Liu}}, \bibinfo {author} {\bibfnamefont
  {K.}~\bibnamefont {Watanabe}}, \bibinfo {author} {\bibfnamefont
  {T.}~\bibnamefont {Taniguchi}}, \bibinfo {author} {\bibfnamefont
  {X.}~\bibnamefont {Xu}}, \ and\ \bibinfo {author} {\bibfnamefont
  {M.}~\bibnamefont {Yankowitz}},\ }\href@noop {} {\bibfield  {journal}
  {\bibinfo  {journal} {Nat. Phys.}\ }\textbf {\bibinfo {volume} {17}},\
  \bibinfo {pages} {26–30} (\bibinfo {year} {2021})}\BibitemShut {NoStop}%
\bibitem [{\citenamefont {Choi}\ and\ \citenamefont
  {Choi}(2021)}]{Choi2021Dichotomy}%
  \BibitemOpen
  \bibfield  {author} {\bibinfo {author} {\bibfnamefont {Y.~W.}\ \bibnamefont
  {Choi}}\ and\ \bibinfo {author} {\bibfnamefont {H.~J.}\ \bibnamefont
  {Choi}},\ }\href@noop {} {\bibfield  {journal} {\bibinfo  {journal} {Phys.
  Rev. Lett.}\ }\textbf {\bibinfo {volume} {127}},\ \bibinfo {pages} {167001}
  (\bibinfo {year} {2021})}\BibitemShut {NoStop}%
\bibitem [{\citenamefont {Liu}\ \emph {et~al.}(2021)\citenamefont {Liu},
  \citenamefont {Chiu}, \citenamefont {Lee}, \citenamefont {Farahi},
  \citenamefont {Watanabe}, \citenamefont {Taniguchi}, \citenamefont
  {Vishwanath},\ and\ \citenamefont {Yazdani}}]{Liu2021}%
  \BibitemOpen
  \bibfield  {author} {\bibinfo {author} {\bibfnamefont {X.}~\bibnamefont
  {Liu}}, \bibinfo {author} {\bibfnamefont {C.-L.}\ \bibnamefont {Chiu}},
  \bibinfo {author} {\bibfnamefont {J.~Y.}\ \bibnamefont {Lee}}, \bibinfo
  {author} {\bibfnamefont {G.}~\bibnamefont {Farahi}}, \bibinfo {author}
  {\bibfnamefont {K.}~\bibnamefont {Watanabe}}, \bibinfo {author}
  {\bibfnamefont {T.}~\bibnamefont {Taniguchi}}, \bibinfo {author}
  {\bibfnamefont {A.}~\bibnamefont {Vishwanath}}, \ and\ \bibinfo {author}
  {\bibfnamefont {A.}~\bibnamefont {Yazdani}},\ }\href@noop {} {\bibfield
  {journal} {\bibinfo  {journal} {Nat. Commun.}\ }\textbf {\bibinfo {volume}
  {12}},\ \bibinfo {pages} {2732} (\bibinfo {year} {2021})}\BibitemShut
  {NoStop}%
\bibitem [{\citenamefont {Zhang}\ \emph {et~al.}(12)\citenamefont {Zhang},
  \citenamefont {Zhu}, \citenamefont {Kahn}, \citenamefont {Li}, \citenamefont
  {Yang}, \citenamefont {Herbig}, \citenamefont {Wu}, \citenamefont {Li},
  \citenamefont {Watanabe}, \citenamefont {Taniguchi}, \citenamefont {Cabrini},
  \citenamefont {Zettl}, \citenamefont {Zaletel}, \citenamefont {Wang},\ and\
  \citenamefont {Crommie}}]{Crommie2021}%
  \BibitemOpen
  \bibfield  {author} {\bibinfo {author} {\bibfnamefont {C.}~\bibnamefont
  {Zhang}}, \bibinfo {author} {\bibfnamefont {T.}~\bibnamefont {Zhu}}, \bibinfo
  {author} {\bibfnamefont {S.}~\bibnamefont {Kahn}}, \bibinfo {author}
  {\bibfnamefont {S.}~\bibnamefont {Li}}, \bibinfo {author} {\bibfnamefont
  {B.}~\bibnamefont {Yang}}, \bibinfo {author} {\bibfnamefont {C.}~\bibnamefont
  {Herbig}}, \bibinfo {author} {\bibfnamefont {X.}~\bibnamefont {Wu}}, \bibinfo
  {author} {\bibfnamefont {H.}~\bibnamefont {Li}}, \bibinfo {author}
  {\bibfnamefont {K.}~\bibnamefont {Watanabe}}, \bibinfo {author}
  {\bibfnamefont {T.}~\bibnamefont {Taniguchi}}, \bibinfo {author}
  {\bibfnamefont {S.}~\bibnamefont {Cabrini}}, \bibinfo {author} {\bibfnamefont
  {A.}~\bibnamefont {Zettl}}, \bibinfo {author} {\bibfnamefont {M.~P.}\
  \bibnamefont {Zaletel}}, \bibinfo {author} {\bibfnamefont {F.}~\bibnamefont
  {Wang}}, \ and\ \bibinfo {author} {\bibfnamefont {M.~F.}\ \bibnamefont
  {Crommie}},\ }\href@noop {} {\bibfield  {journal} {\bibinfo  {journal} {Nat.
  Commun.}\ }\textbf {\bibinfo {volume} {2021}},\ \bibinfo {pages} {2516}
  (\bibinfo {year} {12})}\BibitemShut {NoStop}%
\bibitem [{\citenamefont {Rubio-Verd\'u}\ \emph {et~al.}(2021)\citenamefont
  {Rubio-Verd\'u}, \citenamefont {Turkel}, \citenamefont {Song}, \citenamefont
  {Klebl}, \citenamefont {Samajdar}, \citenamefont {Scheurer}, \citenamefont
  {Venderbos}, \citenamefont {Watanabe}, \citenamefont {Taniguchi},
  \citenamefont {Ochoa}, \citenamefont {Xian}, \citenamefont {Kennes},
  \citenamefont {Fernandes}, \citenamefont {Rubio},\ and\ \citenamefont
  {Pasupathy}}]{Carmen2021}%
  \BibitemOpen
  \bibfield  {author} {\bibinfo {author} {\bibfnamefont {C.}~\bibnamefont
  {Rubio-Verd\'u}}, \bibinfo {author} {\bibfnamefont {S.}~\bibnamefont
  {Turkel}}, \bibinfo {author} {\bibfnamefont {L.}~\bibnamefont {Song}},
  \bibinfo {author} {\bibfnamefont {L.}~\bibnamefont {Klebl}}, \bibinfo
  {author} {\bibfnamefont {R.}~\bibnamefont {Samajdar}}, \bibinfo {author}
  {\bibfnamefont {M.~S.}\ \bibnamefont {Scheurer}}, \bibinfo {author}
  {\bibfnamefont {J.~W.~F.}\ \bibnamefont {Venderbos}}, \bibinfo {author}
  {\bibfnamefont {K.}~\bibnamefont {Watanabe}}, \bibinfo {author}
  {\bibfnamefont {T.}~\bibnamefont {Taniguchi}}, \bibinfo {author}
  {\bibfnamefont {H.}~\bibnamefont {Ochoa}}, \bibinfo {author} {\bibfnamefont
  {L.}~\bibnamefont {Xian}}, \bibinfo {author} {\bibfnamefont {D.}~\bibnamefont
  {Kennes}}, \bibinfo {author} {\bibfnamefont {R.~M.}\ \bibnamefont
  {Fernandes}}, \bibinfo {author} {\bibfnamefont {A.}~\bibnamefont {Rubio}}, \
  and\ \bibinfo {author} {\bibfnamefont {A.~N.}\ \bibnamefont {Pasupathy}},\
  }\href@noop {} {\bibfield  {journal} {\bibinfo  {journal} {Nat. Phys.}\ ,\
  \bibinfo {pages} {https://doi.org/10.1038/s41567}} (\bibinfo {year}
  {2021})}\BibitemShut {NoStop}%
\bibitem [{\citenamefont {Rickhaus}\ \emph {et~al.}(2019)\citenamefont
  {Rickhaus}, \citenamefont {Zheng}, \citenamefont {Lado}, \citenamefont {Lee},
  \citenamefont {Kurzmann}, \citenamefont {Eich}, \citenamefont {Pisoni},
  \citenamefont {Tong}, \citenamefont {Garreis}, \citenamefont {Gold},
  \citenamefont {Masseroni}, \citenamefont {Taniguchi}, \citenamefont
  {Wantanabe}, \citenamefont {Ihn},\ and\ \citenamefont
  {Ensslin}}]{RickhausPeter2019GOiT}%
  \BibitemOpen
  \bibfield  {author} {\bibinfo {author} {\bibfnamefont {P.}~\bibnamefont
  {Rickhaus}}, \bibinfo {author} {\bibfnamefont {G.}~\bibnamefont {Zheng}},
  \bibinfo {author} {\bibfnamefont {J.~L.}\ \bibnamefont {Lado}}, \bibinfo
  {author} {\bibfnamefont {Y.}~\bibnamefont {Lee}}, \bibinfo {author}
  {\bibfnamefont {A.}~\bibnamefont {Kurzmann}}, \bibinfo {author}
  {\bibfnamefont {M.}~\bibnamefont {Eich}}, \bibinfo {author} {\bibfnamefont
  {R.}~\bibnamefont {Pisoni}}, \bibinfo {author} {\bibfnamefont
  {C.}~\bibnamefont {Tong}}, \bibinfo {author} {\bibfnamefont {R.}~\bibnamefont
  {Garreis}}, \bibinfo {author} {\bibfnamefont {C.}~\bibnamefont {Gold}},
  \bibinfo {author} {\bibfnamefont {M.}~\bibnamefont {Masseroni}}, \bibinfo
  {author} {\bibfnamefont {T.}~\bibnamefont {Taniguchi}}, \bibinfo {author}
  {\bibfnamefont {K.}~\bibnamefont {Wantanabe}}, \bibinfo {author}
  {\bibfnamefont {T.}~\bibnamefont {Ihn}}, \ and\ \bibinfo {author}
  {\bibfnamefont {K.}~\bibnamefont {Ensslin}},\ }\href@noop {} {\bibfield
  {journal} {\bibinfo  {journal} {Nano lett.}\ }\textbf {\bibinfo {volume}
  {19}},\ \bibinfo {pages} {8821–8828} (\bibinfo {year} {2019})}\BibitemShut
  {NoStop}%
\bibitem [{\citenamefont {Culchac}\ \emph {et~al.}(2020)\citenamefont
  {Culchac}, \citenamefont {Del~Grande}, \citenamefont {Capaz}, \citenamefont
  {Chico},\ and\ \citenamefont {Morell}}]{CulchacF.J.2020Fbag}%
  \BibitemOpen
  \bibfield  {author} {\bibinfo {author} {\bibfnamefont {F.~J.}\ \bibnamefont
  {Culchac}}, \bibinfo {author} {\bibfnamefont {R.~R.}\ \bibnamefont
  {Del~Grande}}, \bibinfo {author} {\bibfnamefont {R.~B.}\ \bibnamefont
  {Capaz}}, \bibinfo {author} {\bibfnamefont {L.}~\bibnamefont {Chico}}, \ and\
  \bibinfo {author} {\bibfnamefont {E.~S.}\ \bibnamefont {Morell}},\
  }\href@noop {} {\bibfield  {journal} {\bibinfo  {journal} {Nanoscale}\
  }\textbf {\bibinfo {volume} {12}},\ \bibinfo {pages} {5014} (\bibinfo {year}
  {2020})}\BibitemShut {NoStop}%
\bibitem [{\citenamefont {Guinea}\ and\ \citenamefont {Walet}(2018)}]{EE}%
  \BibitemOpen
  \bibfield  {author} {\bibinfo {author} {\bibfnamefont {F.}~\bibnamefont
  {Guinea}}\ and\ \bibinfo {author} {\bibfnamefont {N.~R.}\ \bibnamefont
  {Walet}},\ }\href@noop {} {\bibfield  {journal} {\bibinfo  {journal} {PNAS}\
  }\textbf {\bibinfo {volume} {115}},\ \bibinfo {pages} {13174–13179}
  (\bibinfo {year} {2018})}\BibitemShut {NoStop}%
\bibitem [{\citenamefont {Cea}\ \emph {et~al.}(2019)\citenamefont {Cea},
  \citenamefont {Walet},\ and\ \citenamefont {Guinea}}]{Cea2019}%
  \BibitemOpen
  \bibfield  {author} {\bibinfo {author} {\bibfnamefont {T.}~\bibnamefont
  {Cea}}, \bibinfo {author} {\bibfnamefont {N.~R.}\ \bibnamefont {Walet}}, \
  and\ \bibinfo {author} {\bibfnamefont {F.}~\bibnamefont {Guinea}},\
  }\href@noop {} {\bibfield  {journal} {\bibinfo  {journal} {Phys. Rev. B}\
  }\textbf {\bibinfo {volume} {100}},\ \bibinfo {pages} {205113} (\bibinfo
  {year} {2019})}\BibitemShut {NoStop}%
\bibitem [{\citenamefont {Rademaker}\ \emph {et~al.}(2019)\citenamefont
  {Rademaker}, \citenamefont {Abanin},\ and\ \citenamefont
  {Mellado}}]{Rademaker2019}%
  \BibitemOpen
  \bibfield  {author} {\bibinfo {author} {\bibfnamefont {L.}~\bibnamefont
  {Rademaker}}, \bibinfo {author} {\bibfnamefont {D.~A.}\ \bibnamefont
  {Abanin}}, \ and\ \bibinfo {author} {\bibfnamefont {P.}~\bibnamefont
  {Mellado}},\ }\href@noop {} {\bibfield  {journal} {\bibinfo  {journal} {Phys.
  Rev. B}\ }\textbf {\bibinfo {volume} {100}},\ \bibinfo {pages} {205114}
  (\bibinfo {year} {2019})}\BibitemShut {NoStop}%
\bibitem [{\citenamefont {Goodwin}\ \emph
  {et~al.}(2020{\natexlab{a}})\citenamefont {Goodwin}, \citenamefont {Vitale},
  \citenamefont {Liang}, \citenamefont {Mostofi},\ and\ \citenamefont
  {Lischner}}]{PHD_4}%
  \BibitemOpen
  \bibfield  {author} {\bibinfo {author} {\bibfnamefont {Z.~A.~H.}\
  \bibnamefont {Goodwin}}, \bibinfo {author} {\bibfnamefont {V.}~\bibnamefont
  {Vitale}}, \bibinfo {author} {\bibfnamefont {X.}~\bibnamefont {Liang}},
  \bibinfo {author} {\bibfnamefont {A.~A.}\ \bibnamefont {Mostofi}}, \ and\
  \bibinfo {author} {\bibfnamefont {J.}~\bibnamefont {Lischner}},\ }\href@noop
  {} {\bibfield  {journal} {\bibinfo  {journal} {Electron. Struct.}\ }\textbf
  {\bibinfo {volume} {2}},\ \bibinfo {pages} {034001} (\bibinfo {year}
  {2020}{\natexlab{a}})}\BibitemShut {NoStop}%
\bibitem [{\citenamefont {Calder\'on}\ and\ \citenamefont
  {Bascones}(2020)}]{Bascones2020}%
  \BibitemOpen
  \bibfield  {author} {\bibinfo {author} {\bibfnamefont {M.~J.}\ \bibnamefont
  {Calder\'on}}\ and\ \bibinfo {author} {\bibfnamefont {E.}~\bibnamefont
  {Bascones}},\ }\href@noop {} {\bibfield  {journal} {\bibinfo  {journal}
  {Phys. Rev. B}\ }\textbf {\bibinfo {volume} {102}},\ \bibinfo {pages}
  {155149} (\bibinfo {year} {2020})}\BibitemShut {NoStop}%
\bibitem [{\citenamefont {Klebl}\ \emph {et~al.}(2021)\citenamefont {Klebl},
  \citenamefont {Goodwin}, \citenamefont {Mostofi}, \citenamefont {Kennes},\
  and\ \citenamefont {Lischner}}]{PHD_6}%
  \BibitemOpen
  \bibfield  {author} {\bibinfo {author} {\bibfnamefont {L.}~\bibnamefont
  {Klebl}}, \bibinfo {author} {\bibfnamefont {Z.~A.~H.}\ \bibnamefont
  {Goodwin}}, \bibinfo {author} {\bibfnamefont {A.~A.}\ \bibnamefont
  {Mostofi}}, \bibinfo {author} {\bibfnamefont {D.~M.}\ \bibnamefont {Kennes}},
  \ and\ \bibinfo {author} {\bibfnamefont {J.}~\bibnamefont {Lischner}},\
  }\href@noop {} {\bibfield  {journal} {\bibinfo  {journal} {Phys. Rev. B}\
  }\textbf {\bibinfo {volume} {103}},\ \bibinfo {pages} {195127} (\bibinfo
  {year} {2021})}\BibitemShut {NoStop}%
\bibitem [{\citenamefont {Plimpton}(1995)}]{LAMMPS}%
  \BibitemOpen
  \bibfield  {author} {\bibinfo {author} {\bibfnamefont {S.}~\bibnamefont
  {Plimpton}},\ }\href@noop {} {\bibfield  {journal} {\bibinfo  {journal} {J.
  Comp. Phys.}\ }\textbf {\bibinfo {volume} {117}},\ \bibinfo {pages} {1}
  (\bibinfo {year} {1995})}\BibitemShut {NoStop}%
\bibitem [{\citenamefont {O’Connor}\ \emph {et~al.}(2015)\citenamefont
  {O’Connor}, \citenamefont {Andzelm},\ and\ \citenamefont
  {Robbins}}]{AIREBO}%
  \BibitemOpen
  \bibfield  {author} {\bibinfo {author} {\bibfnamefont {T.~C.}\ \bibnamefont
  {O’Connor}}, \bibinfo {author} {\bibfnamefont {J.}~\bibnamefont {Andzelm}},
  \ and\ \bibinfo {author} {\bibfnamefont {M.~O.}\ \bibnamefont {Robbins}},\
  }\href@noop {} {\bibfield  {journal} {\bibinfo  {journal} {J. Chem. Phys.}\
  }\textbf {\bibinfo {volume} {142}},\ \bibinfo {pages} {024903} (\bibinfo
  {year} {2015})}\BibitemShut {NoStop}%
\bibitem [{\citenamefont {Kolmogorov}\ and\ \citenamefont {Crespi}(2005)}]{KC}%
  \BibitemOpen
  \bibfield  {author} {\bibinfo {author} {\bibfnamefont {A.~N.}\ \bibnamefont
  {Kolmogorov}}\ and\ \bibinfo {author} {\bibfnamefont {V.~H.}\ \bibnamefont
  {Crespi}},\ }\href@noop {} {\bibfield  {journal} {\bibinfo  {journal} {Phys.
  Rev. B}\ }\textbf {\bibinfo {volume} {71}},\ \bibinfo {pages} {235415}
  (\bibinfo {year} {2005})}\BibitemShut {NoStop}%
\bibitem [{\citenamefont {ans Zachary A. H.~Goodwin}\ \emph
  {et~al.}(2020)\citenamefont {ans Zachary A. H.~Goodwin}, \citenamefont
  {Vitale}, \citenamefont {Corsetti}, \citenamefont {Mostofi},\ and\
  \citenamefont {Lischner}}]{Xia2020}%
  \BibitemOpen
  \bibfield  {author} {\bibinfo {author} {\bibfnamefont {X.~L.}\ \bibnamefont
  {ans Zachary A. H.~Goodwin}}, \bibinfo {author} {\bibfnamefont
  {V.}~\bibnamefont {Vitale}}, \bibinfo {author} {\bibfnamefont
  {F.}~\bibnamefont {Corsetti}}, \bibinfo {author} {\bibfnamefont {A.~A.}\
  \bibnamefont {Mostofi}}, \ and\ \bibinfo {author} {\bibfnamefont
  {J.}~\bibnamefont {Lischner}},\ }\href@noop {} {\bibfield  {journal}
  {\bibinfo  {journal} {Phys. Rev. B}\ }\textbf {\bibinfo {volume} {102}},\
  \bibinfo {pages} {155146} (\bibinfo {year} {2020})}\BibitemShut {NoStop}%
\bibitem [{\citenamefont {Slater}\ and\ \citenamefont {Koster}(1954)}]{SK}%
  \BibitemOpen
  \bibfield  {author} {\bibinfo {author} {\bibfnamefont {J.~C.}\ \bibnamefont
  {Slater}}\ and\ \bibinfo {author} {\bibfnamefont {G.~F.}\ \bibnamefont
  {Koster}},\ }\href@noop {} {\bibfield  {journal} {\bibinfo  {journal} {Phys.
  Rev.}\ }\textbf {\bibinfo {volume} {94}},\ \bibinfo {pages} {1498} (\bibinfo
  {year} {1954})}\BibitemShut {NoStop}%
\bibitem [{\citenamefont {Neto}\ \emph {et~al.}(2009)\citenamefont {Neto},
  \citenamefont {Guinea}, \citenamefont {Peres}, \citenamefont {Novoselov},\
  and\ \citenamefont {Geim}}]{EPG}%
  \BibitemOpen
  \bibfield  {author} {\bibinfo {author} {\bibfnamefont {A.~H.~C.}\
  \bibnamefont {Neto}}, \bibinfo {author} {\bibfnamefont {F.}~\bibnamefont
  {Guinea}}, \bibinfo {author} {\bibfnamefont {N.~M.~R.}\ \bibnamefont
  {Peres}}, \bibinfo {author} {\bibfnamefont {K.~S.}\ \bibnamefont
  {Novoselov}}, \ and\ \bibinfo {author} {\bibfnamefont {A.~K.}\ \bibnamefont
  {Geim}},\ }\href@noop {} {\bibfield  {journal} {\bibinfo  {journal} {Rev.
  Mod. Phys.}\ }\textbf {\bibinfo {volume} {81}},\ \bibinfo {pages} {109}
  (\bibinfo {year} {2009})}\BibitemShut {NoStop}%
\bibitem [{\citenamefont {Angeli}\ \emph {et~al.}(2018)\citenamefont {Angeli},
  \citenamefont {Mandelli}, \citenamefont {Valli}, \citenamefont {Amaricci},
  \citenamefont {Capone}, \citenamefont {Tosatti},\ and\ \citenamefont
  {Fabrizio}}]{EDS}%
  \BibitemOpen
  \bibfield  {author} {\bibinfo {author} {\bibfnamefont {M.}~\bibnamefont
  {Angeli}}, \bibinfo {author} {\bibfnamefont {D.}~\bibnamefont {Mandelli}},
  \bibinfo {author} {\bibfnamefont {A.}~\bibnamefont {Valli}}, \bibinfo
  {author} {\bibfnamefont {A.}~\bibnamefont {Amaricci}}, \bibinfo {author}
  {\bibfnamefont {M.}~\bibnamefont {Capone}}, \bibinfo {author} {\bibfnamefont
  {E.}~\bibnamefont {Tosatti}}, \ and\ \bibinfo {author} {\bibfnamefont
  {M.}~\bibnamefont {Fabrizio}},\ }\href@noop {} {\bibfield  {journal}
  {\bibinfo  {journal} {Phys. Rev. B}\ }\textbf {\bibinfo {volume} {98}},\
  \bibinfo {pages} {235137} (\bibinfo {year} {2018})}\BibitemShut {NoStop}%
\bibitem [{\citenamefont {Goodwin}\ \emph
  {et~al.}(2020{\natexlab{b}})\citenamefont {Goodwin}, \citenamefont {Vitale},
  \citenamefont {Corsetti}, \citenamefont {Efetov}, \citenamefont {Mostofi},\
  and\ \citenamefont {Lischner}}]{PHD_3}%
  \BibitemOpen
  \bibfield  {author} {\bibinfo {author} {\bibfnamefont {Z.~A.~H.}\
  \bibnamefont {Goodwin}}, \bibinfo {author} {\bibfnamefont {V.}~\bibnamefont
  {Vitale}}, \bibinfo {author} {\bibfnamefont {F.}~\bibnamefont {Corsetti}},
  \bibinfo {author} {\bibfnamefont {D.}~\bibnamefont {Efetov}}, \bibinfo
  {author} {\bibfnamefont {A.~A.}\ \bibnamefont {Mostofi}}, \ and\ \bibinfo
  {author} {\bibfnamefont {J.}~\bibnamefont {Lischner}},\ }\href@noop {}
  {\bibfield  {journal} {\bibinfo  {journal} {Phys. Rev. B}\ }\textbf {\bibinfo
  {volume} {101}},\ \bibinfo {pages} {165110} (\bibinfo {year}
  {2020}{\natexlab{b}})}\BibitemShut {NoStop}%
\bibitem [{\citenamefont {Stepanov}\ \emph {et~al.}(2020)\citenamefont
  {Stepanov}, \citenamefont {Das}, \citenamefont {Lu}, \citenamefont
  {Fahimniya}, \citenamefont {Watanabe}, \citenamefont {Taniguchi},
  \citenamefont {Koppens}, \citenamefont {Lischner}, \citenamefont {Levitov},\
  and\ \citenamefont {Efetov}}]{Stepanov2020untying}%
  \BibitemOpen
  \bibfield  {author} {\bibinfo {author} {\bibfnamefont {P.}~\bibnamefont
  {Stepanov}}, \bibinfo {author} {\bibfnamefont {I.}~\bibnamefont {Das}},
  \bibinfo {author} {\bibfnamefont {X.}~\bibnamefont {Lu}}, \bibinfo {author}
  {\bibfnamefont {A.}~\bibnamefont {Fahimniya}}, \bibinfo {author}
  {\bibfnamefont {K.}~\bibnamefont {Watanabe}}, \bibinfo {author}
  {\bibfnamefont {T.}~\bibnamefont {Taniguchi}}, \bibinfo {author}
  {\bibfnamefont {F.~H.}\ \bibnamefont {Koppens}}, \bibinfo {author}
  {\bibfnamefont {J.}~\bibnamefont {Lischner}}, \bibinfo {author}
  {\bibfnamefont {L.}~\bibnamefont {Levitov}}, \ and\ \bibinfo {author}
  {\bibfnamefont {D.~K.}\ \bibnamefont {Efetov}},\ }\href@noop {} {\bibfield
  {journal} {\bibinfo  {journal} {Nature}\ }\textbf {\bibinfo {volume} {583}},\
  \bibinfo {pages} {375} (\bibinfo {year} {2020})}\BibitemShut {NoStop}%
\bibitem [{\citenamefont {Saito}\ \emph {et~al.}(2020)\citenamefont {Saito},
  \citenamefont {Ge}, \citenamefont {Watanabe}, \citenamefont {Taniguchi},\
  and\ \citenamefont {Young}}]{Saito2020independent}%
  \BibitemOpen
  \bibfield  {author} {\bibinfo {author} {\bibfnamefont {Y.}~\bibnamefont
  {Saito}}, \bibinfo {author} {\bibfnamefont {J.}~\bibnamefont {Ge}}, \bibinfo
  {author} {\bibfnamefont {K.}~\bibnamefont {Watanabe}}, \bibinfo {author}
  {\bibfnamefont {T.}~\bibnamefont {Taniguchi}}, \ and\ \bibinfo {author}
  {\bibfnamefont {A.~F.}\ \bibnamefont {Young}},\ }\href@noop {} {\bibfield
  {journal} {\bibinfo  {journal} {Nat. Phys.}\ }\textbf {\bibinfo {volume}
  {16}},\ \bibinfo {pages} {926} (\bibinfo {year} {2020})}\BibitemShut
  {NoStop}%
\bibitem [{\citenamefont {Throckmorton}\ and\ \citenamefont
  {Vafek}(2012)}]{MGS}%
  \BibitemOpen
  \bibfield  {author} {\bibinfo {author} {\bibfnamefont {R.~E.}\ \bibnamefont
  {Throckmorton}}\ and\ \bibinfo {author} {\bibfnamefont {O.}~\bibnamefont
  {Vafek}},\ }\href@noop {} {\bibfield  {journal} {\bibinfo  {journal} {Phys.
  Rev. B}\ }\textbf {\bibinfo {volume} {86}},\ \bibinfo {pages} {115447}
  (\bibinfo {year} {2012})}\BibitemShut {NoStop}%
\bibitem [{\citenamefont {Goodwin}\ \emph {et~al.}(2019)\citenamefont
  {Goodwin}, \citenamefont {Corsetti}, \citenamefont {Mostofi},\ and\
  \citenamefont {Lischner}}]{PHD_1}%
  \BibitemOpen
  \bibfield  {author} {\bibinfo {author} {\bibfnamefont {Z.~A.~H.}\
  \bibnamefont {Goodwin}}, \bibinfo {author} {\bibfnamefont {F.}~\bibnamefont
  {Corsetti}}, \bibinfo {author} {\bibfnamefont {A.~A.}\ \bibnamefont
  {Mostofi}}, \ and\ \bibinfo {author} {\bibfnamefont {J.}~\bibnamefont
  {Lischner}},\ }\href@noop {} {\bibfield  {journal} {\bibinfo  {journal}
  {Phys. Rev. B}\ }\textbf {\bibinfo {volume} {100}},\ \bibinfo {pages}
  {121106(R)} (\bibinfo {year} {2019})}\BibitemShut {NoStop}%
\bibitem [{\citenamefont {Laturia}\ \emph {et~al.}(2018)\citenamefont
  {Laturia}, \citenamefont {de~Put},\ and\ \citenamefont
  {Vandenberghe}}]{Laturia2018}%
  \BibitemOpen
  \bibfield  {author} {\bibinfo {author} {\bibfnamefont {A.}~\bibnamefont
  {Laturia}}, \bibinfo {author} {\bibfnamefont {M.~L.~V.}\ \bibnamefont
  {de~Put}}, \ and\ \bibinfo {author} {\bibfnamefont {W.~G.}\ \bibnamefont
  {Vandenberghe}},\ }\href@noop {} {\bibfield  {journal} {\bibinfo  {journal}
  {npj 2D Materials and Applications}\ }\textbf {\bibinfo {volume} {2}},\
  \bibinfo {pages} {6} (\bibinfo {year} {2018})}\BibitemShut {NoStop}%
\bibitem [{\citenamefont {Cea}\ \emph {et~al.}(2020)\citenamefont {Cea},
  \citenamefont {Pantale\'{o}n},\ and\ \citenamefont {Guinea}}]{Cea2020hBN}%
  \BibitemOpen
  \bibfield  {author} {\bibinfo {author} {\bibfnamefont {T.}~\bibnamefont
  {Cea}}, \bibinfo {author} {\bibfnamefont {P.~A.}\ \bibnamefont
  {Pantale\'{o}n}}, \ and\ \bibinfo {author} {\bibfnamefont {F.}~\bibnamefont
  {Guinea}},\ }\href@noop {} {\bibfield  {journal} {\bibinfo  {journal} {Phys.
  Rev. B}\ }\textbf {\bibinfo {volume} {102}},\ \bibinfo {pages} {155136}
  (\bibinfo {year} {2020})}\BibitemShut {NoStop}%
\bibitem [{\citenamefont {Wehling}\ \emph {et~al.}(2011)\citenamefont
  {Wehling}, \citenamefont {\c{S}a\c{s}ıo\u{g}lu}, \citenamefont {Friedrich},
  \citenamefont {Lichtenstein}, \citenamefont {Katsnelson},\ and\ \citenamefont
  {Bl\"{u}gel}}]{SECI}%
  \BibitemOpen
  \bibfield  {author} {\bibinfo {author} {\bibfnamefont {T.~O.}\ \bibnamefont
  {Wehling}}, \bibinfo {author} {\bibfnamefont {E.}~\bibnamefont
  {\c{S}a\c{s}ıo\u{g}lu}}, \bibinfo {author} {\bibfnamefont {C.}~\bibnamefont
  {Friedrich}}, \bibinfo {author} {\bibfnamefont {A.~I.}\ \bibnamefont
  {Lichtenstein}}, \bibinfo {author} {\bibfnamefont {M.~I.}\ \bibnamefont
  {Katsnelson}}, \ and\ \bibinfo {author} {\bibfnamefont {S.}~\bibnamefont
  {Bl\"{u}gel}},\ }\href@noop {} {\bibfield  {journal} {\bibinfo  {journal}
  {Phys. Rev. Lett.}\ }\textbf {\bibinfo {volume} {106}},\ \bibinfo {pages}
  {236805} (\bibinfo {year} {2011})}\BibitemShut {NoStop}%
\bibitem [{\citenamefont {Prentice}(2020)}]{onetep_2020}%
  \BibitemOpen
  \bibfield  {author} {\bibinfo {author} {\bibfnamefont {J.~C. A.~t.}\
  \bibnamefont {Prentice}},\ }\href@noop {} {\bibfield  {journal} {\bibinfo
  {journal} {J. Chem. Phys.}\ }\textbf {\bibinfo {volume} {152}},\ \bibinfo
  {pages} {174111} (\bibinfo {year} {2020})}\BibitemShut {NoStop}%
\bibitem [{\citenamefont {Ratcliff}\ \emph {et~al.}(2018)\citenamefont
  {Ratcliff}, \citenamefont {Conduit}, \citenamefont {Hine},\ and\
  \citenamefont {Haynes}}]{Ratclif_PRB98_2018}%
  \BibitemOpen
  \bibfield  {author} {\bibinfo {author} {\bibfnamefont {L.~E.}\ \bibnamefont
  {Ratcliff}}, \bibinfo {author} {\bibfnamefont {G.~J.}\ \bibnamefont
  {Conduit}}, \bibinfo {author} {\bibfnamefont {N.~D.~M.}\ \bibnamefont
  {Hine}}, \ and\ \bibinfo {author} {\bibfnamefont {P.~D.}\ \bibnamefont
  {Haynes}},\ }\href@noop {} {\bibfield  {journal} {\bibinfo  {journal} {Phys.
  Rev. B}\ }\textbf {\bibinfo {volume} {98}},\ \bibinfo {pages} {125123}
  (\bibinfo {year} {2018})}\BibitemShut {NoStop}%
\bibitem [{\citenamefont {Perdew}\ \emph {et~al.}(1996)\citenamefont {Perdew},
  \citenamefont {Burke},\ and\ \citenamefont {Ernzerhof}}]{PBE_PRL77}%
  \BibitemOpen
  \bibfield  {author} {\bibinfo {author} {\bibfnamefont {J.~P.}\ \bibnamefont
  {Perdew}}, \bibinfo {author} {\bibfnamefont {K.}~\bibnamefont {Burke}}, \
  and\ \bibinfo {author} {\bibfnamefont {M.}~\bibnamefont {Ernzerhof}},\
  }\href@noop {} {\bibfield  {journal} {\bibinfo  {journal} {Phys. Rev. Lett.}\
  }\textbf {\bibinfo {volume} {77}},\ \bibinfo {pages} {3865} (\bibinfo {year}
  {1996})}\BibitemShut {NoStop}%
\bibitem [{\citenamefont {Bl\"ochl}(1994)}]{PAW_PRB50}%
  \BibitemOpen
  \bibfield  {author} {\bibinfo {author} {\bibfnamefont {P.~E.}\ \bibnamefont
  {Bl\"ochl}},\ }\href@noop {} {\bibfield  {journal} {\bibinfo  {journal}
  {Phys. Rev. B}\ }\textbf {\bibinfo {volume} {50}},\ \bibinfo {pages} {17953}
  (\bibinfo {year} {1994})}\BibitemShut {NoStop}%
\bibitem [{\citenamefont {Jollet}\ \emph {et~al.}(2014)\citenamefont {Jollet},
  \citenamefont {Torrent},\ and\ \citenamefont {Holzwarth}}]{JOLLET20141246}%
  \BibitemOpen
  \bibfield  {author} {\bibinfo {author} {\bibfnamefont {F.}~\bibnamefont
  {Jollet}}, \bibinfo {author} {\bibfnamefont {M.}~\bibnamefont {Torrent}}, \
  and\ \bibinfo {author} {\bibfnamefont {N.}~\bibnamefont {Holzwarth}},\
  }\href@noop {} {\bibfield  {journal} {\bibinfo  {journal} {Comput. Phys.
  Commun.}\ }\textbf {\bibinfo {volume} {185}},\ \bibinfo {pages} {1246 }
  (\bibinfo {year} {2014})}\BibitemShut {NoStop}%
\bibitem [{\citenamefont {Garrity}\ \emph {et~al.}(2014)\citenamefont
  {Garrity}, \citenamefont {Bennett}, \citenamefont {Rabe},\ and\ \citenamefont
  {Vanderbilt}}]{GARRITY2014446}%
  \BibitemOpen
  \bibfield  {author} {\bibinfo {author} {\bibfnamefont {K.~F.}\ \bibnamefont
  {Garrity}}, \bibinfo {author} {\bibfnamefont {J.~W.}\ \bibnamefont
  {Bennett}}, \bibinfo {author} {\bibfnamefont {K.~M.}\ \bibnamefont {Rabe}}, \
  and\ \bibinfo {author} {\bibfnamefont {D.}~\bibnamefont {Vanderbilt}},\
  }\href@noop {} {\bibfield  {journal} {\bibinfo  {journal} {Computational
  Materials Science}\ }\textbf {\bibinfo {volume} {81}},\ \bibinfo {pages}
  {446} (\bibinfo {year} {2014})}\BibitemShut {NoStop}%
\bibitem [{\citenamefont {Ruiz-Serrano}\ and\ \citenamefont
  {Skylaris}(2013)}]{Serrano_JCP139_2013}%
  \BibitemOpen
  \bibfield  {author} {\bibinfo {author} {\bibfnamefont {{\`A}.}~\bibnamefont
  {Ruiz-Serrano}}\ and\ \bibinfo {author} {\bibfnamefont {C.-K.}\ \bibnamefont
  {Skylaris}},\ }\href@noop {} {\bibfield  {journal} {\bibinfo  {journal} {J.
  Chem. Phys.}\ }\textbf {\bibinfo {volume} {139}},\ \bibinfo {pages} {054107}
  (\bibinfo {year} {2013})}\BibitemShut {NoStop}%
\bibitem [{\citenamefont {Marzari}\ \emph {et~al.}(1997)\citenamefont
  {Marzari}, \citenamefont {Vanderbilt},\ and\ \citenamefont
  {Payne}}]{Marzari_PRL79_1997}%
  \BibitemOpen
  \bibfield  {author} {\bibinfo {author} {\bibfnamefont {N.}~\bibnamefont
  {Marzari}}, \bibinfo {author} {\bibfnamefont {D.}~\bibnamefont {Vanderbilt}},
  \ and\ \bibinfo {author} {\bibfnamefont {M.~C.}\ \bibnamefont {Payne}},\
  }\href@noop {} {\bibfield  {journal} {\bibinfo  {journal} {Phys. Rev. Lett.}\
  }\textbf {\bibinfo {volume} {79}},\ \bibinfo {pages} {1337} (\bibinfo {year}
  {1997})}\BibitemShut {NoStop}%
\bibitem [{\citenamefont {Lewandowski}\ \emph {et~al.}(2021)\citenamefont
  {Lewandowski}, \citenamefont {Nadj-Perge},\ and\ \citenamefont
  {Chowdhury}}]{Lewandowski2021}%
  \BibitemOpen
  \bibfield  {author} {\bibinfo {author} {\bibfnamefont {C.}~\bibnamefont
  {Lewandowski}}, \bibinfo {author} {\bibfnamefont {S.}~\bibnamefont
  {Nadj-Perge}}, \ and\ \bibinfo {author} {\bibfnamefont {D.}~\bibnamefont
  {Chowdhury}},\ }\href@noop {} {\bibfield  {journal} {\bibinfo  {journal} {npj
  Quantum Mater.}\ }\textbf {\bibinfo {volume} {6}},\ \bibinfo {pages} {82}
  (\bibinfo {year} {2021})}\BibitemShut {NoStop}%
\bibitem [{\citenamefont {Cea}\ and\ \citenamefont {Guinea}(2020)}]{Cea2020}%
  \BibitemOpen
  \bibfield  {author} {\bibinfo {author} {\bibfnamefont {T.}~\bibnamefont
  {Cea}}\ and\ \bibinfo {author} {\bibfnamefont {F.}~\bibnamefont {Guinea}},\
  }\href@noop {} {\bibfield  {journal} {\bibinfo  {journal} {Phys. Rev. B}\
  }\textbf {\bibinfo {volume} {102}},\ \bibinfo {pages} {045107} (\bibinfo
  {year} {2020})}\BibitemShut {NoStop}%
\bibitem [{\citenamefont {Cea}\ and\ \citenamefont {Guinea}(2021)}]{Cea2021}%
  \BibitemOpen
  \bibfield  {author} {\bibinfo {author} {\bibfnamefont {T.}~\bibnamefont
  {Cea}}\ and\ \bibinfo {author} {\bibfnamefont {F.}~\bibnamefont {Guinea}},\
  }\href@noop {} {\bibfield  {journal} {\bibinfo  {journal} {PNAS}\ }\textbf
  {\bibinfo {volume} {118}},\ \bibinfo {pages} {e2107874118} (\bibinfo {year}
  {2021})}\BibitemShut {NoStop}%
\bibitem [{\citenamefont {Fischer}\ \emph {et~al.}(2021)\citenamefont
  {Fischer}, \citenamefont {Goodwin}, \citenamefont {Mostofi}, \citenamefont
  {Lischner}, \citenamefont {Kennes},\ and\ \citenamefont
  {Klebl}}]{Fischer_TTLG}%
  \BibitemOpen
  \bibfield  {author} {\bibinfo {author} {\bibfnamefont {A.}~\bibnamefont
  {Fischer}}, \bibinfo {author} {\bibfnamefont {Z.~A.~H.}\ \bibnamefont
  {Goodwin}}, \bibinfo {author} {\bibfnamefont {A.~A.}\ \bibnamefont
  {Mostofi}}, \bibinfo {author} {\bibfnamefont {J.}~\bibnamefont {Lischner}},
  \bibinfo {author} {\bibfnamefont {D.~M.}\ \bibnamefont {Kennes}}, \ and\
  \bibinfo {author} {\bibfnamefont {L.}~\bibnamefont {Klebl}},\ }\href@noop {}
  {\bibfield  {journal} {\bibinfo  {journal} {arXiv:2104.10176}\ } (\bibinfo
  {year} {2021})}\BibitemShut {NoStop}%
\bibitem [{\citenamefont {Pantale\'on}\ \emph {et~al.}(2021)\citenamefont
  {Pantale\'on}, \citenamefont {Cea}, \citenamefont {Walet},\ and\
  \citenamefont {Guinea}}]{Pierre2020}%
  \BibitemOpen
  \bibfield  {author} {\bibinfo {author} {\bibfnamefont {P.~A.}\ \bibnamefont
  {Pantale\'on}}, \bibinfo {author} {\bibfnamefont {T.}~\bibnamefont {Cea}},
  \bibinfo {author} {\bibfnamefont {R.~B. N.~R.}\ \bibnamefont {Walet}}, \ and\
  \bibinfo {author} {\bibfnamefont {F.}~\bibnamefont {Guinea}},\ }\href@noop {}
  {\bibfield  {journal} {\bibinfo  {journal} {2D Mater}\ }\textbf {\bibinfo
  {volume} {8}},\ \bibinfo {pages} {044006} (\bibinfo {year}
  {2021})}\BibitemShut {NoStop}%
\bibitem [{\citenamefont {Choi}\ \emph {et~al.}(2021)\citenamefont {Choi},
  \citenamefont {Kim}, \citenamefont {Lewandowski}, \citenamefont {Peng},
  \citenamefont {Thomson}, \citenamefont {Polski}, \citenamefont {Zhang},
  \citenamefont {Watanabe}, \citenamefont {Taniguchi}, \citenamefont {Alicea},\
  and\ \citenamefont {Nadj-Perge}}]{Choi2021driven}%
  \BibitemOpen
  \bibfield  {author} {\bibinfo {author} {\bibfnamefont {Y.}~\bibnamefont
  {Choi}}, \bibinfo {author} {\bibfnamefont {H.}~\bibnamefont {Kim}}, \bibinfo
  {author} {\bibfnamefont {C.}~\bibnamefont {Lewandowski}}, \bibinfo {author}
  {\bibfnamefont {Y.}~\bibnamefont {Peng}}, \bibinfo {author} {\bibfnamefont
  {A.}~\bibnamefont {Thomson}}, \bibinfo {author} {\bibfnamefont
  {R.}~\bibnamefont {Polski}}, \bibinfo {author} {\bibfnamefont
  {Y.}~\bibnamefont {Zhang}}, \bibinfo {author} {\bibfnamefont
  {K.}~\bibnamefont {Watanabe}}, \bibinfo {author} {\bibfnamefont
  {T.}~\bibnamefont {Taniguchi}}, \bibinfo {author} {\bibfnamefont
  {J.}~\bibnamefont {Alicea}}, \ and\ \bibinfo {author} {\bibfnamefont
  {S.}~\bibnamefont {Nadj-Perge}},\ }\href@noop {} {\bibfield  {journal}
  {\bibinfo  {journal} {Nat. Phys.}\ }\textbf {\bibinfo {volume} {17}},\
  \bibinfo {pages} {1375} (\bibinfo {year} {2021})}\BibitemShut {NoStop}%
\bibitem [{\citenamefont {Santos}\ and\ \citenamefont {Kaxiras}(
  902)}]{Santos2013}%
  \BibitemOpen
  \bibfield  {author} {\bibinfo {author} {\bibfnamefont {E.~J.}\ \bibnamefont
  {Santos}}\ and\ \bibinfo {author} {\bibfnamefont {E.}~\bibnamefont
  {Kaxiras}},\ }\href@noop {} {\bibfield  {journal} {\bibinfo  {journal} {Nano
  Lett.}\ }\textbf {\bibinfo {volume} {13}},\ \bibinfo {pages} {2013} (\bibinfo
  {year} {898--902})}\BibitemShut {NoStop}%
\bibitem [{\citenamefont {Tepliakov}\ \emph {et~al.}(2021)\citenamefont
  {Tepliakov}, \citenamefont {Wu},\ and\ \citenamefont {Yazyev}}]{Nikita2021}%
  \BibitemOpen
  \bibfield  {author} {\bibinfo {author} {\bibfnamefont {N.~V.}\ \bibnamefont
  {Tepliakov}}, \bibinfo {author} {\bibfnamefont {Q.}~\bibnamefont {Wu}}, \
  and\ \bibinfo {author} {\bibfnamefont {O.~V.}\ \bibnamefont {Yazyev}},\
  }\href@noop {} {\bibfield  {journal} {\bibinfo  {journal} {Nano Lett.}\
  }\textbf {\bibinfo {volume} {21}},\ \bibinfo {pages} {4636} (\bibinfo {year}
  {2021})}\BibitemShut {NoStop}%
\end{thebibliography}%

\end{document}